\documentclass[11pt]{article}

\usepackage{amsmath}
\usepackage{graphicx}
\usepackage{indentfirst}
\usepackage{amssymb}
\usepackage{cite}
\usepackage{color}
\usepackage{subfigure}
\usepackage{mathrsfs}
\usepackage{varwidth}
\usepackage{subfigure}
\usepackage[colorlinks=true, linkcolor=red, citecolor=blue, urlcolor=magenta]{hyperref}
\usepackage{xcolor}
\usepackage{bm}

\begin{document}
\title{Optical images of Kerr-Sen black hole illuminated by thick accretion disks}

\date{}
\maketitle

\begin{center}
	\author{Yu-Kang Wang,}$^{a}$\footnote{E-mail: yukang\_\_wang@163.com}
	\author{Chen-Yu Yang,}$^{b}$\footnote{E-mail: chenyu\_yang2024@163.com}
	\author{Xiao-Xiong Zeng}$^{c}$\footnote{E-mail: xxzengphysics@163.com (Corresponding author)}
	\\
	
	\vskip 0.25in
	$^{a}$\it{School of Material Science and Engineering, Chongqing Jiaotong University, Chongqing 400074, People's Republic of China}\\
	$^{b}$\it{Department of Mechanics, Chongqing Jiaotong University, Chongqing 4000, People's Republic of China}\\
	$^{c}$\it{College of Physics and Optoelectronic Engineering, Chongqing Normal University, Chongqing 401331, People's Republic of China}\\
\end{center}
\vskip 0.6in
{\abstract
	{This paper investigates the shadow and polarization images of a Kerr-Sen black hole illuminated by geometrically thick and optically thin accretion disks. We adopt two classes of accretion models, namely the phenomenological radiatively inefficient accretion flow (RIAF) model and the analytical ballistic approximation accretion flow (BAAF) model. Based on radiative transfer theory, we examine the effects of the spin parameter $a$, black hole charge $Q$, and observer inclination angle $\theta$ on the shadow images. Both models show that, as the charge $Q$ increases, the photon rings and the central dark regions shrink simultaneously. Meanwhile, frame dragging gives rise to a pronounced brightness asymmetry, which becomes more significant with increasing $a$ and $\theta$. The main difference between isotropic and anisotropic radiation is that, in the latter case, the higher order images are brighter in the upper and lower polar regions. For the BAAF model, because the conical approximation renders certain regions geometrically thinner, the spatial extent of the higher order images is narrower than that in the RIAF model, and the separation between the direct image and the higher order images is more distinct. In the polarization images, the spatial distribution of the polarization vector directions is mainly determined by gravitational lensing and frame dragging, whereas the intensity near the photon ring and the scale of the higher order images are significantly influenced by $Q$.
}}

\thispagestyle{empty}
\newpage
\setcounter{page}{1}

\tableofcontents
\newpage

\section{Introduction}
Since Schwarzschild presented the first exact solution of general relativity in 1915~\cite{Schwarzschild}, the study of black holes has continued to advance. Black holes reside in extreme gravitational environments and exhibit remarkable physical properties that often defy intuition. Investigating such extreme astrophysical objects is of fundamental importance for the development of both theoretical physics and observational astronomy.

With the advancement of astronomical observational techniques, black hole shadows have become a central topic of interest in contemporary physics. The  black hole shadow is a dark region formed by the deflection of light in the strong gravitational field of a black hole. Its shape and size are primarily determined by the underlying spacetime geometry, allowing one to infer the physical properties of black holes through detailed analysis of shadow morphology and scale. Continued progress in experimental physics and observational astronomy has further enriched our understanding of black holes. Extensive studies have been carried out on the size, shape, and observational signatures of shadows associated with various black hole spacetimes and exotic compact objects, both within general relativity and in modified gravity theories~\cite{2024saq,Li:2022eue,Yunusov:2024xzu,He:2024qka,Dey:2020bgo,Ali:2025beh,Luo:2024nul}. In particular, since the Event Horizon Telescope (EHT) collaboration released the first image of a black hole in 2019~\cite{EventHorizonTelescope:2019dse,EventHorizonTelescope:2022wkp,Lu:2023bbn}, observations of supermassive black holes have opened new avenues for exploring black hole physics and advanced the field into a new stage. These developments have further revealed that black hole shadow images are influenced not only by gravity but also by the surrounding accreting matter. Beyond EHT observations, other research groups have also employed diverse methods to investigate black holes. In 2025, the National Astronomical Observatories of China reported a rapid and luminous X-ray transient, EP250702a, detected by the Einstein Probe mission, which was interpreted as a tidal disruption event involving a white dwarf and an intermediate-mass black hole~\cite{EinsteinProbe}. These observational advances have significantly stimulated interest in accretion flow imaging, particularly in systematic studies of polarized radiation within general relativity~\cite{Hou:2023bep,Zhang:2024lsf,Zhang:2025} and in various modified gravity frameworks~\cite{Yang:2025usj,Zeng:2025pch,Wang2025,zeng2025pnb,Yang2025sap,Chen:2025ysv,Li:2025knj,Hou:2022eev,wang2025oio,Wan:2025gbm}.

It is well known that when matter is captured by the gravitational field of a rotating supermassive black hole, frame dragging effects force the particles to corotate with the black hole. As a result, the infalling matter forms a hot, magnetized plasma that emits thermal synchrotron radiation, giving rise to a bright accretion disk. For rapidly spinning black holes, electromagnetic extraction of energy may further power relativistic jets~\cite{1977Electromagnetic}, commonly referred to as funnel wall jets. The base of the jet is surrounded by the jet sheath and produces significant thermal synchrotron emission, so that the observed radiation in black hole images may originate from both the accretion flow and the jet. Substantial progress has been made in the study of black hole shadows, including spherical accretion models~\cite{2019tso,2020sap,Heydari_Fard_2023}, optically and geometrically thin disk models~\cite{2024Image,he2024,Zeng:2021dlj,2021sar,Yang:2026bgi,Yang:2024nin}, and   holographic Einstein rings~\cite{2020ibh,2024her,2025sab}. In particular, since 2019, EHT observations have been widely used to constrain black hole parameters and test accretion flow models~\cite{hio,2024oao,2024ioq,Hou_2025,2024iaf}. Although accretion disks are often assumed to be optically and geometrically thin, EHT observations suggest that accretion flows may become geometrically thick due to suppressed vertical cooling and matter compression~\cite{EventHorizonTelescope:2019dse,EventHorizonTelescope:2022wkp,EventHorizonTelescope:2019pgp,Ho:1999ss,Narayan_1994}. Motivated by this scenario, the present work investigates the impact of thick accretion flows on shadow images by incorporating the electron number density, electron temperature, and magnetic field structure. In addition to the standard RIAF model, we also consider a recently proposed accretion flow model by Hou et al.~\cite{Hou:2023bep,Zhang:2024lsf}, namely the BAAF model. This model assumes that gravity dominates the fluid acceleration near the event horizon and provides explicit expressions for thermodynamic quantities and magnetic field configurations. Therefore, it offers a physically consistent description of the morphology and dynamics of geometrically thick accretion flows and facilitates the study of near horizon polarization properties.

The synchrotron radiation emitted by electrons in black hole accretion flows exhibits distinct polarization signatures. The polarization vector is orthogonal to both the magnetic field and the wave vector, and aligned with the electric field vector. In the strong gravitational field of a black hole, the polarization vector undergoes parallel transport along null geodesics and is subsequently projected onto the observer’s celestial sphere, forming the polarized image of the accretion flow on the observer’s image plane. Polarimetric observations thus provide valuable insights into the plasma dynamics and magnetic field structure in the vicinity of black holes~\cite{2024ioq}. Recent polarization images released by EHT have revealed prominent polarization structures within emission rings. In particular, the linear polarization maps display a characteristic spiral pattern in the electric vector position angles (EVPAs)~\cite{EventHorizonTelescope:2021bee,EventHorizonTelescope:2024hpu}. In 2021, EHT analyses reproduced both the EVPA distribution and the relative polarization intensity of M87* using the approximate analytic ray-tracing formalism developed by Beloborodov~\cite{Beloborodov2002,2021tpi}. Polarization signatures of other classes of compact objects, including alternative black hole solutions and horizonless ultra compact objects, have also been widely investigated~\cite{Yin:2025rao,2021Polarized,2024Polarized,deliyski2023pio,zeng2025pio}. These studies contribute to a deeper understanding of accretion physics and the underlying spacetime geometry.

Advances in experimental and observational techniques now enable increasingly precise and high resolution tests of general relativity~\cite{Abbott_2020,2020Gravitational}, while also motivating extensions of general relativity and the exploration of alternative theories of gravity. In this context, Sen constructed in 1992 a four dimensional rotating charged black hole solution by applying solution generating techniques to the Kerr metric, namely the Kerr–Sen black hole~\cite{Sen1992rcb,Sen1994bhs}.  The Kerr–Sen black hole is a nonvacuum and non-algebraically special solution. In addition to the metric, it includes three non-gravitational fields: an antisymmetric rank-3 tensor field, an Abelian gauge field, and a dilaton scalar field, with nontrivial coupling between the latter two~\cite{2016dkb,2013hag}. Though both Kerr–Sen and Kerr–Newman black holes carry electric charge, their physical origins differ fundamentally. The charge of a Kerr–Newman black hole arises from the classical Maxwell field, whereas the charge of a Kerr–Sen black hole originates from a $U(1)$ gauge field in string theory and is intrinsically coupled to a dilaton field. Now, Various properties of Kerr–Sen black holes have been extensively studied~\cite{wang2025oio,Larranaga2011,Chen2010,zhang2024tto}. In this work, we investigate the optical appearance of a Kerr–Sen black hole illuminated by a thick accretion flow. We   adopt   the RIAF model and  BAAF model. The images are obtained using general relativistic radiative transfer (GRRT) equation. We analyze the effects of the charge parameter, observer inclination, and spin parameter on the black hole shadows. Within the BAAF framework, we further examine polarization properties and explore how these parameters influence the polarization structure near the event horizon.

This paper is organized as follows. In Sec. 2, we briefly review the Kerr–Sen spacetime  and present the null geodesic equations. Sec. 3 and Sec. 4 describe the electron emission models and accretion flow models, respectively. The emission models include both isotropic and anisotropic radiation, while the accretion models consist of the RIAF and BAAF frameworks. In Sec. 5, we outline the theoretical framework for polarization imaging within the BAAF model. Numerical results are presented and analyzed in Sec. 6. Finally, Sec. 7 summarizes our conclusions and discusses. In this work, we have set the fundamental constants $c$ and $G$ to unity, where $c$ is the speed of light in vacuum and $G$ is the gravitational constant, unless otherwise specified, we will work in the convention $(-,+,+,+)$.

\section{The Kerr-Sen Spacetime and Photon Sphere}\label{sec2}
We consider the Kerr-Sen  spacetime, whose axisymmetric rotating black hole solution in Boyer--Lindquist coordinates can be written as~\cite{kerr-sen1,kerr-sen2}
\begin{align}
	ds^{2} = & -\left(1 - \frac{2Mr}{\Sigma}\right)dt^{2} 
	+ \frac{\Sigma}{\Delta}dr^{2} 
	+ \Sigma d\theta^{2} 
	- \frac{4aMr}{\Sigma}\sin^{2}\theta \, dt \, d\phi \nonumber\\
	& + \sin^{2}\theta\left[ r(r + r_{0}) + a^{2} 
	+ \frac{2Mra^{2}\sin^{2}\theta}{\Sigma} \right] d\phi^{2},
	\label{kerr-sen}
\end{align}
where
\begin{align}
	\Sigma &= r(r+r_{0}) + a^{2}\cos^{2}\theta, \quad
	\Delta = r(r+r_{0}) - 2Mr + a^{2},
\end{align}
and
\begin{align}
	r_{0} = \frac{Q^{2}}{M}.
\end{align}
Here $M$ is the black hole mass, $a$ is the spin parameter, $r_{0}$ is the expansion parameter, and $Q$ is the black hole charge. When $Q=0$, the solution reduces to the Kerr black hole. The event horizons are determined by $\Delta/\Sigma=0$, i.e.
\begin{align}
	r(r+r_{0}) - 2Mr + a^{2} = 0,
\end{align}
which yields
\begin{align}
	r_h = M - \frac{r_{0}}{2} + \sqrt{\left(M - \frac{r_{0}}{2}\right)^{2} - a^{2}} ,\\
	r_c = M - \frac{r_{0}}{2} - \sqrt{\left(M - \frac{r_{0}}{2}\right)^{2} - a^{2}} .
\end{align}
The $r_h$ and $r_c$ correspond to the event horizon and the Cauchy horizon, respectively. In addition, the rotating Kerr-Sen black hole ~\eqref{kerr-sen} possesses time translational and rotational invariance isometries, implying the existence of two Killing vector fields, $\left(\frac{\partial}{\partial t}\right)^{\mu}$ and $\left(\frac{\partial}{\partial \phi}\right)^{\mu}$. To facilitate black hole shadow imaging, we adopt the zero angular momentum observer (ZAMO), defined as a stationary observer with vanishing angular momentum at infinity. However, due to frame dragging, the ZAMO still has a position dependent angular velocity $\omega$. In the Kerr-Sen spacetime, it is given by
\begin{align}
	\omega = \frac{d\phi}{dt} = -\frac{g_{t\phi}}{g_{\phi\phi}} = \frac{2Mar}{\Sigma\left[r(r+r_{0})+a^{2}\right] + 2Mra^{2}\sin^{2}\theta}.
\end{align}
As the observer approaches the black hole, $\omega$ increases and reaches its maximum at the event horizon
\begin{align}
	\Omega = \omega\big|_{r=r_{h}} = \frac{a}{r_{h}(r_{h}+r_{0}) + a^{2}}.
\end{align}
Here $\Omega$ denotes the angular velocity of the black hole. At this location, the observer's angular velocity $\omega$ coincides with $\Omega$, indicating a state of corotation. When $Q=0$, i.e., $r_{0}=0$, the above expression reduces to
\begin{align}
	\Omega = \frac{a}{r_{h}^{2} + a^{2}},
\end{align}
which is identical to the angular velocity of a Kerr black hole~\cite{Chandrasekhar}.

Next, we investigate photon motion in the vicinity of a rotating black hole in the KS spacetime. The null geodesic equations in the spacetime~\eqref{kerr-sen} are given by~\cite{zubair2023}
\begin{align}
	\Sigma\dot{t} &= \frac{E\left[r(r+r_{0})+a^{2}\right]^{2}-2aMrL}{\Delta}-Ea^{2}\sin^{2}\theta, \\
	\Sigma\dot{r} &= \sqrt{\bm{R}(r)},\\
	\Sigma\dot{\theta} &= \sqrt{\bm{\Theta}(\theta)},  \\
	\Sigma\dot{\phi} &= L\csc^{2}\theta+\frac{a(2MrE-aL)}{\Delta}.
\end{align}
where
\begin{align}
	\bm{R}(r) &= \left[r(r+r_{0})+a^{2}\right]^{2}E^{2} + a^{2}L^{2} - \Delta\mathcal{K} - 4aMrEL, \\
	\bm{\Theta}(\theta) &= \mathcal{K} - a^{2}E^{2}\sin^{2}\theta - L^{2}\csc^{2}\theta.
\end{align}
Here $\bm{R}(r)$ and $\bm{\Theta}(\theta)$ denote the radial and angular potentials, respectively. The overdot represents differentiation with respect to an affine parameter. The quantities $E$ and $L$ are conserved quantities associated with the Killing vector fields, and $\mathcal{K}$ is the separation constant. On the equatorial plane,  where $ \theta = \dfrac{\pi}{2}$, the radial equation can be rewritten as $\dot{r}^{2} + V_{\text{eff}}(r)=0$, where the effective potential $V_{\text{eff}}(r)$ is given by
\begin{align}
	V_{\text{eff}}(r) = -\frac{\bm{R}(r)}{[r(r+r_0)]^{2}} = -\frac{\left[r(r+r_{0})+a^{2}\right]^{2}E^{2} + a^{2}L^{2} - \Delta\mathcal{K} - 4aMrEL}{[r(r+r_0)]^{2}} .
\end{align}
For convenience, we introduce the dimensionless impact parameters of photons
\begin{align}
	\xi &\equiv \frac{L}{E}, \quad
	\eta \equiv \frac{\mathcal{K}}{E^2}.
\end{align}
It should be noted that the boundary of the black hole shadow is determined by a luminous spherical region composed of massless particles, namely the photon sphere~\cite{Luminet:1979nyg,1966The,Cunningham1972TheOA}. Photons entering the photon sphere eventually fall into the event horizon, while only those outside it can reach the observer. The photon  sphere radius $r_p$ can be determined from the radial potential and its derivative as
\begin{align}
	\label{photon-sphere-condition}
	\left.\bm{R}(r)\right|_{r=r_{p}}=0,\qquad
	\left.\frac{\partial\bm{R}(r)}{\partial r}\right|_{r=r_{p}}=0.
\end{align}
Considering a photon captured by the black hole, if it lies on a stable orbit, it remains bound for a long time; if it lies on an unstable orbit, it may escape after orbiting the black hole for some time. The unstable orbit must satisfy~\cite{ZUBAIR2023101334}
\begin{align}
	\label{unstable-orbit-condition}
	\left.\frac{\partial^{2}V_{\text{eff}}(r)}{\partial r^{2}}\right|_{r=r_{p}} < 0.
\end{align}
Solving Eqs.~\eqref{photon-sphere-condition} and~\eqref{unstable-orbit-condition}, one obtains the critical impact parameters corresponding to unstable photon orbits
\begin{align}
	\xi_{c} &= a + \frac{r(r+r_{0})}{a}, \\
	\eta_{c} &= 2\left[r(r+r_{0}) + a^{2}\right].
\end{align}
Based on these theoretical foundations, the shadow of this black hole illuminated by a thin accretion disk has been extensively studied~\cite{wang2025oio}. In this work, we focus on the characteristics of the shadow images under illumination by a thick accretion disk and compare them with those obtained by a thin accretion disk.

\section{Electron Radiation Model}\label{sec3}
In the thick accretion flow model, the particle number density, electron temperature, and magnetic field structure are essential quantities. In principle, these physical variables can be obtained by solving the general relativistic magnetohydrodynamic (GRMHD) equations. However, given the substantial analytical complexity of the GRMHD equations, we adopt numerical methods together with appropriate simplifications.

\subsection{Radiative Transfer Equation}
We first consider the unpolarized case. The covariant form of the radiative transfer equation is
\begin{align}
	\label{eq:radiative_transfer}
	\frac{d}{d\lambda} I = J - \alpha I.
\end{align}
Here, $I$, $J$, and $\alpha$ are Lorentz invariant quantities. For an arbitrary observer at a given spacetime point, let $\nu$ denote the photon frequency measured in the observer's frame. The relations between these invariants and the corresponding physical quantities are
\begin{align}
	\label{eq:invariant_relations}
	I = \frac{I_\nu}{\nu^3}, \quad J = \frac{j_\nu}{\nu^2}, \quad \alpha = \nu \alpha_\nu.
\end{align}
Here, $I_\nu$ is the specific intensity, $j_\nu$ is the emissivity, and $\alpha_\nu$ is the absorption coefficient. The solution of Eq.~\eqref{eq:radiative_transfer}   is
\begin{align}
	I(\lambda) = I(\lambda_0) + \int_{\lambda_0}^{\lambda} d\lambda' \, J(\lambda') \exp\left( -\int_{\lambda'}^{\lambda} d\lambda'' \, \alpha(\lambda'') \right).
\end{align}
To ensure consistency with the unit system in Eq.~\eqref{eq:invariant_relations}, we convert the equation into the Centimeter -Gram -Second (CGS) system of units. We rescale the affine parameter $\lambda$ in Eq.~\eqref{eq:radiative_transfer} as $\frac{d}{d\lambda} \rightarrow \frac{1}{C}\frac{d}{d\lambda}$, where $C = \frac{r_g}{\nu_0}$, here, $r_g = \frac{GM}{c^2}$ denotes the gravitational radius, and $\nu_0$ is the photon frequency measured at infinity. The equation then becomes as
\begin{align}
	\frac{1}{C} \frac{d}{d\lambda} I = J - \alpha I,
\end{align}
with the corresponding solution
\begin{align}
	\label{eq:specific_intensity}
	I_\nu = g^3 I_{\nu_0} + r_g \int_{\lambda_0}^{\lambda} d\lambda' \, g^2 j_\nu(\lambda') \exp\left( -r_g \int_{\lambda'}^{\lambda} d\lambda'' \, \frac{\alpha_\nu(\lambda'')}{g} \right).
\end{align}
Here, $g = \nu_0 / \nu$ is the redshift factor, and $\nu$ is the photon frequency measured in the local rest frame. We now provide the explicit expression for $g$. Let the fluid four velocity be $u^{\zeta}$ and the photon four-momentum be $k_{\zeta}$ (with $k_t = -1$), then
\begin{align}
	g = \frac{k_\zeta (\partial_t)^\zeta}{k_\zeta u^\zeta} = \frac{k_t}{k_\zeta u^\zeta} = - \frac{1}{k_\zeta u^\zeta}.
\end{align}
From the above relations, it follows that both the emissivity and the absorption coefficient must be specified in order to compute the observed intensity.

\subsection{Synchrotron Radiation}
We note that the radiative coefficients $j_{\nu}$ and $\alpha_\nu$ in Eq.~\eqref{eq:invariant_relations} depend on the underlying emission mechanism, with different processes yielding different forms for $j_{\nu}$ and $\alpha_\nu$. Here, we consider synchrotron radiation from electrons in the extreme relativistic regime (in CGS units). In this section, $c$ denotes the speed of light, $h$ the Planck constant, $e$ the elementary charge,  $k_B$ the Boltzmann constant, and the local magnetic field is denoted by $b^\zeta$.

In a plasma system, synchrotron radiation is primarily contributed by electrons, and its emissivity $j_{\nu}$ plays a key role in thick disk imaging, given explicitly by
\begin{align}
	\label{eq:emissivity}
	j_\nu = \frac{\sqrt{3} e^3 B \sin\theta_B}{4\pi m_e c^2} \int_0^\infty d\gamma \, N(\gamma) F\left( \frac{\nu}{\nu_s} \right).
\end{align}
Here, $\gamma = \frac{1}{\sqrt{1 - \beta^2}}$ is the Lorentz factor of the charged particle, $N(\gamma)$ is the electron distribution function, and $F(x)$ is defined as
\begin{align}
	\label{eq:F_function}
	F(x) = x \int_x^\infty dy\, K_{5/3}(y),
\end{align}
where $K_n(x)$ is the modified Bessel function of the second kind of order $n$. The angle $\theta_B$ between the spatial projection of the photon four-momentum $e^{\zeta}_{(k)}$ and the magnetic field direction $e^{\zeta}_{(b)}$ is
\begin{align}
	\theta_B = \arccos\left(e_{(b)}^\zeta \cdot e_{(k)}^\zeta\right) = \arccos\left[\frac{g}{B}(b_\zeta k^\zeta)\right],
\end{align}
with
\begin{align}
	\label{eq:four_vectors}
	e^\zeta_{(k)} = -\left( \frac{k^\zeta}{u^{\nu} k_\nu} + u^\zeta \right), \quad e^\zeta_{(b)} = \frac{b^\zeta}{B}.
\end{align}
Here, $B = \sqrt{b_\zeta b^\zeta}$ is the magnitude of the local magnetic field. In Eq.~\eqref{eq:emissivity}, the characteristic frequency $\nu_s$ is
\begin{align}
	\nu_s = \frac{3 e B \gamma^2 \sin\theta_B}{4 \pi m_e c}.
\end{align}
Different electron distributions yield different emissivities, for a thermal electron distribution, the distribution function $N(\gamma)$ takes the form
\begin{align}
	\label{eq:thermal_distribution}
	N(\gamma) = \frac{n_e \beta \gamma^2}{\theta_e K_2(\theta_e^{-1})} e^{-\gamma / \theta_e},
\end{align}
where $n_e$ is the electron number density, $\theta_e = k_B T_e / m_e c^2$ is the dimensionless electron temperature, and $T_e$ is the thermodynamic temperature of electrons. In the extreme relativistic limit, $\beta \approx 1$ and $\theta_e \gg 1$, the asymptotic form $K_2(\theta_e^{-1}) \approx 2 \theta_e^2$ holds. Defining $z = \gamma / \theta_e$, Eq.~\eqref{eq:emissivity} becomes as
\begin{align}
	j_\nu = \frac{\sqrt{3} n_e e^3 B \sin\theta_B}{8 \pi m_e c^2} \int_0^\infty z^2 e^{-z} F\left( \frac{\nu}{\nu_s} \right) dz.
\end{align}
Defining $x = (\nu / \nu_s) z^2$, the emissivity can be expressed as
\begin{align}
	\label{eq:anisotropic_emissivity}
	j_\nu = \frac{\sqrt{3} n_e e^2 \nu}{6 c \theta_e^2} \mathcal{I}(x), \quad x = \frac{\nu}{\nu_c}, \quad \nu_c = \frac{3 e B \theta_e^2 \sin\theta_B}{4 \pi m_e c},
\end{align}
where the dimensionless function $\mathcal{I}(x)$ is defined by
\begin{align}
	\mathcal{I}(x) = \frac{1}{x} \int_0^\infty z^2 e^{-z} F\left( \frac{x}{z^2} \right) dz.
\end{align}
Since this function does not have a closed form expression in terms of elementary functions, it is approximated using a fitting formula.

In this paper, we consider two radiation models: isotropic and anisotropic radiation. We first discuss the isotropic radiation model. For isotropic radiation, only the magnitude of the magnetic field is considered, ignoring its direction. The angle averaged synchrotron emissivity is defined by
\begin{align}
	\label{eq:angle_averaged_emissivity}
	\overline{j}_{\nu} = \frac{1}{2} \int_{0}^{\pi} j_{\nu} \sin\theta_B \, d\theta_B.
\end{align}
Its corresponding fitting formula, as provided in Ref.~\cite{Mahadevan:1996cc}, reads
\begin{align}
	\overline{j}_{\nu} = \frac{n e^{2} \nu}{2\sqrt{3} c \theta_{e}^{2}} \mathcal{M}(x), \quad x = \frac{\nu}{\nu_c}, \quad \nu_c = \frac{3 e B \theta_e^2}{4\pi m_e c},
\end{align}
where the dimensionless function $\mathcal{M}(x)$ is given by
\begin{align}
	\mathcal{M}(x) = \frac{4.0505}{x^{1/6}} \left(1 + \frac{0.4}{x^{1/4}} + \frac{0.5316}{x^{1/2}}\right) \exp\left(-1.8899 x^{1/3}\right).
\end{align}

Next, we discuss the anisotropic radiation model. In this case, we assume that the magnetic field is a mixture of toroidal and poloidal components, so that the magnetic four vector can be expressed as
\begin{align}
	b^\zeta \sim (l, 0, 0, 1),
\end{align}
with
\begin{align}
	l = -\frac{u_\phi}{u_t}, \quad u_\nu = g_{\zeta\nu} u^\zeta = (u_t, u_r, u_\theta, u_\phi).
\end{align}
The magnetic field is perpendicular to the fluid four velocity, satisfying $u^\zeta b_\zeta = 0$. The emissivity for the anisotropic radiation model is given by Eq.~\eqref{eq:anisotropic_emissivity}, i.e.,
\begin{align*}
	j_\nu = \frac{n_e e^2 \nu}{2\sqrt{3} c \theta_e^2} \mathcal{I}(x), \quad x = \frac{\nu}{\nu_c}, \quad \nu_c = \frac{3 e B \theta_e^2 \sin\theta_B}{4 \pi m_e c},
\end{align*}
where the dimensionless function $\mathcal{I}(x)$ is given in Ref.~\cite{2011Numerical} as
\begin{align}
	\mathcal{I}(x) = 2.5651 \left( 1 + 1.92 x^{-1/3} + 0.9977 x^{-2/3} \right) \exp\left( -1.8899 x^{1/3} \right).
\end{align}
For a thermal electron distribution, the absorption process obeys Kirchhoff's law, so that the absorption coefficient $\alpha_\nu$ satisfies
\begin{align}
	\label{eq:blackbody}
	\alpha_\nu = \frac{j_\nu}{\mathscr{B}_\nu}, \quad \mathscr{B}_\nu = \frac{2 h \nu^3}{c^2} \frac{1}{\exp\left( \frac{h \nu}{k_B T_e} \right) - 1}.
\end{align}
Here, $\mathscr{B}_\nu$ is the Planck blackbody function.

For numerical simulations, we define the following five constants
\begin{align}
	C_{1} = \frac{\sqrt{3} e^{2} n_{h} \nu_{h}}{6 \theta_{h}^{2} c}, \quad
	C_{2} = \frac{4 \pi c m_{e} \nu_{h}}{3 e B_{h} \theta_{h}^{2}}, \quad
	C_{3} = \frac{h \nu_{h}}{m_{e} \theta_{h} c^{2}}, \quad
	C_{4} = \frac{2 h \nu_{h}^{3}}{c^{2}}, \quad
	C_{5} = \sqrt{c^{2} n_{h} m_{p}}.
\end{align}
Here, $n_h$, $\theta_h$, $\nu_h$, and $B_h$ denote the values of the electron number density, the dimensionless electron temperature, the photon frequency, and the local magnetic field strength at the event horizon, respectively. In particular, we set $\nu_h = 10^9~\mathrm{Hz} = 1~\mathrm{GHz}$ and $B_h = 1$. Under this parameterization, the emissivity and the blackbody function can be expressed as
\begin{align}
	\label{eq:parameterized}
	j_\nu = \frac{C_1 \hat{n}_e \hat{\nu} I(x)}{\hat{\theta}_e^2}, \qquad
	x = \frac{C_2 \hat{\nu}}{\hat{B} \hat{\theta}_e^2 \sin\theta_B}, \qquad
	\mathscr{B}_\nu = \frac{C_4 \hat{\nu}^3}{\exp\left( \frac{C_3 \hat{\nu}}{\hat{\theta}_e} \right) - 1},
\end{align}
where $\hat{\nu} = \nu / \nu_h$, $\hat{n}_e = n_e / n_h$, $\hat{B} = B / B_h$, and $\hat{\theta}_e = \theta_e / \theta_h$. In principle, using Eqs.~\eqref{eq:blackbody} and \eqref{eq:parameterized}, one can compute the intensity in Eq.~\eqref{eq:specific_intensity}. However, the electron number density and temperature in these formulas remain to be determined. The next section discusses how these parameters are set for different accretion flow models.

\section{Accretion Flow Models}\label{sec4}
We first consider unpolarized imaging. In  this case, we examine two geometrically thick and optically thin accretion flow models, namely the RIAF model~\cite{Broderick:2010kx} and the BAAF model~\cite{Zhang:2024lsf,Hou:2023bep}. The RIAF model is highly consistent with GRMHD simulations and has achieved significant success in reproducing the overall morphology of M87*~\cite{EventHorizonTelescope:2019pgp}. However, its application in polarization studies is limited because it neglects outflows, non-thermal particles, and the full GRMHD dynamics. To address this limitation, we consider the BAAF model, which assumes that fluid acceleration near the event horizon is primarily governed by gravity and provides explicit expressions for the thermodynamic variables and magnetic field configuration. This allows for a more accurate description of the morphology and dynamics of geometrically thick accretion flows in the near horizon region of the black hole.
\subsection{RIAF Model}
We adopt a cylindrical coordinate system, where the cylindrical radius is given by $R = r \sin\theta$ and the height above the equatorial plane ($\theta = \pi/2$) is $z = r \cos\theta$. Following the construction of the RIAF model in Ref.~\cite{Broderick:2010kx}, the density and temperature profiles can be defined as
\begin{align}
	n_e = n_h \left( \frac{r}{r_+} \right)^2 \exp\left( -\frac{z^2}{2 R^2} \right), \quad
	T_e = T_h \left( \frac{r}{r_+} \right),
\end{align}
where $n_h$ and $T_h$ are the electron number density and temperature at the outer horizon, respectively. The magnetic field magnitude is defined via the cold magnetization parameter $\sigma = \frac{b^2}{\rho} = \frac{b^2}{n_e m_p c^2}$ as $b = \sqrt{\sigma \rho}$, with $\rho = n_e m_p c^2$ denoting the fluid mass density. For the accretion flow models considered here, $\sigma$ is of order $\sigma \sim 0.1$~\cite{Pu:2016qak}. We consider both isotropic radiation~\eqref{eq:angle_averaged_emissivity} and anisotropic radiation~\eqref{eq:anisotropic_emissivity} in the phenomenological model.

Regarding the motion of the accretion flow, we adopt the ballistic approximation, in which the fluid moves along geodesics. Under this assumption, the four velocity components are given by
\begin{align}
	\label{eq:four_velocity}
	u^{t} &= \frac{E \left( r^{2} + r r_0 + a^{2} \right)^2 - 2 a M r L}{\Delta \Sigma} - \frac{E a^2 \sin^2 \theta}{\Sigma}, \nonumber\\
	u^{r} &= -\frac{\sqrt{R}}{\Sigma}, \nonumber\\
	u^{\theta} &= 0, \nonumber\\
	u^{\phi} &= \frac{2 a M r E - a^2 L}{\Delta \Sigma} + \frac{L}{\sin^2 \theta \, \Sigma},
\end{align}
where
\begin{align}
	R = \left[ r (r + r_0) + a^2 \right]^2 E^2 + a^2 L^2 - \Delta \mathcal{K} - 4 a M r E L - r (r + r_0) m^2 \Delta,
\end{align}
and
\begin{align}
	L = \pm_L a \sqrt{E^2 - m^2} \sin^2 \theta, \quad 
	\mathcal{K} = a^2 E^2 + a^2 (m^2 - E^2) \cos 2\theta.
\end{align}

\subsection{BAAF Model}
Next, we discuss the BAAF model. The BAAF model, proposed by Hou et al., is a steady state, axisymmetric black hole accretion flow model~\cite{Zhang:2024lsf,Hou:2023bep}. In this model, the fluid is considered electrically neutral, with the plasma fully ionized into electrons and protons. The accreting matter is constrained to constant-$\theta$ surfaces, i.e., $u^\theta \equiv 0$. The continuity equation then reads
\begin{align}
	\frac{d}{dr} \left( \sqrt{-g} \, \rho \, u^r \right) = 0,
\end{align}
with the solution
\begin{align}
	\rho = \rho_0 \frac{\sqrt{-g} \, u^r \big|_{r = r_0}}{\sqrt{-g} \, u^r},
\end{align}
where $\rho_0 = \rho(r_0)$ is the mass density at a reference radius. The projection of the energy momentum tensor along $u^\mu$ satisfies
\begin{align}
	\label{eq:energy_momentum}
	dU = \frac{U + p}{\rho} d\rho,
\end{align}
where $U$ is the internal energy of the fluid. Defining $k = T_p / T_e$ as the proton to electron temperature ratio, the internal energy under this approximation is
\begin{align}
	\label{eq:internal_energy}
	U = \rho + \frac{3 (k + 2) \rho m_e \theta_e}{2 m_p},
\end{align}
with $\theta_e = k_B T_e / m_e c^2$ denoting the dimensionless electron temperature. Using the ideal gas law, the pressure is
\begin{align}
	\label{eq:pressure}
	p = n k_B (T_p + T_e) = \frac{(1 + k) \rho m_e \theta_e}{m_p}.
\end{align}
Substituting Eqs.~\eqref{eq:internal_energy} and \eqref{eq:pressure} into Eq.~\eqref{eq:energy_momentum} and integrating yields
\begin{align}
	\theta_e = (\theta_e)_0 \left( \frac{\rho}{\rho_0} \right)^{\frac{2(1+k)}{3(2+k)}},
\end{align}
where $(\theta_e)_0$ is the reference temperature at $r_+$.

For computational convenience, we assume that $\rho(r_+,\theta)$ follows a Gaussian distribution in the $\theta$ direction and, in the conical solution, take $\theta_e(r_+,\theta)$ to be constant
\begin{align}
	\rho(r_+,\theta) = \rho_h \exp\left[-\left( \frac{\sin\theta - \sin\mu_\theta}{\sigma_\theta} \right)^2 \right], \quad
	\theta_e(r_+,\theta) = \theta_h,
\end{align}
where $\mu_\theta$ denotes the mean position in the $\theta$ direction and $\sigma_\theta$ is the standard deviation of the distribution. For M87$^\ast$, observations indicate $\rho_h \simeq 1.5 \times 10^3~\mathrm{g\,cm^{-1}\,s^{-2}}$ and $\Theta_h \simeq 16.86$, corresponding to an electron number density $n_h = 10^6~\mathrm{cm^{-3}}$ and temperature $T_h = 10^{11}~\mathrm{K}$~\cite{Vincent_2022}.  

In a steady state, axisymmetric spacetime, the general configuration of the magnetic field can be written as
\begin{align}
	b^\mu = \frac{\Phi}{\sqrt{-g} u^r} \left[ \left( u_t + \Omega_b u_\phi \right) u^\mu + \delta_t^\mu + \Omega_b \delta_\phi^\mu \right],
\end{align}
where $\Phi = F_{\theta \phi}$ is a component of the electromagnetic field tensor. Here, we adopt a separable monopole solution
\begin{align}
	\Phi = \Phi_0 \, \mathrm{sign}(\cos\theta) \, \sin\theta.
\end{align}
The angular velocity of the magnetic field is taken as $\Omega_b = 0.3 \, \Omega_h$, with $\Omega_h = a / (2 r_+)$ denoting the spin angular velocity of the black hole. For the BAAF model, we consider only anisotropic radiation~\eqref{eq:anisotropic_emissivity}. The fluid four velocity in the BAAF model is still described by the ballistic approximation~\eqref{eq:four_velocity}, i.e., the fluid moves along geodesics.

\section{Polarized Imaging}\label{sec5}
For polarized imaging, we consider only the BAAF model with anisotropic radiation. Under the WKB approximation, the propagation of light obeys the covariant radiative transfer equation
\begin{align}
	\label{eq:covariant_rte}
	k^\gamma \nabla_\gamma \mathcal{S}^{\alpha\beta} = J^{\alpha\beta} + H^{\alpha\beta\gamma\tau} \mathcal{S}_{\gamma\tau}.
\end{align}
Here, $k^\gamma$ denotes the photon's wave vector, $\mathcal{S}^{\alpha\beta}$ is the polarization tensor describing the polarization state of the radiation, $J^{\alpha\beta}$ specifies the emission properties of the source, and $H^{\alpha\beta\gamma\tau}$ encodes the response of the propagation medium, including absorption and Faraday rotation effects~\cite{Huang:2024bar}. The polarization tensor $\mathcal{S}^{\alpha\beta}$ is proportional to the photon's polarization density matrix. Consequently, it is Hermitian, satisfying $\mathcal{S}^{\alpha\beta} = \overline{\mathcal{S}}^{\beta\alpha}$, where the overline denotes complex conjugation, and it is gauge invariant. Using this gauge invariance, the computation can be performed in a simple parallel transported tetrad. The covariant radiative transfer equation~\eqref{eq:covariant_rte} can then be decomposed into two parts. The first part,
\begin{align}
	k^\gamma \nabla_\gamma f^\beta = 0, \quad f_\beta k^\beta = 0,
\end{align}
reflects the gravitational effects, where $f^\beta$ is a normalized spacelike vector orthogonal to $k^\beta$. The second part corresponds to the radiative transfer along the ray
\begin{align}
	\frac{d}{d\lambda} S = R(\varphi) J - R(\varphi) M R(-\varphi) S,
\end{align}
with the matrices defined as
\begin{align}
	S = \begin{pmatrix}
		\mathcal{I} \\
		\mathcal{Q} \\
		\mathcal{U} \\
		\mathcal{V}
	\end{pmatrix},\qquad 
	J &= \frac{1}{\nu^{2}}\begin{pmatrix}
		j_{I} \\
		j_{Q} \\
		j_{U} \\
		j_{V}
	\end{pmatrix},\qquad 
	M = \nu\begin{pmatrix}
		a_{I} & a_{Q} & a_{U} & a_{V} \\
		a_{Q} & a_{I} & r_{V} & -r_{U} \\
		a_{U} & -r_{V} & a_{I} & r_{Q} \\
		a_{V} & r_{U} & -r_{Q} & a_{I}
	\end{pmatrix},\nonumber\\
	{R}(\varphi) &=
	\begin{pmatrix}
		1 & 0 & 0 & 0 \\
		0 & \cos(2\varphi) & -\sin(2\varphi) & 0 \\
		0 & \sin(2\varphi) & \cos(2\varphi) & 0 \\
		0 & 0 & 0 & 1
	\end{pmatrix}.
\end{align}
where $R(\varphi)$ is a rotation matrix. The rotation angle $\varphi$ is the angle between the reference vector $f^\mu$ and the local magnetic field $b^\mu$ in the transverse plane of the light ray, calculated as~\cite{Zhou:2025moa}
\begin{align}
	\varphi = \mathrm{sign}(\epsilon_{\mu\nu\rho\sigma} u^\mu f^\nu b^\rho k^\sigma) 
	\arccos \left( \frac{P^{\mu\nu} f_\mu b_\nu}{\sqrt{ (P^{\mu\nu} f_\mu f_\nu) (P^{\alpha\beta} b_\alpha b_\beta) }} \right),
\end{align}
with $P^{\mu\nu}$ being the projector onto the transverse subspace.

At the observer, the Stokes parameters are projected onto the observer's screen, again using a rotation matrix. The corresponding rotation angle is
\begin{align}
	\varphi_0 = \mathrm{sign}(\epsilon_{\alpha\beta\mu\nu} u^\alpha f^\beta d^\mu k^\nu) 
	\arccos \left( \frac{P^{\mu\nu} f_\mu d_\nu}{\sqrt{ (P^{\mu\nu} f_\mu f_\nu) (P^{\alpha\beta} d_\alpha d_\beta) }} \right),
\end{align}
where $d^\mu$ is chosen along the $y$-axis of the screen, $d^\mu = -\partial_\theta^\mu$. The projected Stokes parameters are then
\begin{align}
	\mathcal{I}_o = \mathcal{I}, \quad
	\mathcal{Q}_o = \mathcal{Q} \cos\chi_o - \mathcal{U} \sin\chi_o, \quad
	\mathcal{U}_o = \mathcal{Q} \sin\chi_o + \mathcal{U} \cos\chi_o, \quad
	\mathcal{V}_o = \mathcal{V},
\end{align}
where $\mathcal{I}_o$ represents the intensity. The Stokes parameters $\mathcal{Q}_o$ and $\mathcal{U}_o$ are related to the electric field $\vec{E} = (E_x, E_y)$ by
\begin{align}
	\mathcal{Q}_o = E_x^2 - E_y^2, \quad \mathcal{U}_o = 2 E_x E_y.
\end{align}
In general, if $\mathcal{U}_o$ is positive, $E_x$ and $E_y$ have the same sign, and $\vec{E}$ lies in the first or third quadrant; if $\mathcal{U}_o$ is negative, $E_x$ and $E_y$ have opposite signs, and $\vec{E}$ lies in the second or fourth quadrant. The sign of $\mathcal{Q}_o$ indicates whether $\vec{E}$ is aligned closer to the line $y = x$ or $y = -x$. The parameter $\mathcal{V}_o$ describes circular polarization. A positive value corresponds to left hand circular polarization, while a negative value corresponds to right hand circular polarization. From the Stokes parameters, the magnitude and direction of the projected linear polarization vector $\vec{f}$ on the observer's screen can be determined. Its magnitude corresponds to the linear polarization degree, and its angle corresponds to the electric vector position angle (EVPA)
\begin{align}
	|\vec{f}| = \mathcal{P}_o = \frac{\sqrt{\mathcal{Q}_o^2 + \mathcal{U}_o^2}}{\mathcal{I}_o}, \qquad
	\arg(\vec{f}) = \Phi_{\text{EVPA}} = \frac{1}{2} \arctan \left( \frac{\mathcal{U}_o}{\mathcal{Q}_o} \right).
\end{align}
Using this framework, the Stokes parameters and the linear polarization vector $\vec{f}$ can be computed so that we  can obtain the complete polarization properties.

\section{Numerical Results}\label{sec6}
We employ a ray-tracing method combined with a ZAMO frame and celestial coordinates to establish the mapping between pixel coordinates on the projection screen and celestial coordinates. This provides a solid computational foundation for black hole imaging studies. Using numerical simulations, we have generated several synthetic images, which are analyzed and discussed in the following. Detailed procedures and implementation can be found in Refs.~\cite{Hou:2022eev,Zhong2021,wang2025sas,Hu2021qeo,he2025tsa}.

\begin{figure}[!htbp]
	\centering
	
	\subfigure[$a=0.5,\theta=0.001^\circ$]{\includegraphics[scale=0.35]{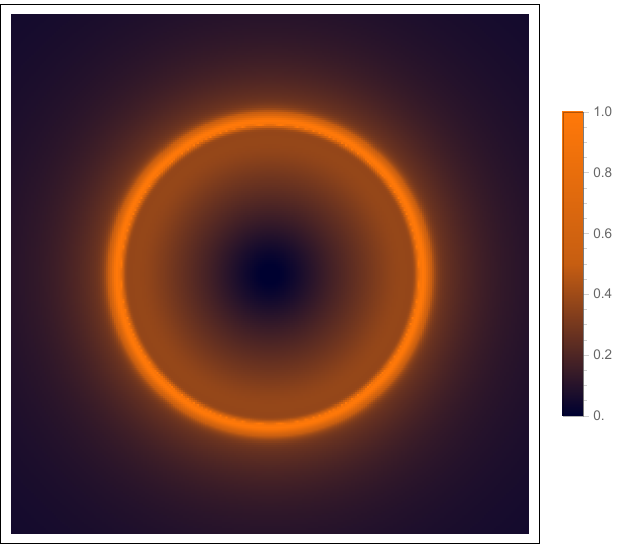}}
	\subfigure[$a=0.65,\theta=0.001^\circ$]{\includegraphics[scale=0.35]{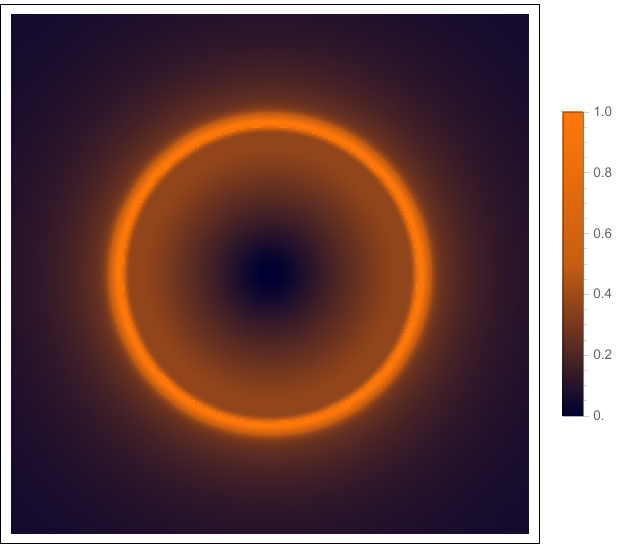}}
	\subfigure[$a=0.8,\theta=0.001^\circ$]{\includegraphics[scale=0.35]{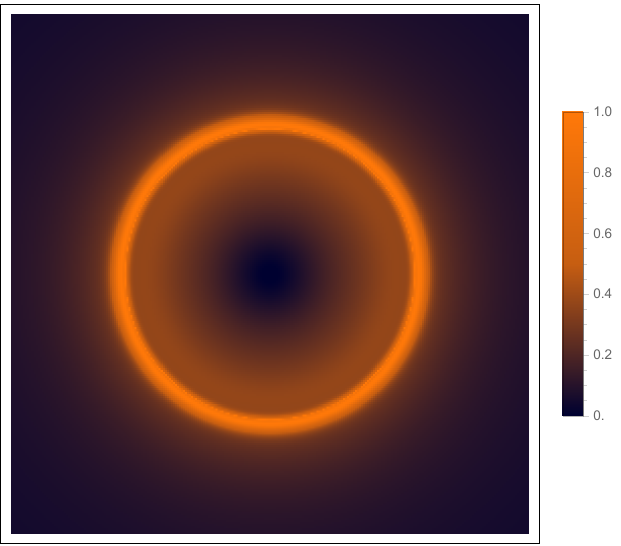}}
	\subfigure[$a=0.95,\theta=0.001^\circ$]{\includegraphics[scale=0.35]{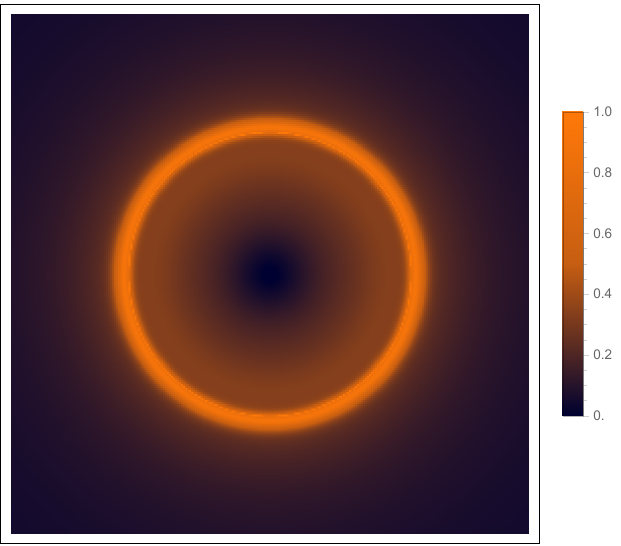}}
	
	\subfigure[$a=0.5,\theta=17^\circ$]{\includegraphics[scale=0.35]{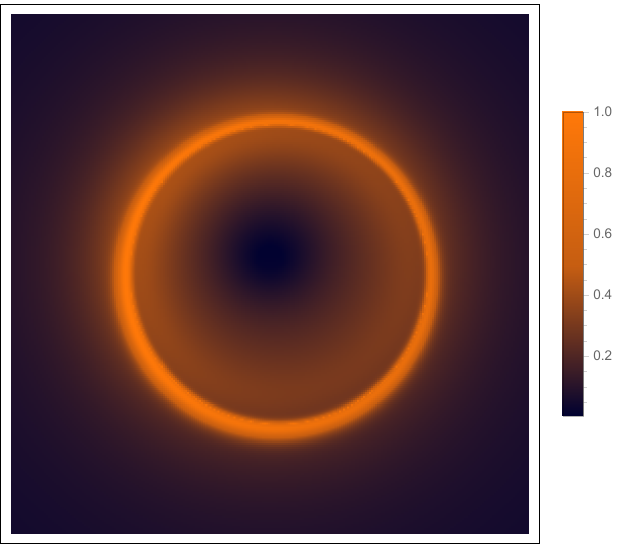}}
	\subfigure[$a=0.65,\theta=17^\circ$]{\includegraphics[scale=0.35]{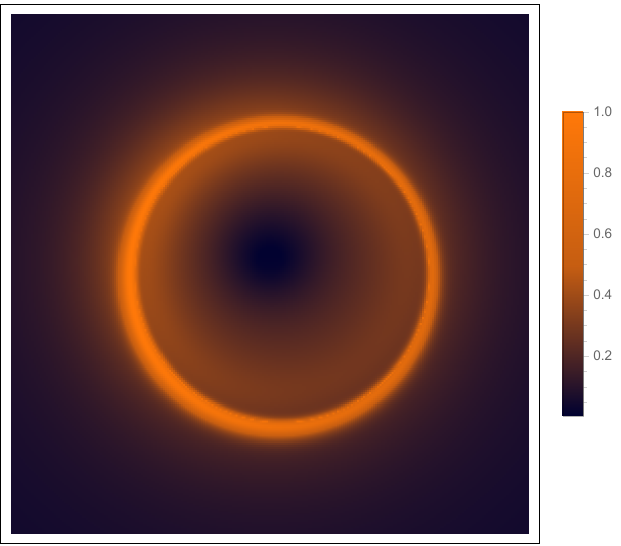}}
	\subfigure[$a=0.8,\theta=17^\circ$]{\includegraphics[scale=0.35]{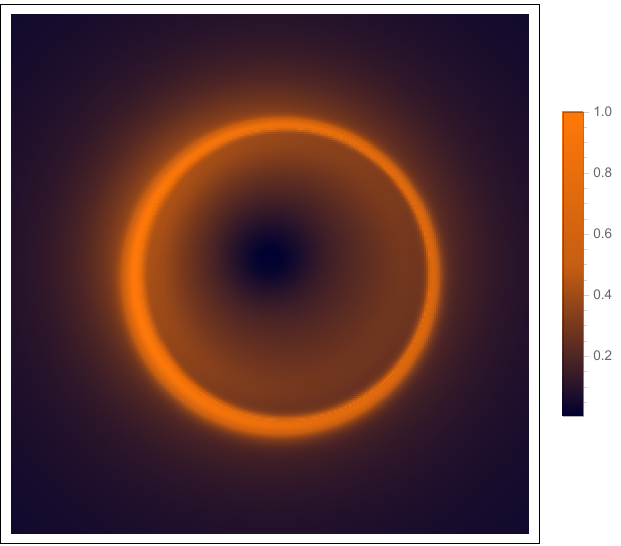}}
	\subfigure[$a=0.95,\theta=17^\circ$]{\includegraphics[scale=0.35]{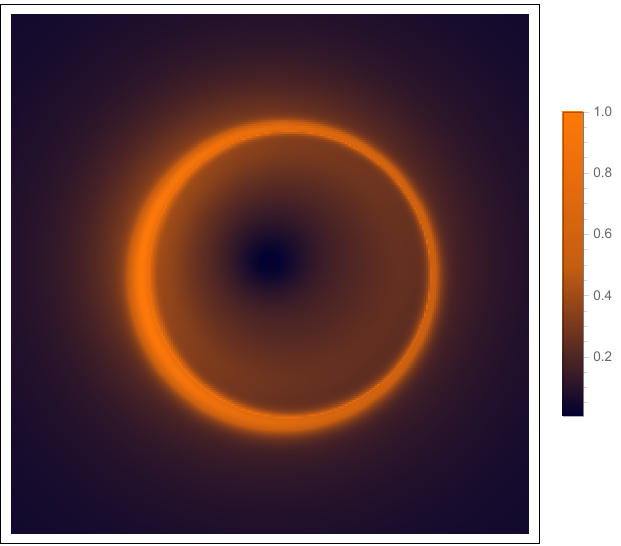}}
	
	\subfigure[$a=0.5,\theta=60^\circ$]{\includegraphics[scale=0.35]{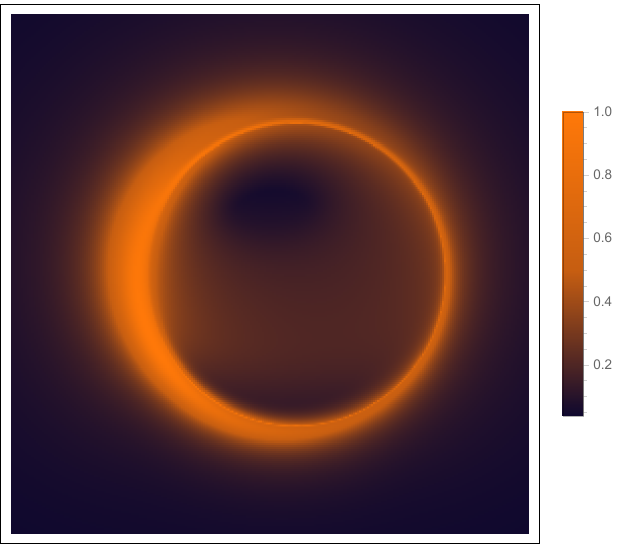}}
	\subfigure[$a=0.65,\theta=60^\circ$]{\includegraphics[scale=0.35]{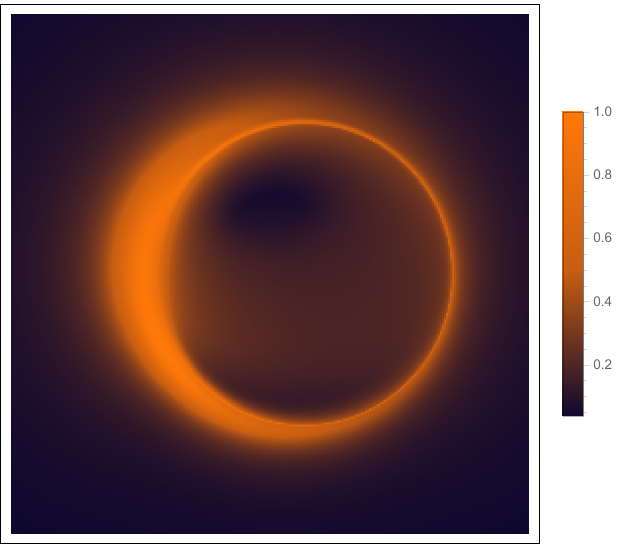}}
	\subfigure[$a=0.8,\theta=60^\circ$]{\includegraphics[scale=0.35]{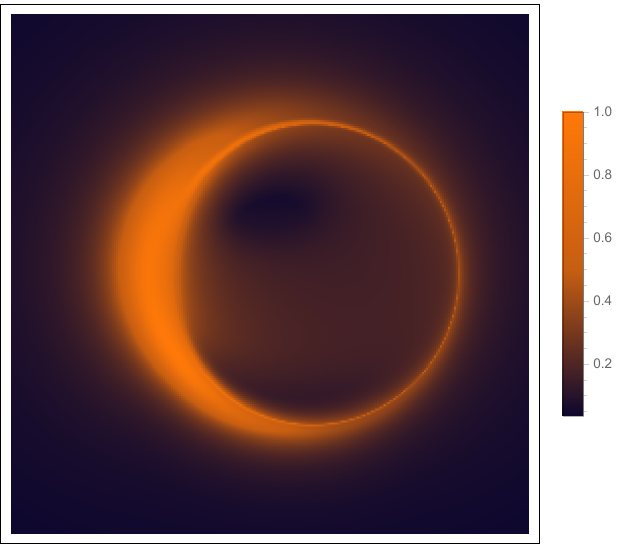}}
	\subfigure[$a=0.95,\theta=60^\circ$]{\includegraphics[scale=0.35]{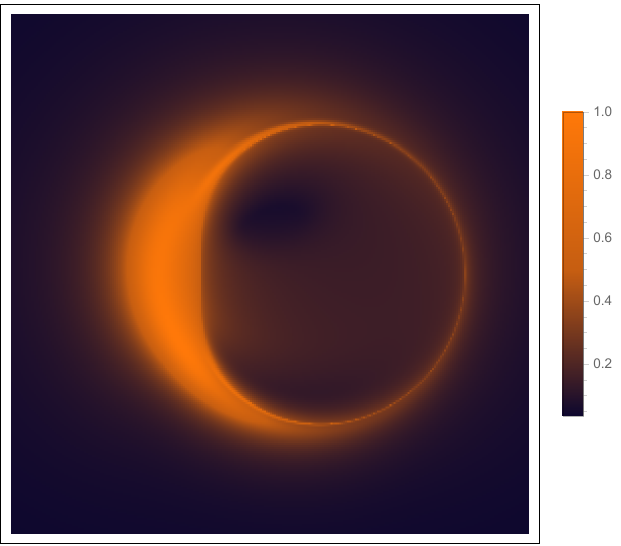}}
	
	\subfigure[$a=0.5,\theta=85^\circ$]{\includegraphics[scale=0.35]{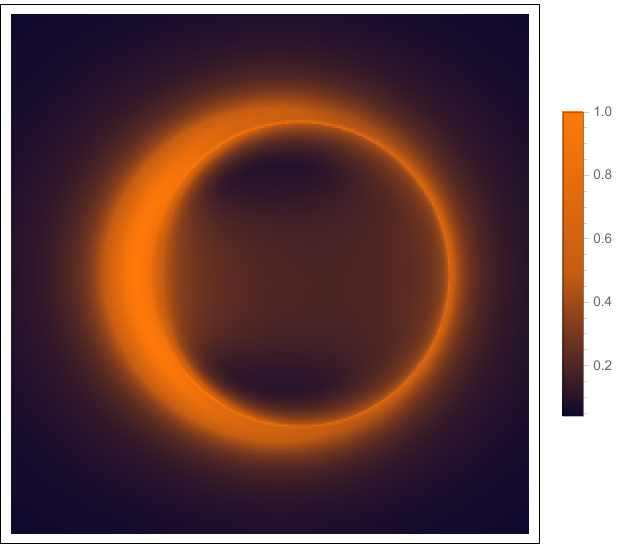}}
	\subfigure[$a=0.65,\theta=85^\circ$]{\includegraphics[scale=0.35]{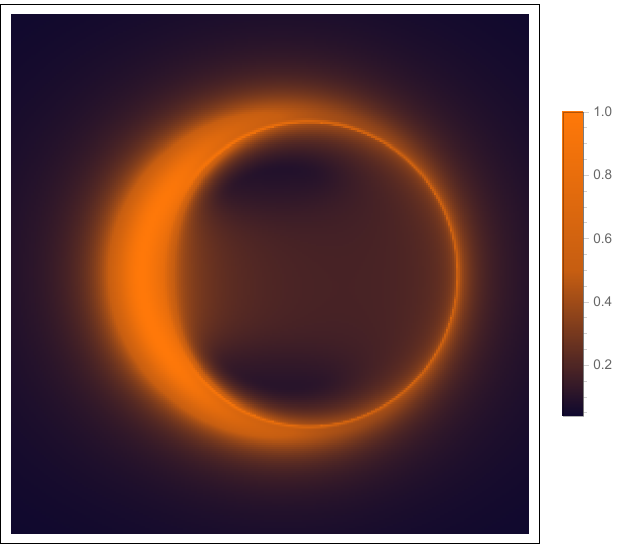}}
	\subfigure[$a=0.8,\theta=85^\circ$]{\includegraphics[scale=0.35]{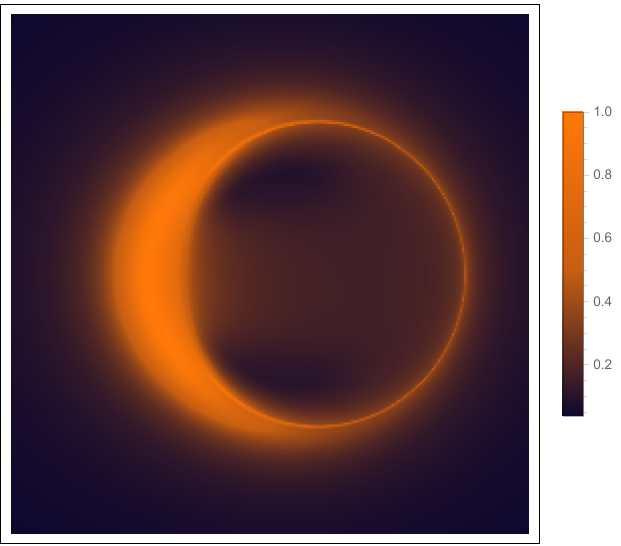}}
	\subfigure[$a=0.95,\theta=85^\circ$]{\includegraphics[scale=0.35]{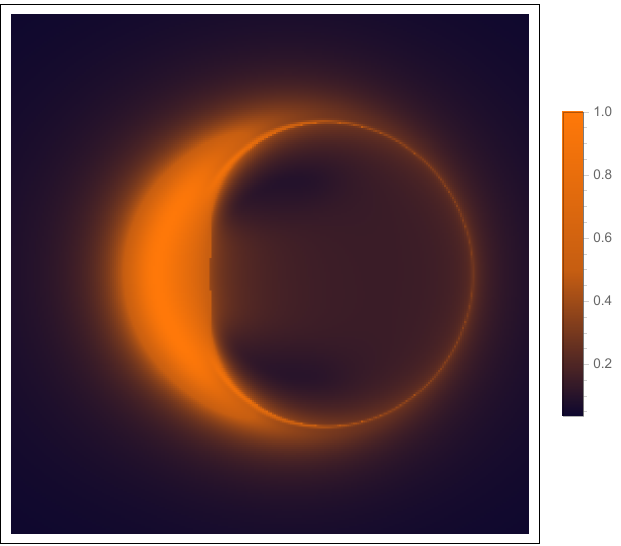}}
	
	\caption{Black hole shadow images of the RIAF model under isotropic radiation, with $Q^2 = 0.1$.}
	
\end{figure}

\begin{figure}[!htbp]
	\centering 
	
	\subfigure[$a=0.5,\theta=85^\circ$]{\includegraphics[scale=0.35]{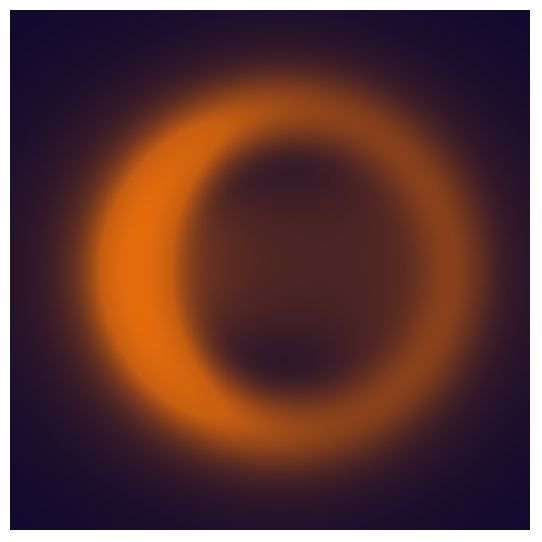}}
	\subfigure[$a=0.65,\theta=85^\circ$]{\includegraphics[scale=0.35]{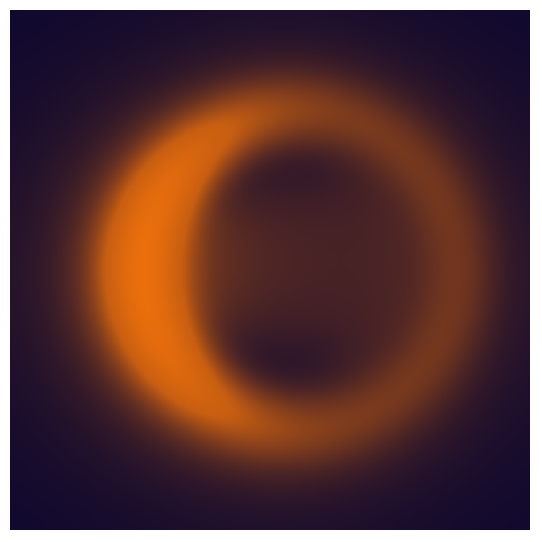}}
	\subfigure[$a=0.8,\theta=85^\circ$]{\includegraphics[scale=0.35]{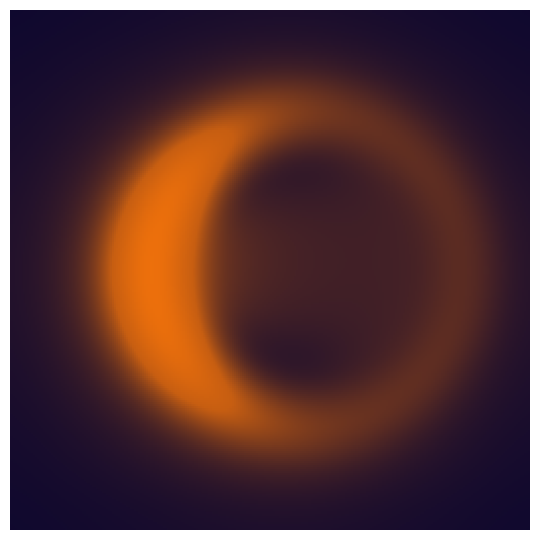}}
	\subfigure[$a=0.95,\theta=85^\circ$]{\includegraphics[scale=0.35]{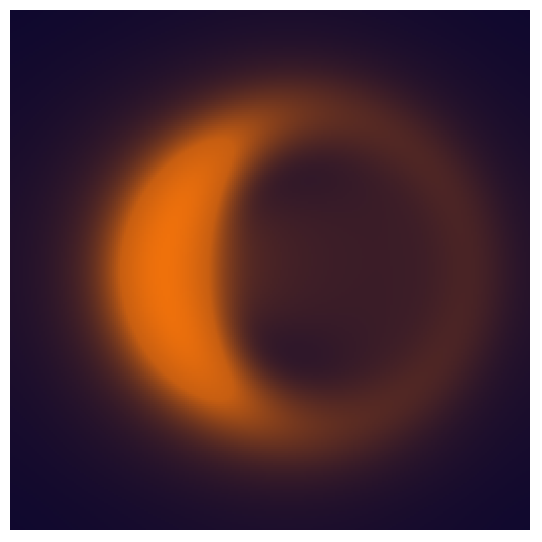}}
	
	\caption{Blurred images obtained using a Gaussian filter with a standard deviation of $1/12$ of the field of view $\gamma_{\rm fov}$. The parameters are the same as those used in Figure~1.}
\end{figure}

\subsection{RIAF Model under Isotropic Radiation}

We first examine the effects of the spin parameter $a$ and the observer inclination $\theta$ on the black hole shadow images of the RIAF model under isotropic radiation. The observation frequency is fixed at 230~GHz, the black hole charge is set to $Q^2 = 0.1$, and the accretion flow follows the ballistic approximation. Figure~1 presents the corresponding numerical results. Analysis of the images shows that all images exhibit a bright ring structure, corresponding to higher-order images, in which photons orbit the black hole one or multiple times before reaching the observer. This feature is a direct manifestation of strong gravitational lensing. The black hole’s gravitational field bends the paths of light, allowing some photons to circumvent the black hole and reach the observer. In addition to this structure, there are regions with nonzero intensity corresponding to the primary image, where photons travel directly from the accretion flow to the observer without orbiting the black hole. We note that, regardless of parameter variations, there are regions of reduced intensity inside the higher-order images. This is due to the inability of light to escape from the black hole’s event horizon. In geometrically thin accretion disks, the accreting matter is confined to the equatorial plane, so the event horizon appears as a well-defined dark region (the “inner shadow”) in the image, which the EHT may capture~\cite{Chael:2021rjo}. In contrast, for geometrically thick accretion disks, radiation from outside the equatorial plane may partially obscure this region, making it less distinguishable. Compared to thin disks, thick disks are more physically realistic, which explains why direct imaging of black hole event horizons remains challenging.

As shown in Figure~1, with increasing spin parameter $a$, the higher-order images of the black hole gradually deform into a “D”-shape at large observer inclinations, with significantly enhanced intensity on the left side. When $a = 0.95$, a crescent-shaped bright region appears on the left side of the image, originating from the frame-dragging effect of the rotating black hole. As $a$ increases, the frame-dragging effect becomes stronger, and the asymmetry of the higher-order images correspondingly increases. Given the current limited resolution of the EHT, Figure~2 presents the blurred images obtained via Gaussian smoothing, with the Gaussian standard deviation set to $1/12$ of the field of view $\gamma_{\rm fov}$, while other parameters are the same as in Figure~1. In Figure~2, the distinction between higher-order and primary images becomes blurred, and the horizon’s outline is less discernible. This further illustrates that, under current observational conditions, direct verification of the black hole event horizon remains challenging.  

Furthermore, Figure~1 shows that at polar viewing angles, the bright ring (higher-order images) and the dark region remain centered and isotropic. At an inclination of $17^\circ$, notable vertical asymmetry appears within the bright ring. When the inclination increases to $60^\circ$, two distinct dark regions emerge within the ring, with the upper region slightly darker. At $85^\circ$, the image is nearly symmetric in the vertical direction, although the left-right intensity remains higher than the top-bottom intensity. This vertical intensity dependence reflects the equatorial symmetry of the thick disk: for observers near the equatorial plane, high-latitude radiation partially fills the dark regions, while for near-polar observers, photons reaching the observer are relatively insufficient.

\begin{figure}[!htbp]
	\centering

	\subfigure[$Q^2=0.1,\theta=0.001^\circ$]{\includegraphics[scale=0.45]{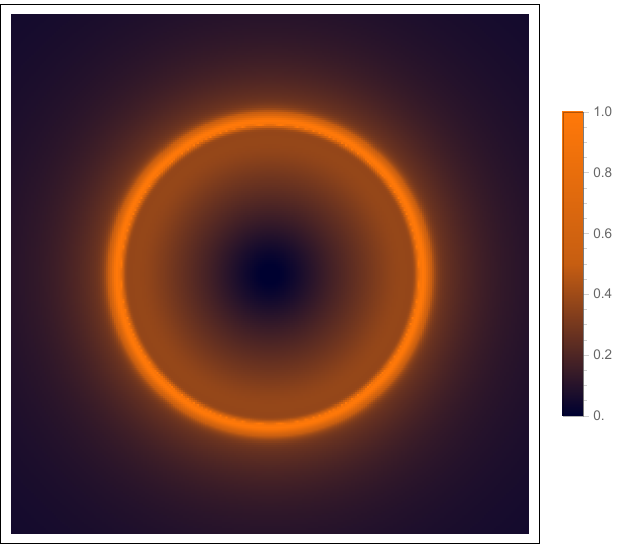}}
	\subfigure[$Q^2=0.5,\theta=0.001^\circ$]{\includegraphics[scale=0.45]{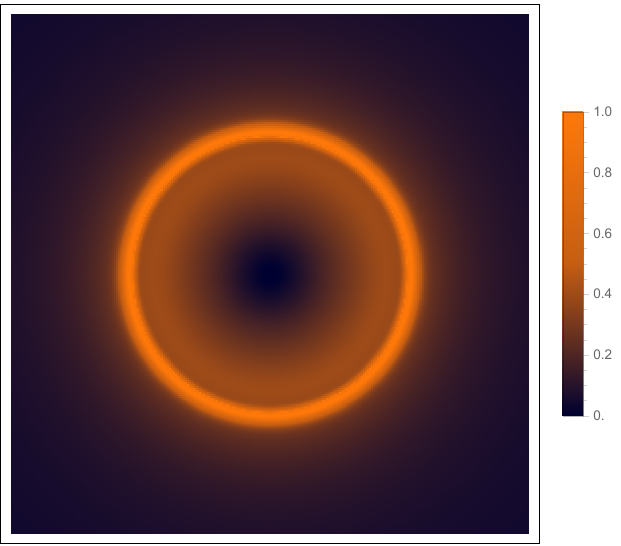}}
	\subfigure[$Q^2=1.0,\theta=0.001^\circ$]{\includegraphics[scale=0.45]{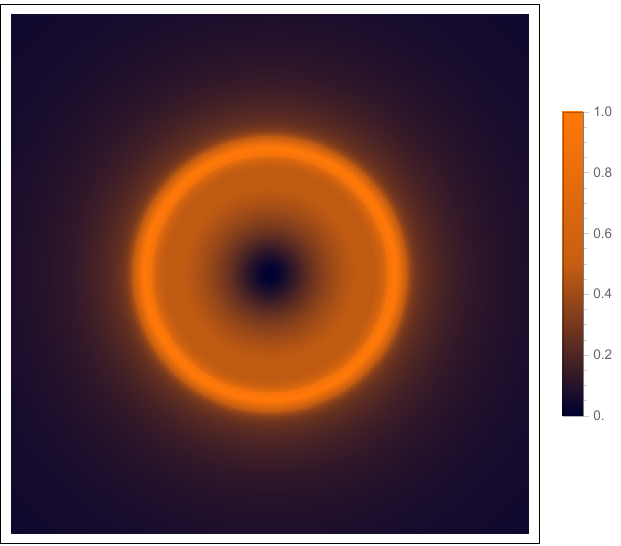}}
	
	\subfigure[$Q^2=0.1,\theta=17^\circ$]{\includegraphics[scale=0.45]{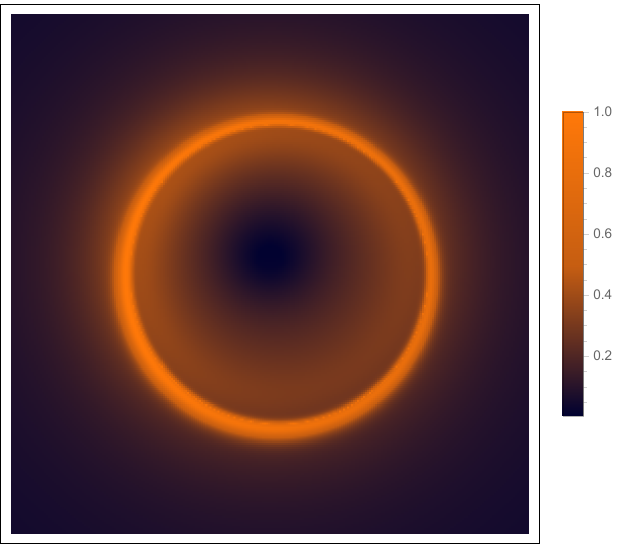}}
	\subfigure[$Q^2=0.5,\theta=17^\circ$]{\includegraphics[scale=0.45]{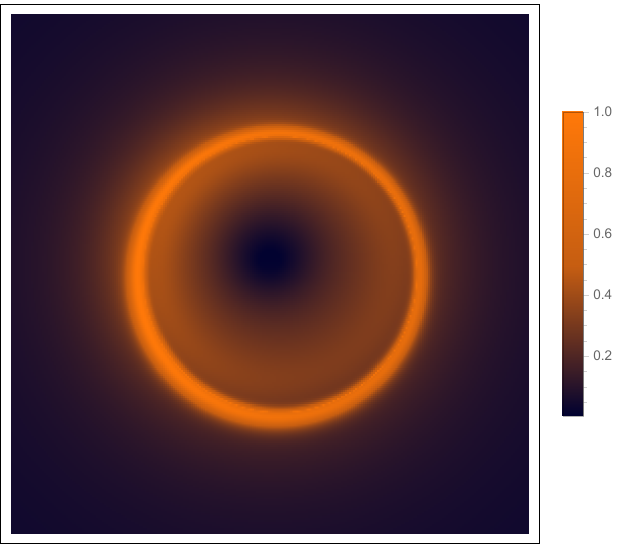}}
	\subfigure[$Q^2=1.0,\theta=17^\circ$]{\includegraphics[scale=0.45]{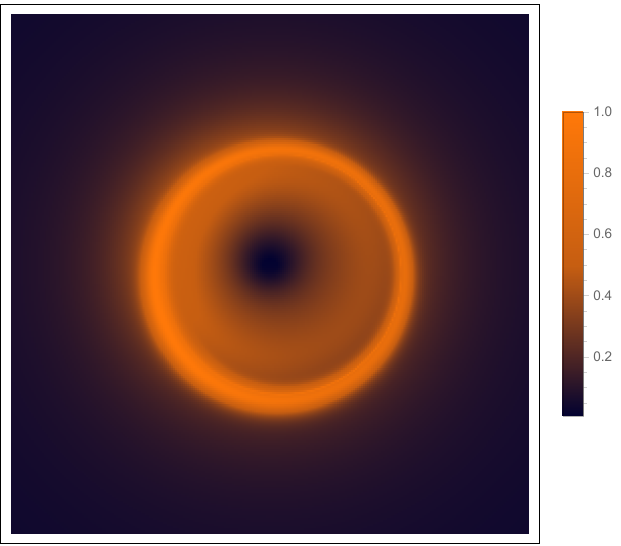}}
	
	\subfigure[$Q^2=0.1,\theta=60^\circ$]{\includegraphics[scale=0.45]{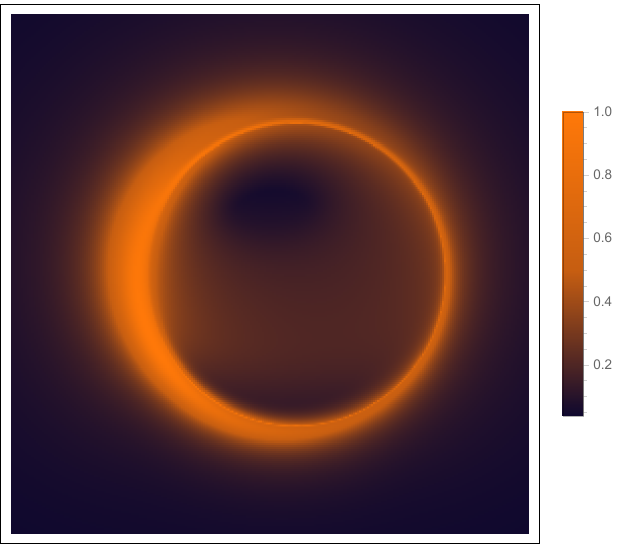}}
	\subfigure[$Q^2=0.5,\theta=60^\circ$]{\includegraphics[scale=0.45]{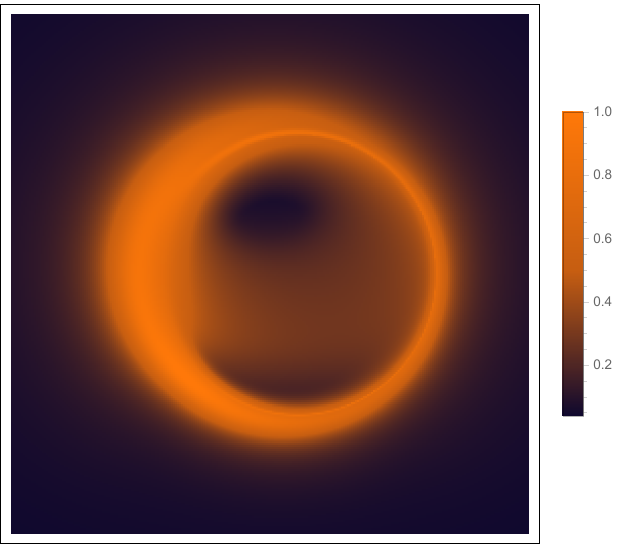}}
	\subfigure[$Q^2=1.0,\theta=60^\circ$]{\includegraphics[scale=0.45]{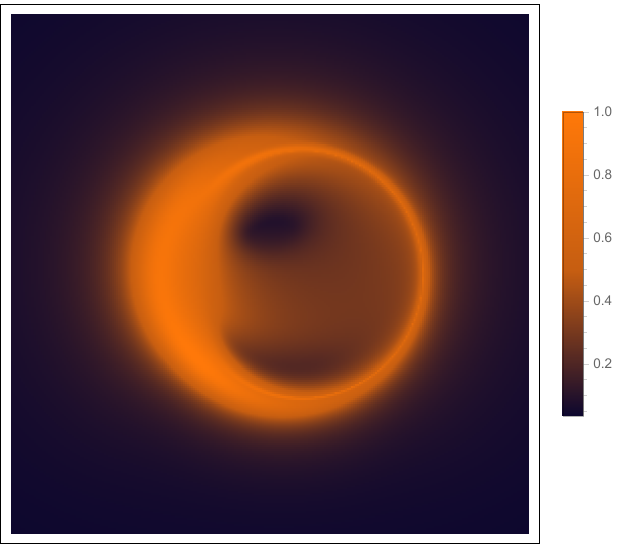}}
	
	\subfigure[$Q^2=0.1,\theta=85^\circ$]{\includegraphics[scale=0.45]{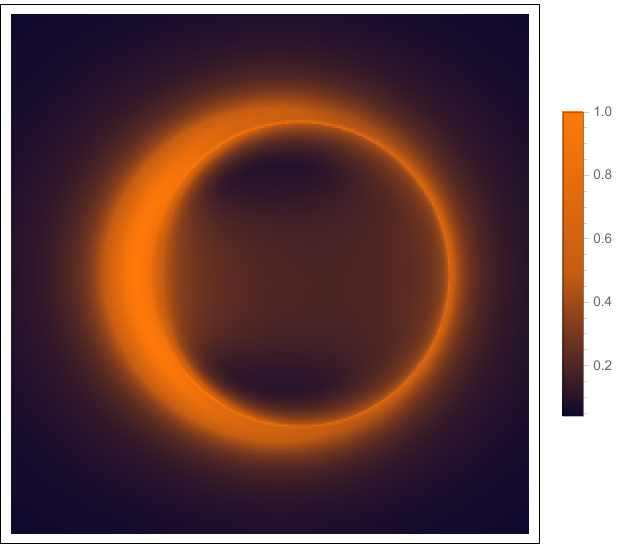}}
	\subfigure[$Q^2=0.5,\theta=85^\circ$]{\includegraphics[scale=0.45]{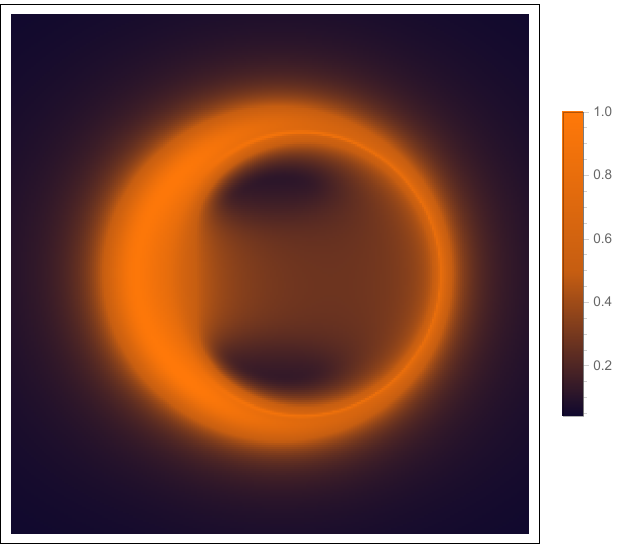}}
	\subfigure[$Q^2=1.0,\theta=85^\circ$]{\includegraphics[scale=0.45]{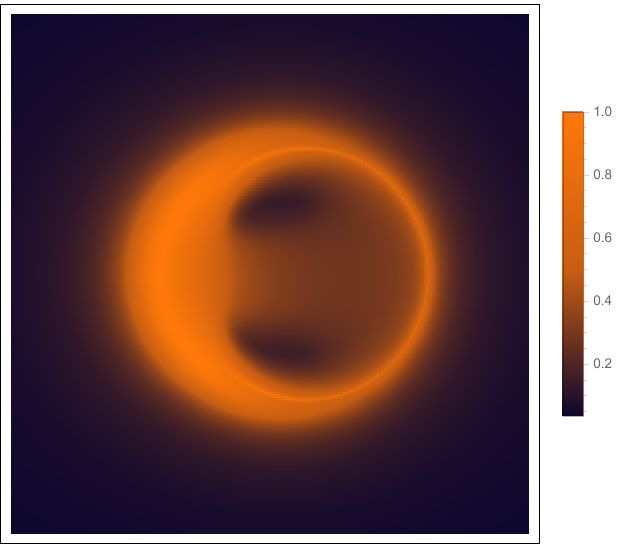}}
	
	\caption{Black hole shadow images of the RIAF model under isotropic radiation, with $a = 0.5$.}
	
\end{figure}

\begin{figure}[!htbp]
	\centering 
	
	\subfigure[$Q^2=0.1,\theta=85^\circ$]{\includegraphics[scale=0.45]{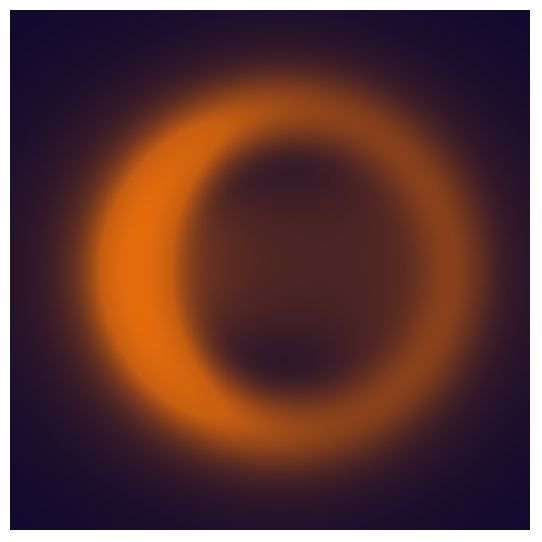}}
	\subfigure[$Q^2=0.5,\theta=85^\circ$]{\includegraphics[scale=0.45]{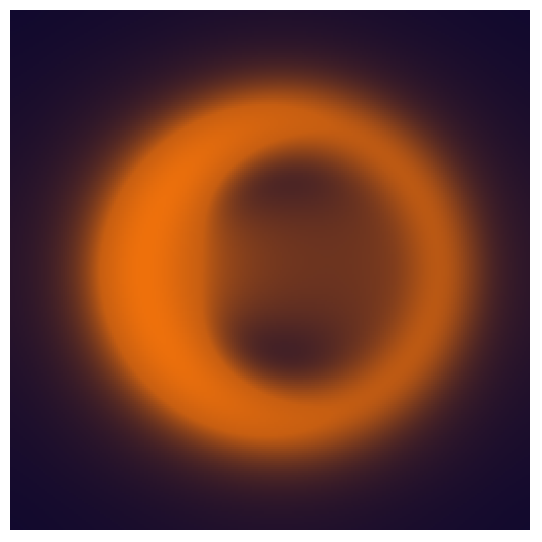}}
	\subfigure[$Q^2=1.0,\theta=85^\circ$]{\includegraphics[scale=0.45]{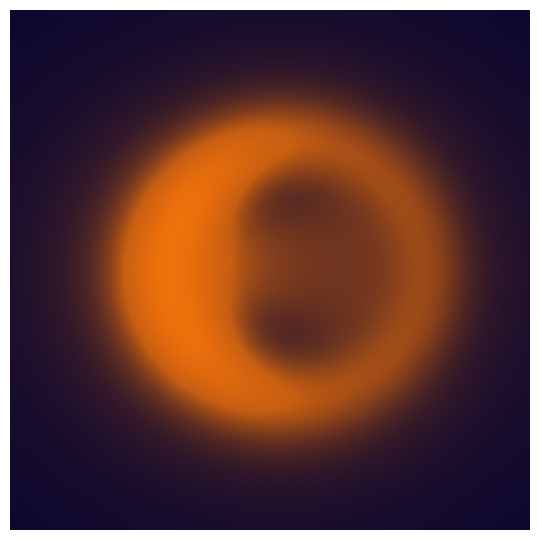}}
	
	\caption{Blurred images obtained using a Gaussian filter with a standard deviation of $1/12$ of the field of view $\gamma_{\rm fov}$. The parameters are the same as those used in Figure~3.}
\end{figure}

Figure~3 illustrates the effects of the black hole charge $Q$ on the shadow images, with the spin parameter fixed at $a = 0.5$ and the observation frequency at 230~GHz. The images show that as the black hole charge increases, both the size of the higher-order images and the inner shadow decrease, while the width of the higher-order images increases. Due to the frame-dragging effect of the black hole, the intensity on the left side of the higher-order images remains greater than on the right side. All images exhibit two dark regions within the higher-order images, with the upper region slightly darker than the lower, a phenomenon caused by gravitational lensing. Considering the limited resolution of the EHT, Figure~4 presents the corresponding blurred images, using the same parameters as in Figure~3. The blurred images are obtained via Gaussian smoothing, with the Gaussian standard deviation set to $1/12$ of the field of view $\gamma_{\rm fov}$. As in the previous cases, the distinction between higher-order and primary images becomes less clear in the blurred images, and the horizon’s outline remains difficult to discern.

\begin{figure}[!htbp]
	\centering

	\subfigure[$a=0.5,\theta=0.001^\circ$]{\includegraphics[scale=0.35]{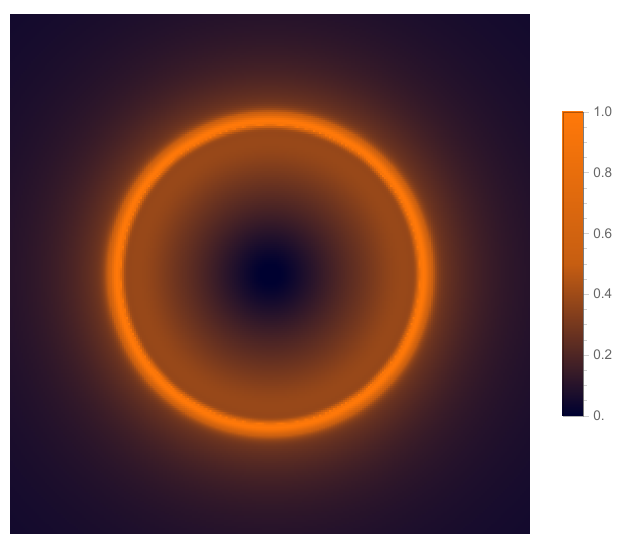}}
	\subfigure[$a=0.65,\theta=0.001^\circ$]{\includegraphics[scale=0.35]{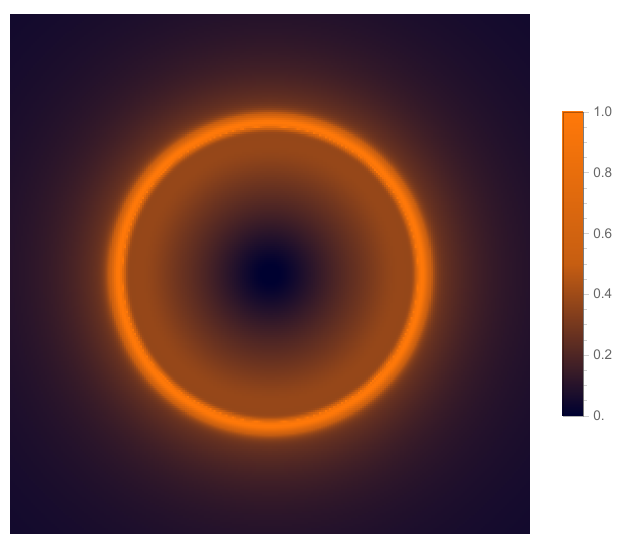}}
	\subfigure[$a=0.8,\theta=0.001^\circ$]{\includegraphics[scale=0.35]{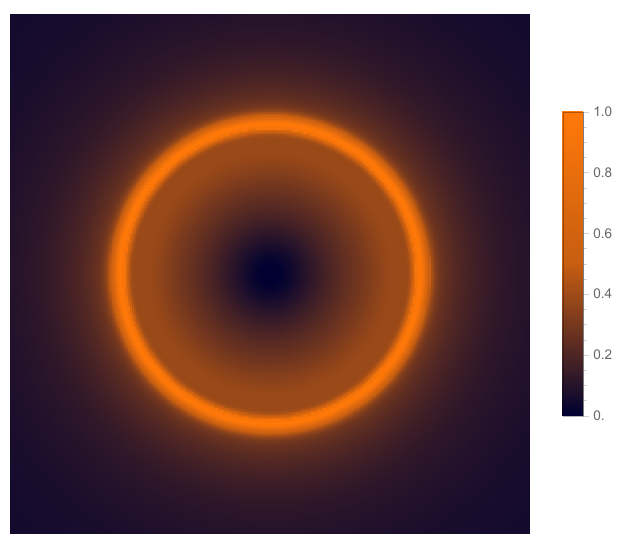}}
	\subfigure[$a=0.95,\theta=0.001^\circ$]{\includegraphics[scale=0.35]{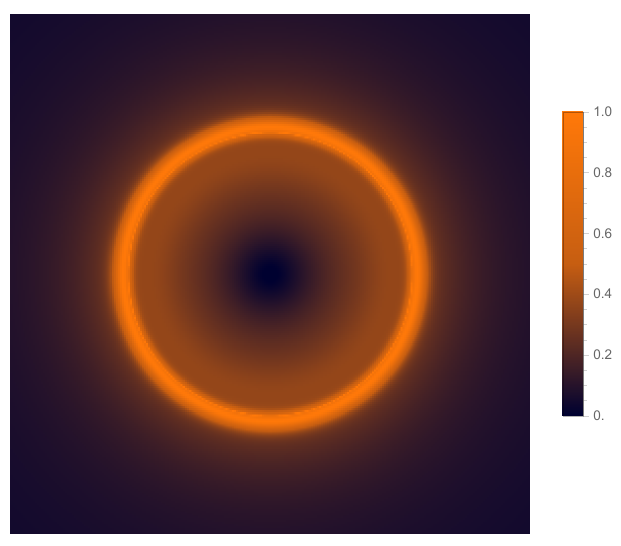}}
	
	\subfigure[$a=0.5,\theta=17^\circ$]{\includegraphics[scale=0.35]{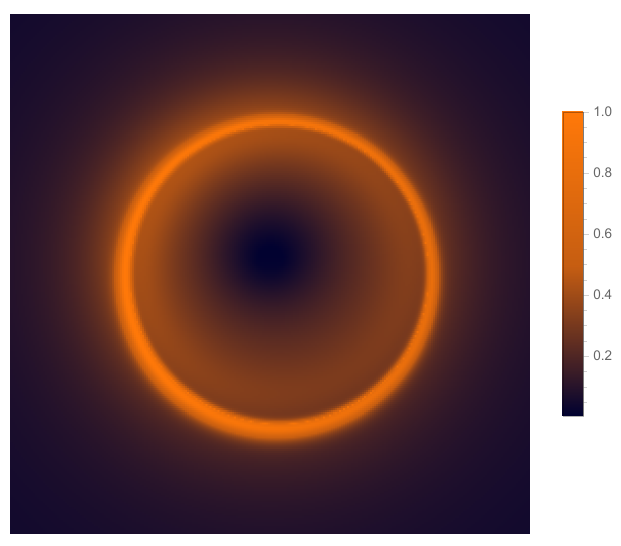}}
	\subfigure[$a=0.65,\theta=17^\circ$]{\includegraphics[scale=0.35]{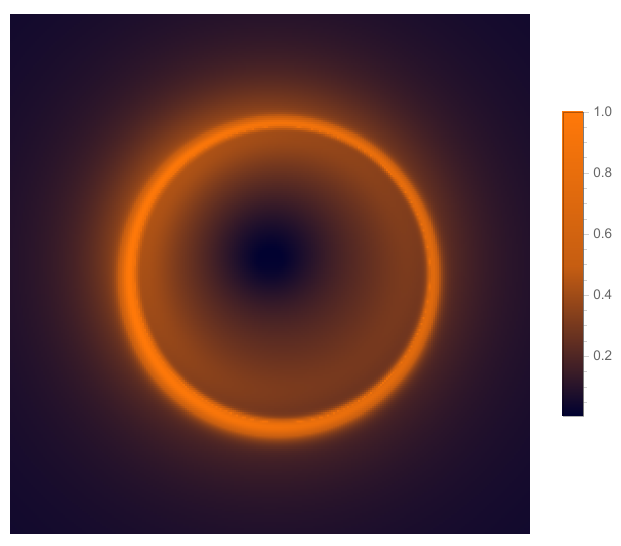}}
	\subfigure[$a=0.8,\theta=17^\circ$]{\includegraphics[scale=0.35]{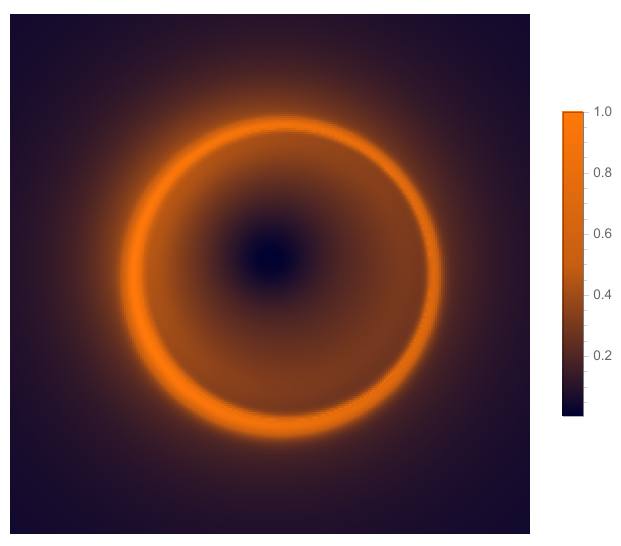}}
	\subfigure[$a=0.95,\theta=17^\circ$]{\includegraphics[scale=0.35]{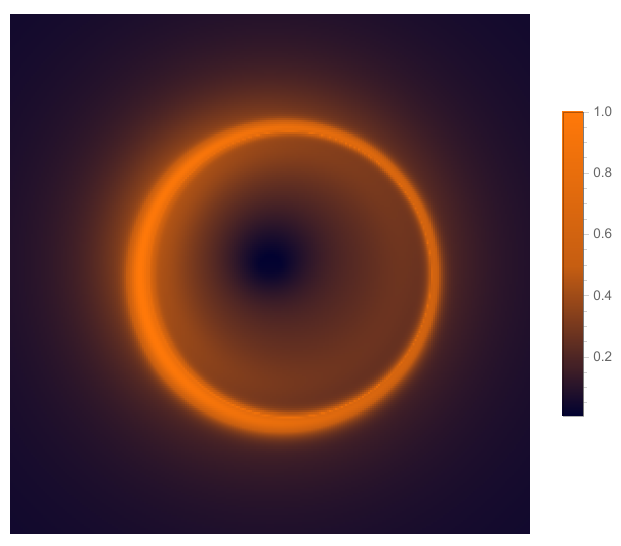}}
	
	\subfigure[$a=0.5,\theta=60^\circ$]{\includegraphics[scale=0.35]{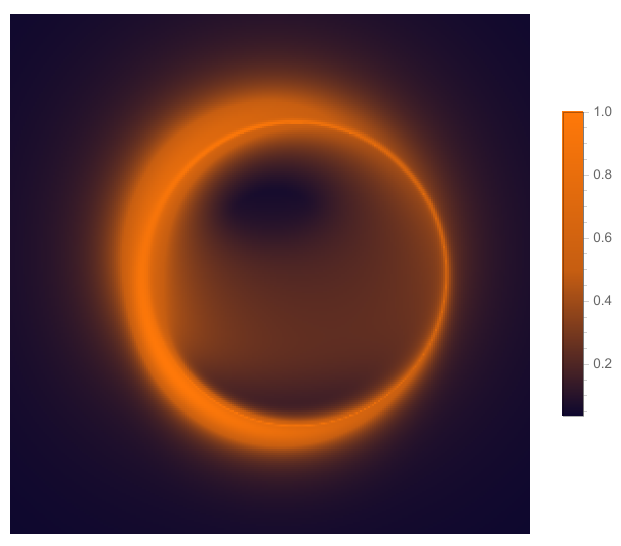}}
	\subfigure[$a=0.65,\theta=60^\circ$]{\includegraphics[scale=0.35]{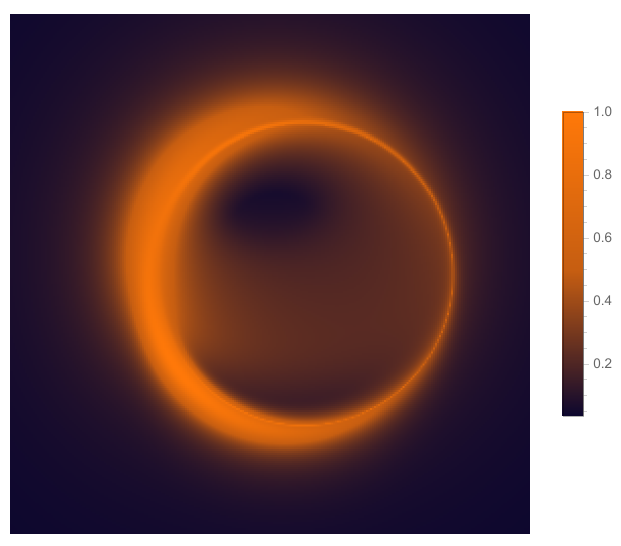}}
	\subfigure[$a=0.8,\theta=60^\circ$]{\includegraphics[scale=0.35]{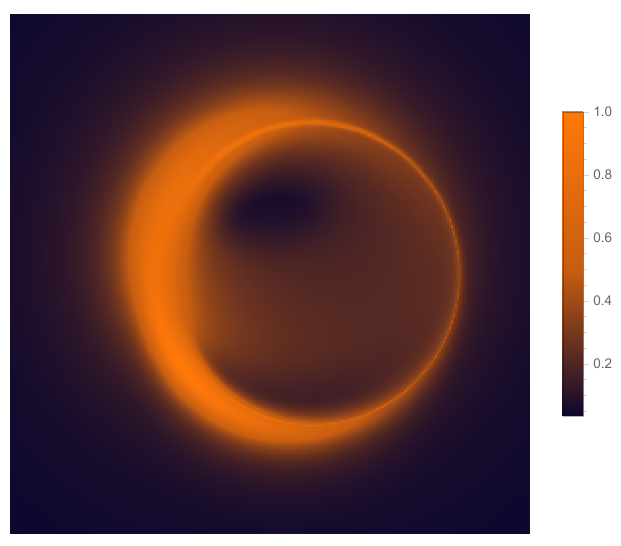}}
	\subfigure[$a=0.95,\theta=60^\circ$]{\includegraphics[scale=0.35]{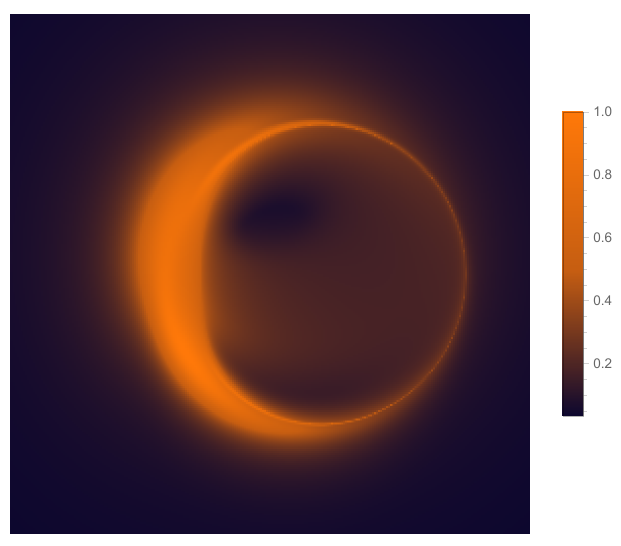}}
	
	\subfigure[$a=0.5,\theta=85^\circ$]{\includegraphics[scale=0.35]{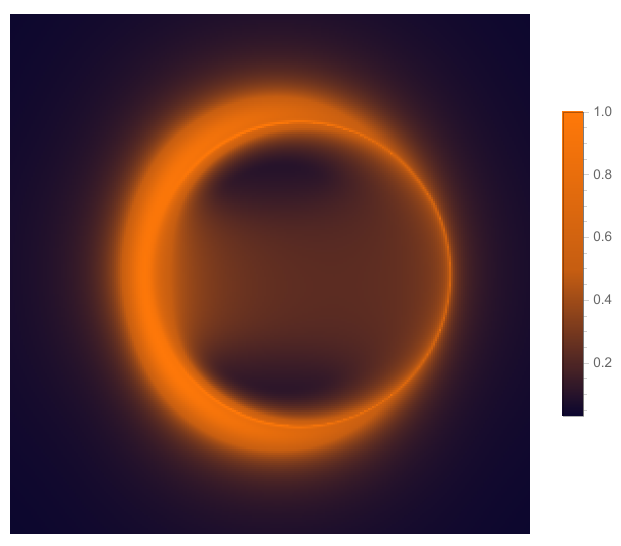}}
	\subfigure[$a=0.65,\theta=85^\circ$]{\includegraphics[scale=0.35]{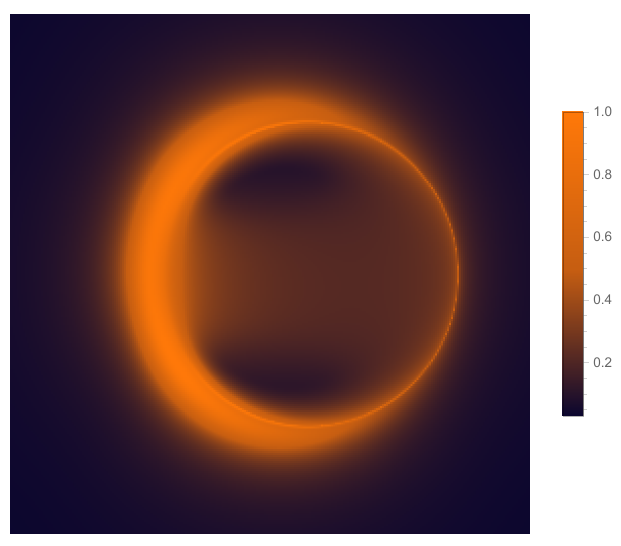}}
	\subfigure[$a=0.8,\theta=85^\circ$]{\includegraphics[scale=0.35]{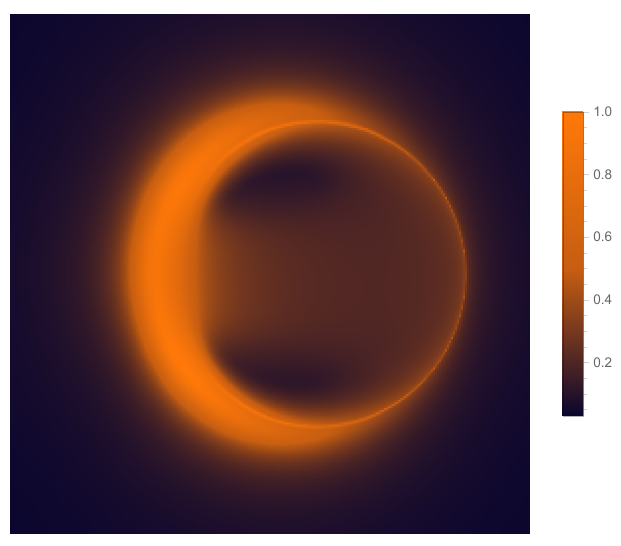}}
	\subfigure[$a=0.95,\theta=85^\circ$]{\includegraphics[scale=0.35]{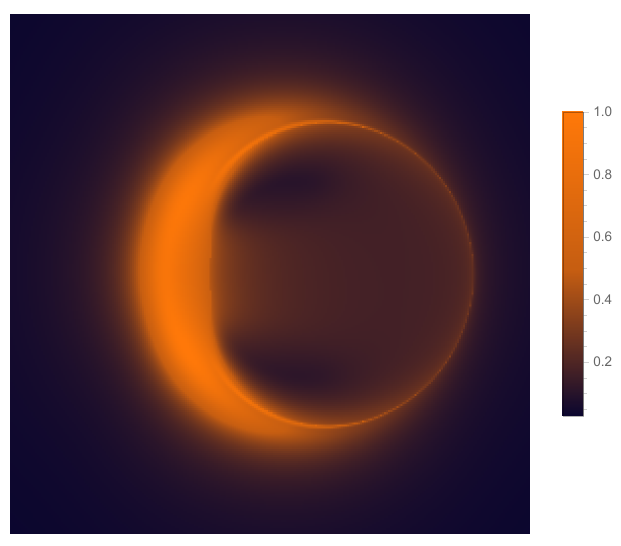}}
	
	\caption{Black hole shadow images of the RIAF model under anisotropic radiation, with $Q^2 = 0.1$.}
	
\end{figure}

\begin{figure}[!htbp]
	\centering 
	
	\subfigure[$a=0.5,\theta=85^\circ$]{\includegraphics[scale=0.35]{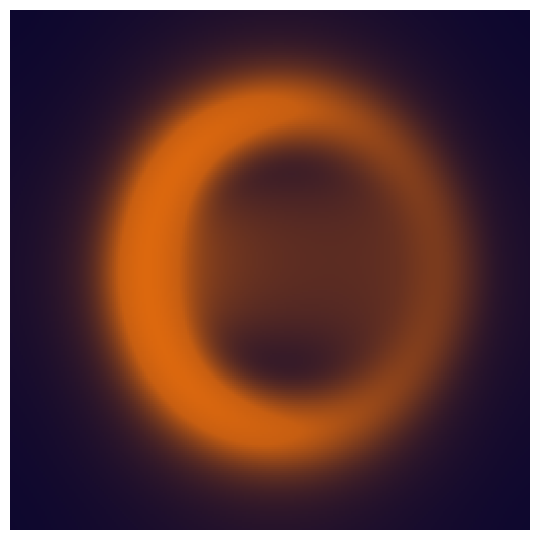}}
	\subfigure[$a=0.65,\theta=85^\circ$]{\includegraphics[scale=0.35]{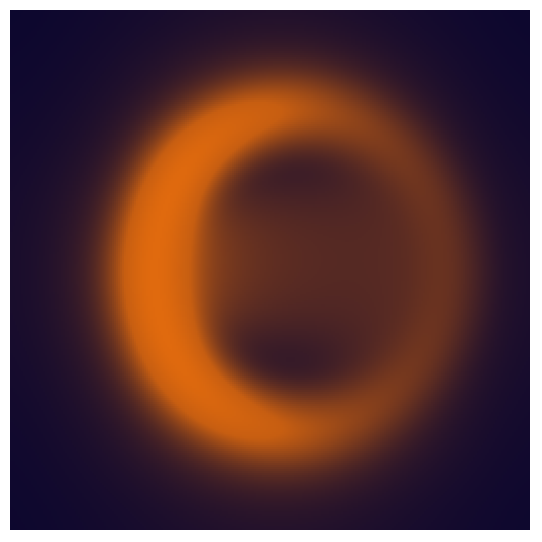}}
	\subfigure[$a=0.8,\theta=85^\circ$]{\includegraphics[scale=0.35]{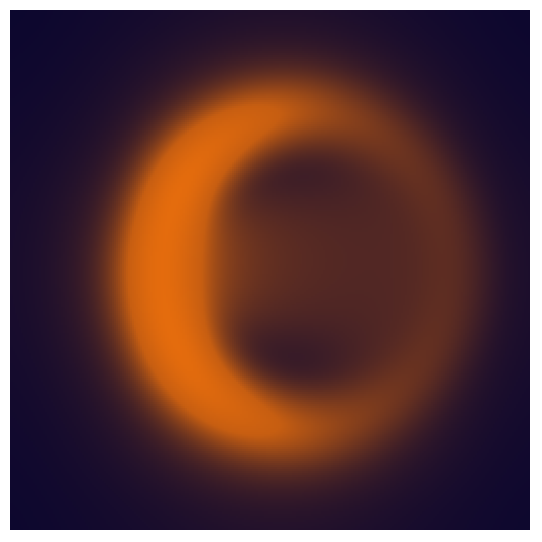}}
	\subfigure[$a=0.95,\theta=85^\circ$]{\includegraphics[scale=0.35]{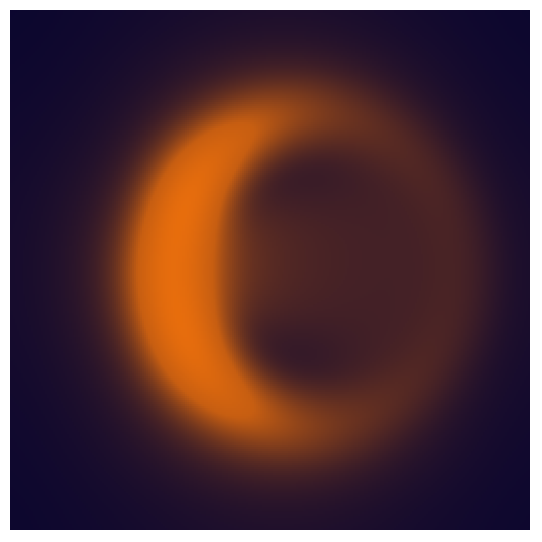}}
	
	\caption{Blurred images obtained using a Gaussian filter with a standard deviation of $1/12$ of the field of view $\gamma_{\rm fov}$. The parameters are the same as those used in Figure~5.}
\end{figure}

\subsection{RIAF Model under Anisotropic Radiation}   
Figures~5 and 6 show the effects of the spin parameter $a$ and observer inclination $\theta$ on the black hole shadow images of the RIAF model under anisotropic radiation, along with the corresponding blurred images. Analysis of these images indicates that at high inclinations, compared with isotropic radiation, the brightness distribution becomes highly uneven, with two pronounced dark regions appearing within the bright ring. Additionally, the higher-order images exhibit enhanced intensity near the polar regions. This asymmetry arises from the angular dependence of the synchrotron emissivity: photons emitted from the upper and lower regions of the accretion disk propagate nearly perpendicular to the magnetic field, enhancing emission in these regions and producing a vertically elongated bright ring. Similar to Figure~1, increasing the spin parameter $a$ results in higher intensity on the left side of the shadow. In the blurred images of Figure~6, the distinction between higher-order and primary images becomes less clear, and the horizon’s outline is less discernible.

\begin{figure}[!htbp]
	\centering 
	
	\subfigure[$Q^2=0.1,\theta=0.001^\circ$]{\includegraphics[scale=0.45]{theta=0du,a=0.5yi.pdf}}
	\subfigure[$Q^2=0.5,\theta=0.001^\circ$]{\includegraphics[scale=0.45]{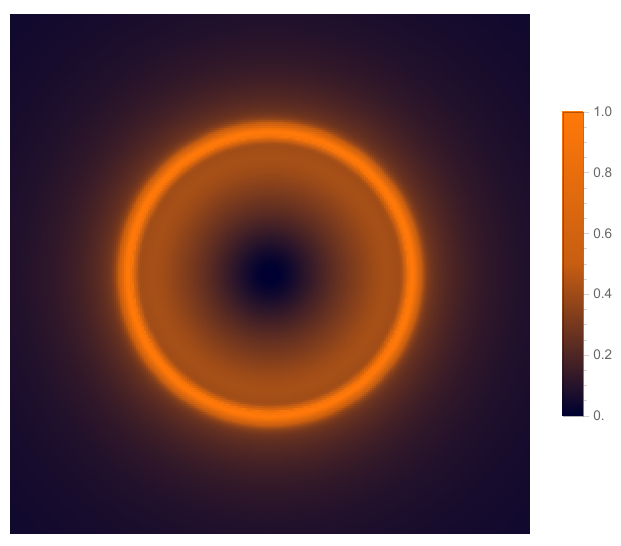}}
	\subfigure[$Q^2=1.0,\theta=0.001^\circ$]{\includegraphics[scale=0.45]{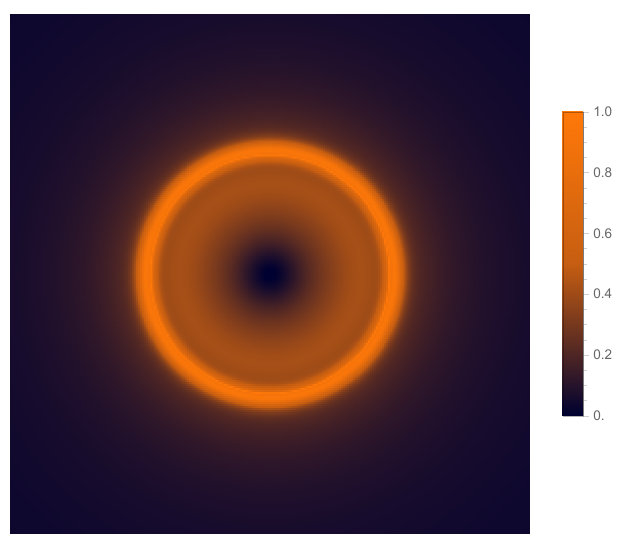}}

	\subfigure[$Q^2=0.1,\theta=17^\circ$]{\includegraphics[scale=0.45]{theta=17du,a=0.5yi.pdf}}
	\subfigure[$Q^2=0.5,\theta=17^\circ$]{\includegraphics[scale=0.45]{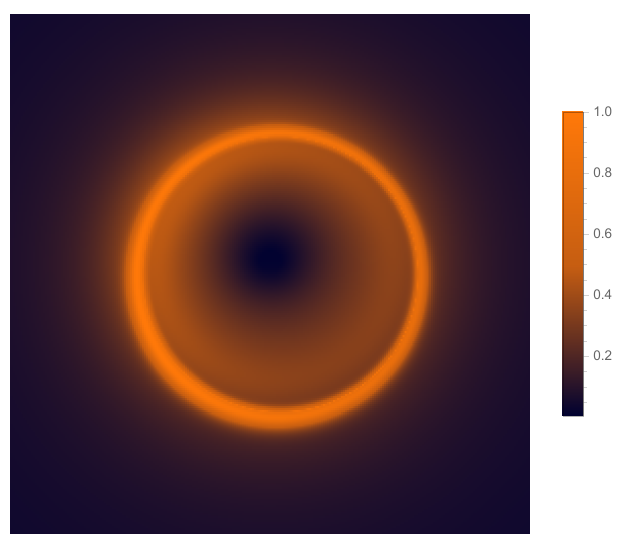}}
	\subfigure[$Q^2=1.0,\theta=17^\circ$]{\includegraphics[scale=0.45]{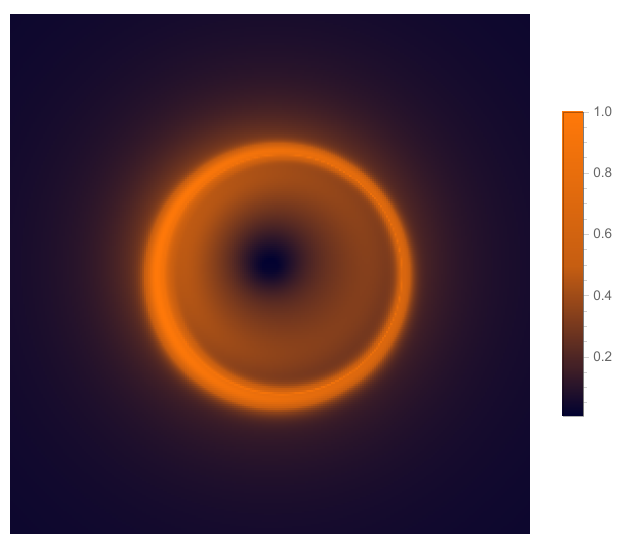}}

	\subfigure[$Q^2=0.1,\theta=60^\circ$]{\includegraphics[scale=0.45]{theta=60du,a=0.5yi.pdf}}
	\subfigure[$Q^2=0.5,\theta=60^\circ$]{\includegraphics[scale=0.45]{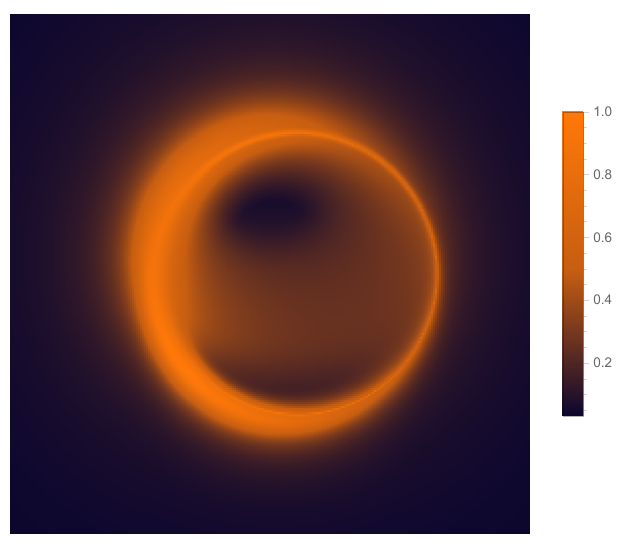}}
	\subfigure[$Q^2=1.0,\theta=60^\circ$]{\includegraphics[scale=0.45]{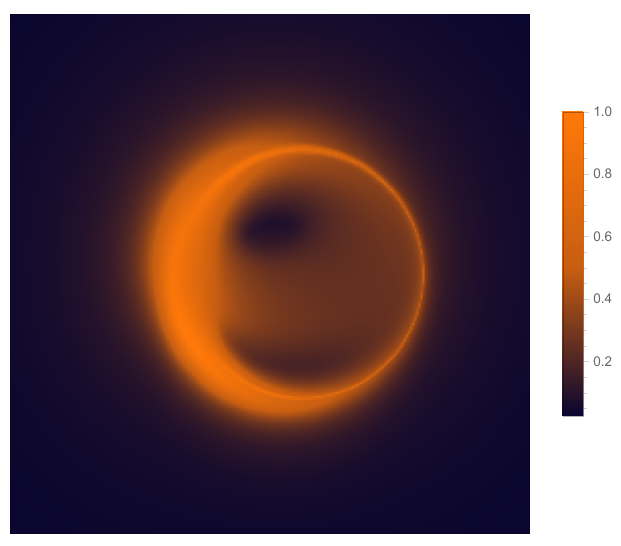}}
	
	\subfigure[$Q^2=0.1,\theta=85^\circ$]{\includegraphics[scale=0.45]{theta=85du,a=0.5yi.pdf}}
	\subfigure[$Q^2=0.5,\theta=85^\circ$]{\includegraphics[scale=0.45]{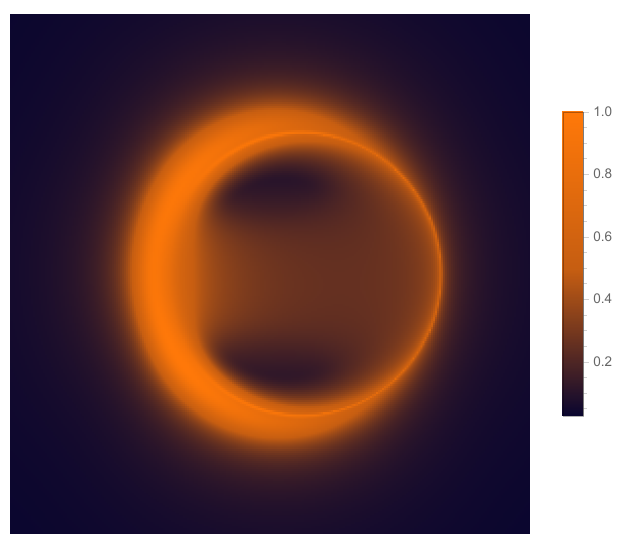}}
	\subfigure[$Q^2=1.0,\theta=85^\circ$]{\includegraphics[scale=0.45]{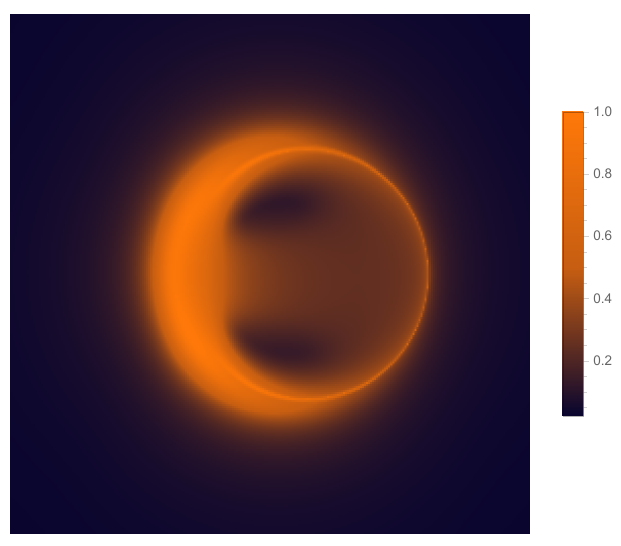}}
	
	\caption{Black hole shadow images of the RIAF model under anisotropic radiation, with $a = 0.5$.}
	
\end{figure}

\begin{figure}[!htbp]
	\centering 
	
	\subfigure[$Q^2=0.1,\theta=85^\circ$]{\includegraphics[scale=0.45]{theta=85du,a=0.5yivague.pdf}}
	\subfigure[$Q^2=0.5,\theta=85^\circ$]{\includegraphics[scale=0.45]{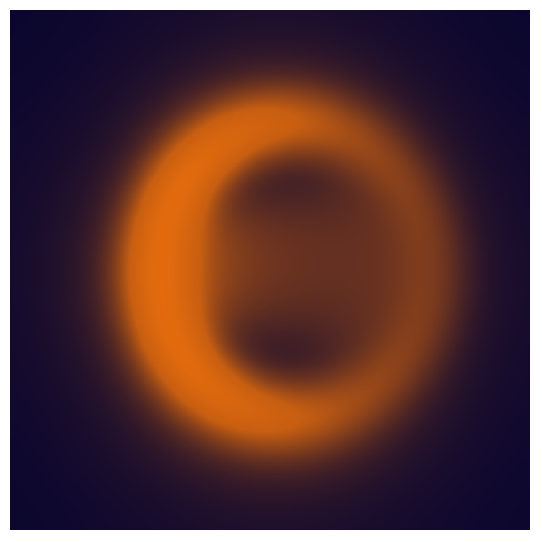}}
	\subfigure[$Q^2=1.0,\theta=85^\circ$]{\includegraphics[scale=0.45]{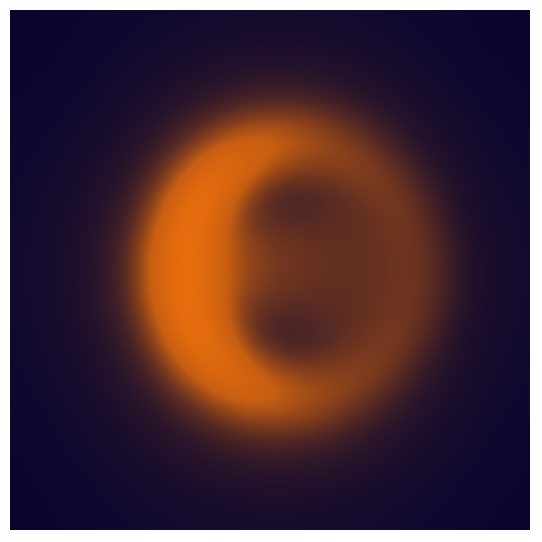}}
	
	\caption{Blurred images obtained using a Gaussian filter with a standard deviation of $1/12$ of the field of view $\gamma_{\rm fov}$. The parameters are the same as those used in Figure~7.}
\end{figure}

Figures~7 and 8 illustrate the effects of the black hole charge $Q$ on the shadow images, along with the corresponding blurred images. Observing these images, we find that, as in the case of isotropic radiation, under anisotropic radiation increasing the charge $Q$ leads to a decrease in the size of both the higher-order images and the inner shadow, while the width of the higher-order images increases.

\subsection{BAAF model}

\begin{figure}[!htbp]
	\centering

	\subfigure[$a=0.5,\theta=0.001^\circ$]{\includegraphics[scale=0.35]{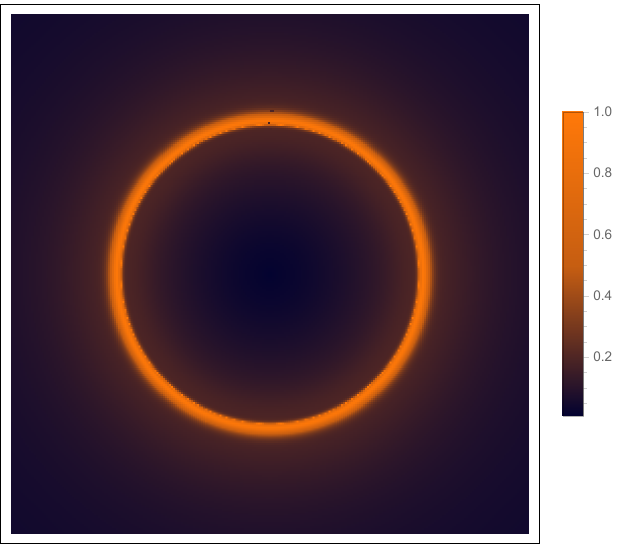}}
	\subfigure[$a=0.65,\theta=0.001^\circ$]{\includegraphics[scale=0.35]{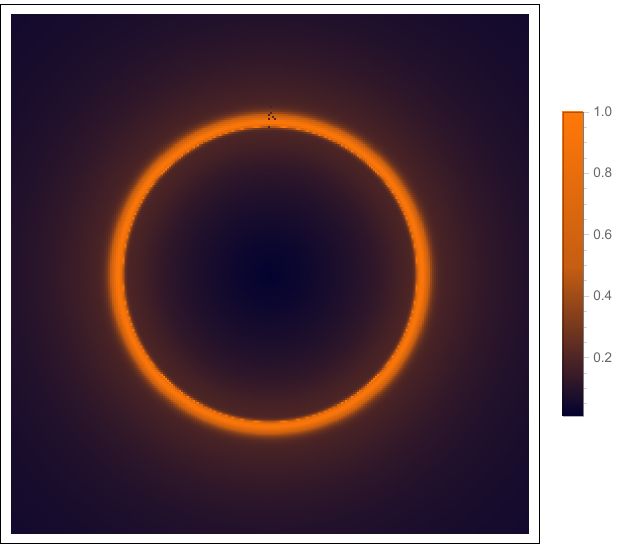}}
	\subfigure[$a=0.8,\theta=0.001^\circ$]{\includegraphics[scale=0.35]{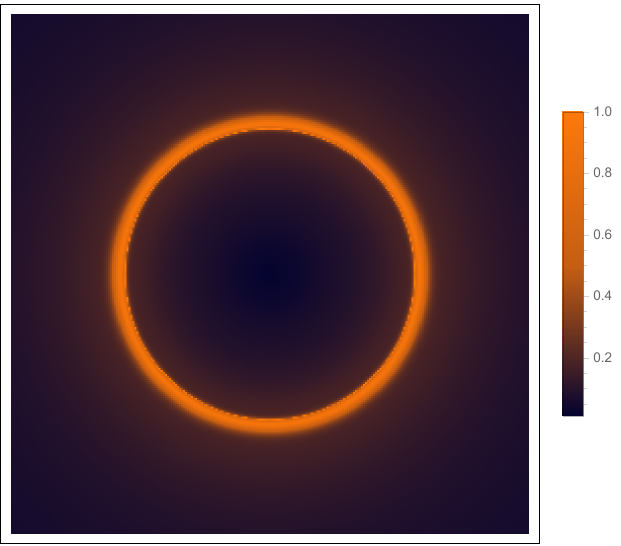}}
	\subfigure[$a=0.95,\theta=0.001^\circ$]{\includegraphics[scale=0.35]{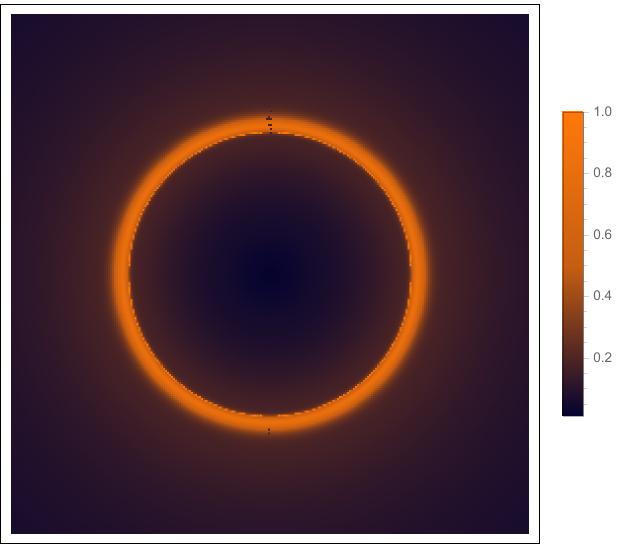}}
	
	\subfigure[$a=0.5,\theta=17^\circ$]{\includegraphics[scale=0.35]{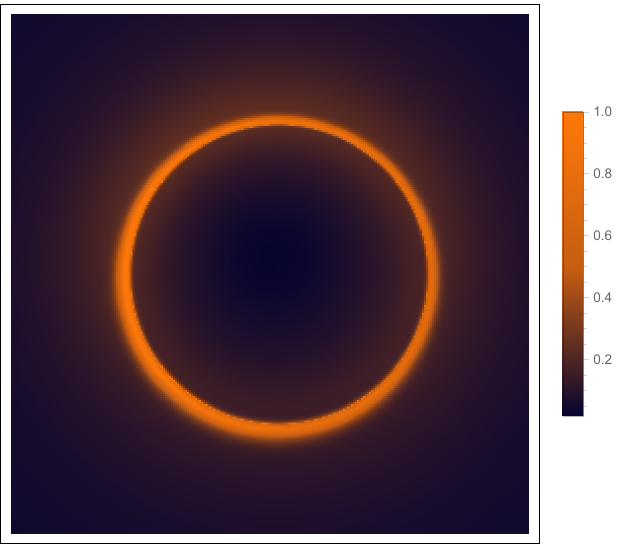}}
	\subfigure[$a=0.65,\theta=17^\circ$]{\includegraphics[scale=0.35]{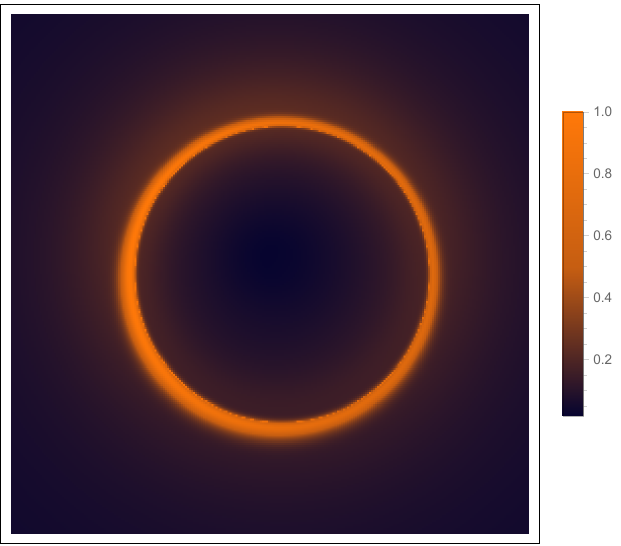}}
	\subfigure[$a=0.8,\theta=17^\circ$]{\includegraphics[scale=0.35]{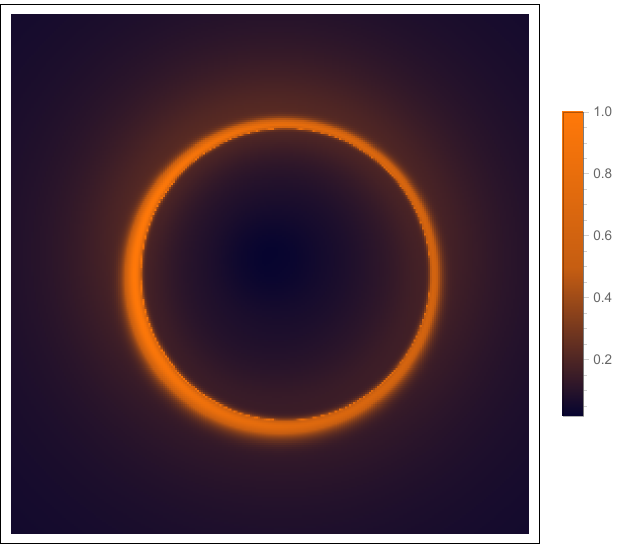}}
	\subfigure[$a=0.95,\theta=17^\circ$]{\includegraphics[scale=0.35]{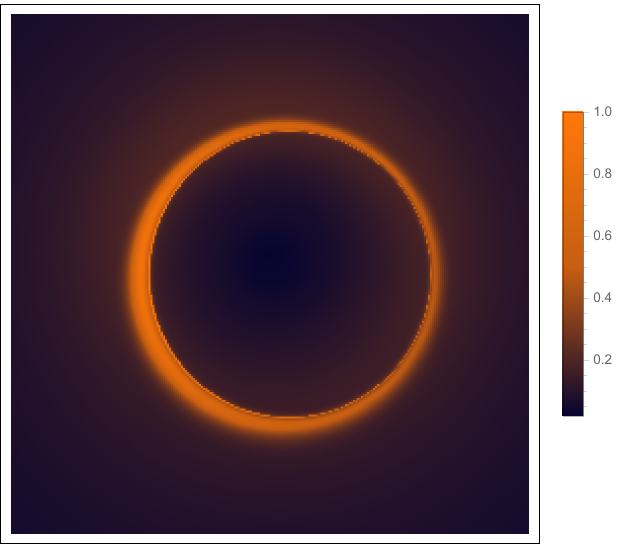}}
	
	\subfigure[$a=0.5,\theta=60^\circ$]{\includegraphics[scale=0.35]{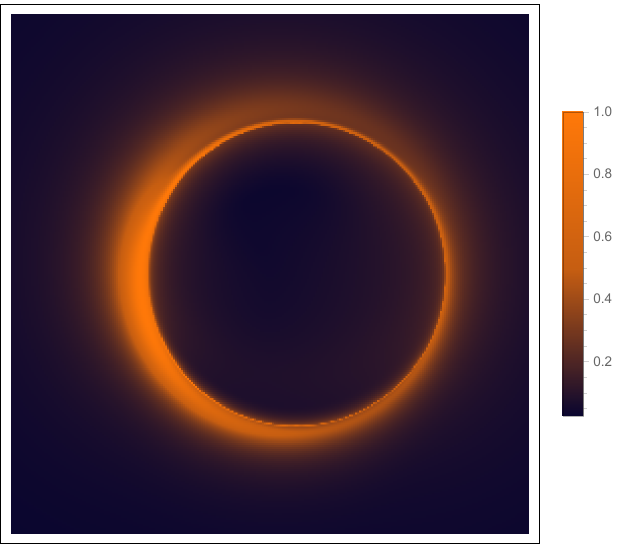}}
	\subfigure[$a=0.65,\theta=60^\circ$]{\includegraphics[scale=0.35]{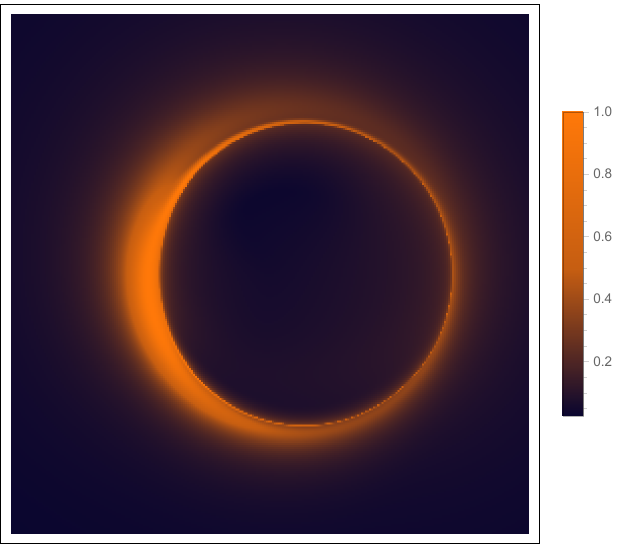}}
	\subfigure[$a=0.8,\theta=60^\circ$]{\includegraphics[scale=0.35]{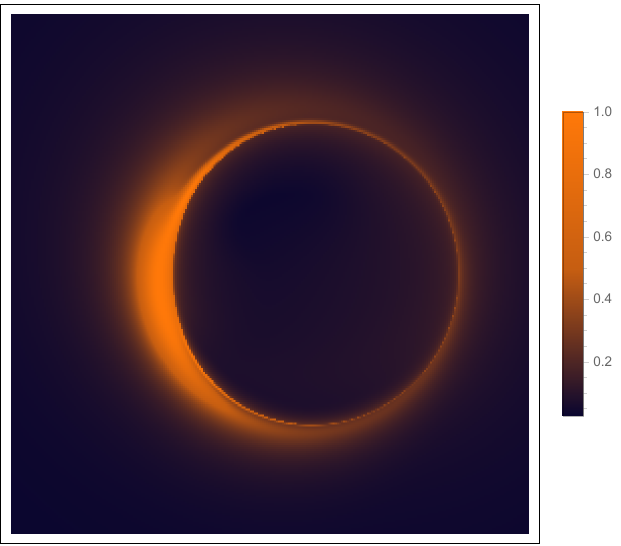}}
	\subfigure[$a=0.95,\theta=60^\circ$]{\includegraphics[scale=0.35]{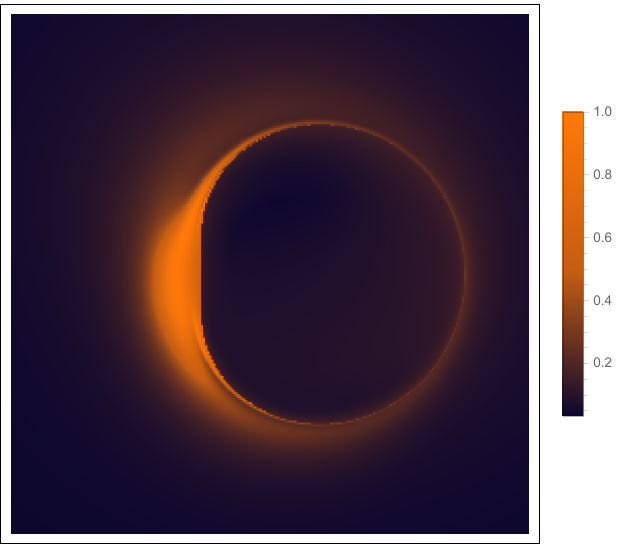}}
	
	\subfigure[$a=0.5,\theta=85^\circ$]{\includegraphics[scale=0.35]{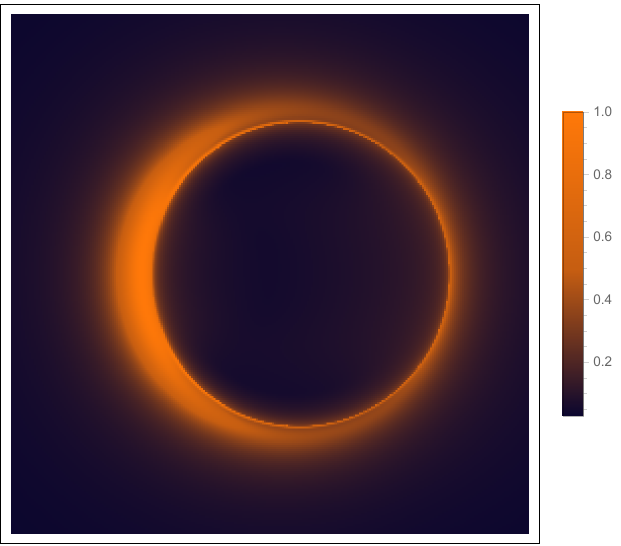}}
	\subfigure[$a=0.65,\theta=85^\circ$]{\includegraphics[scale=0.35]{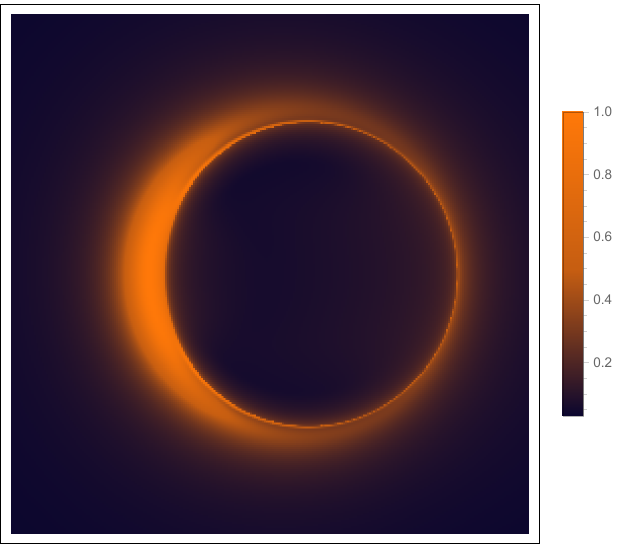}}
	\subfigure[$a=0.8,\theta=85^\circ$]{\includegraphics[scale=0.35]{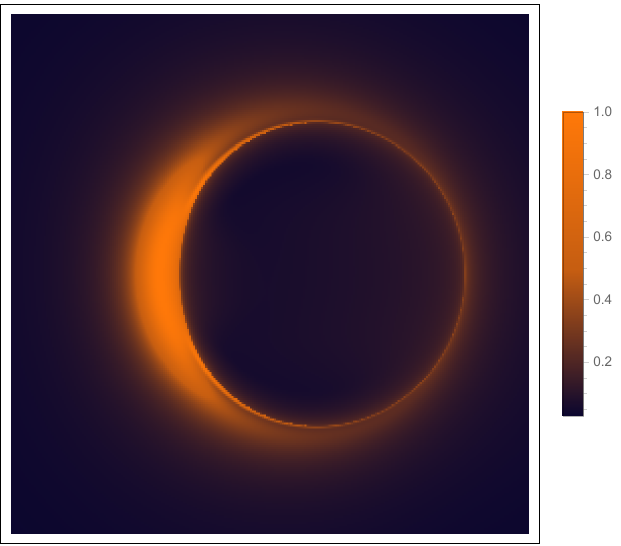}}
	\subfigure[$a=0.95,\theta=85^\circ$]{\includegraphics[scale=0.35]{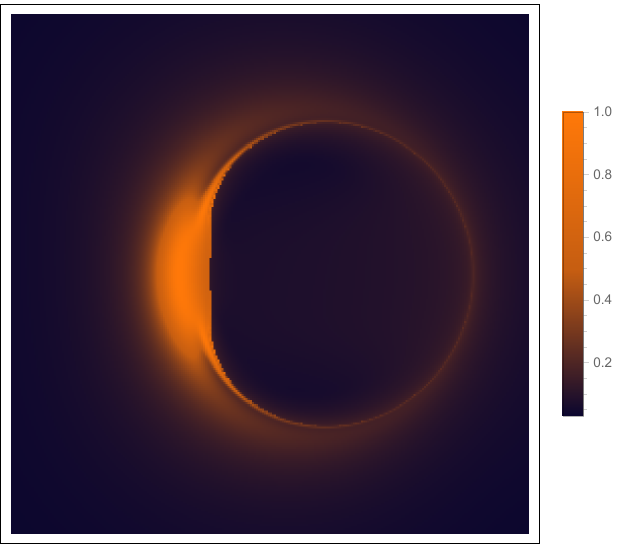}}
	
	\caption{Black hole shadow images of the BAAF model, with $Q^2 = 0.1$.}
	
\end{figure}

\begin{figure}[!htbp]
	\centering 
	
	\subfigure[$a=0.5,\theta=85^\circ$]{\includegraphics[scale=0.35]{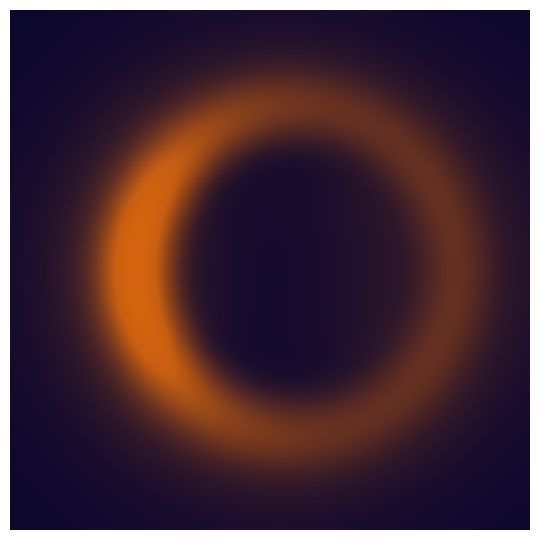}}
	\subfigure[$a=0.65,\theta=85^\circ$]{\includegraphics[scale=0.35]{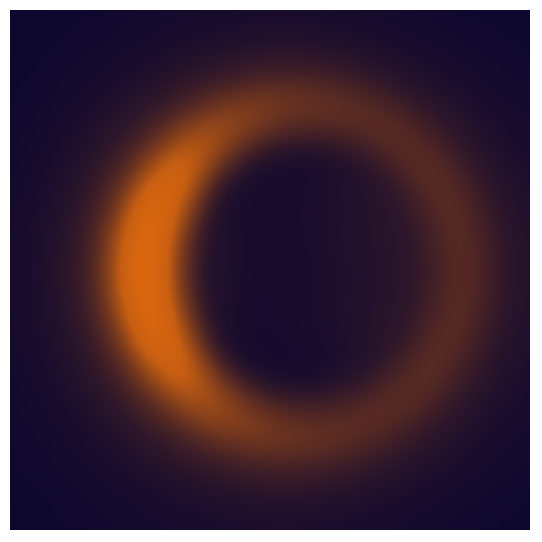}}
	\subfigure[$a=0.8,\theta=85^\circ$]{\includegraphics[scale=0.35]{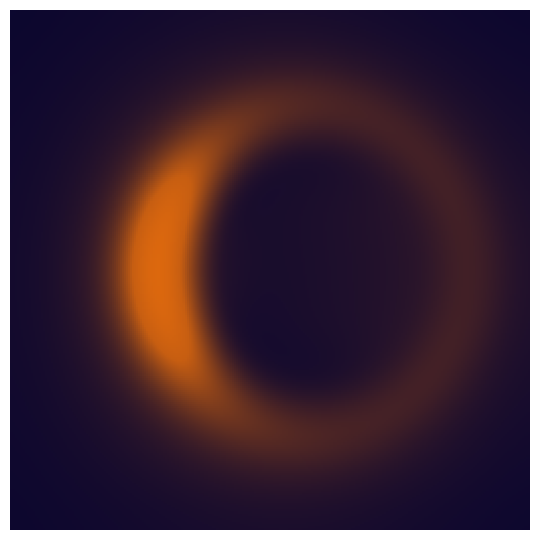}}
	\subfigure[$a=0.95,\theta=85^\circ$]{\includegraphics[scale=0.35]{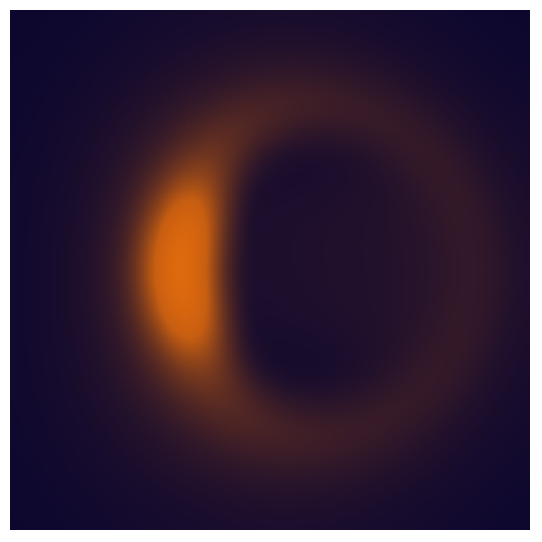}}
	
	\caption{Blurred images obtained using a Gaussian filter with a standard deviation of $1/12$ of the field of view $\gamma_{\rm fov}$. The parameters are the same as those used in Figure~9.}
\end{figure}

Figure~9 shows the effects of the spin parameter $a$ and the observer inclination $\theta$ on the BAAF model, while Figure~10 presents the corresponding blurred images. Overall, the dependence of the image morphology on the black hole spin $a$ and inclination $\theta$ in the BAAF model is qualitatively similar to that observed in the RIAF model. As in the isotropic RIAF case, for small values of $a$, the higher-order images appear nearly circular. However, compared with the RIAF model, the bright ring in the BAAF model is typically narrower, and the separation between the primary and higher-order images is more distinct. Moreover, in the BAAF model, a continuous region of reduced intensity appears within the higher-order images. At large inclinations, the higher-order images in the BAAF model do not exhibit the two distinct dark regions seen in the RIAF model. This indicates that, in the RIAF model, radiation from regions away from the equatorial plane more effectively obscures the outline of the event horizon. These differences likely originate from the fact that, for the chosen parameters, the BAAF flow treated under the conical approximation is structurally thinner than the corresponding RIAF configuration. A comparison between the BAAF and RIAF models shows that the former is more consistent with observations from the EHT.

\begin{figure}[!htbp]
	\centering 
	
	\subfigure[$Q^2=0.1,\theta=0.001^\circ$]{\includegraphics[scale=0.45]{theta=0du,a=0.5hou.pdf}}
	\subfigure[$Q^2=0.5,\theta=0.001^\circ$]{\includegraphics[scale=0.45]{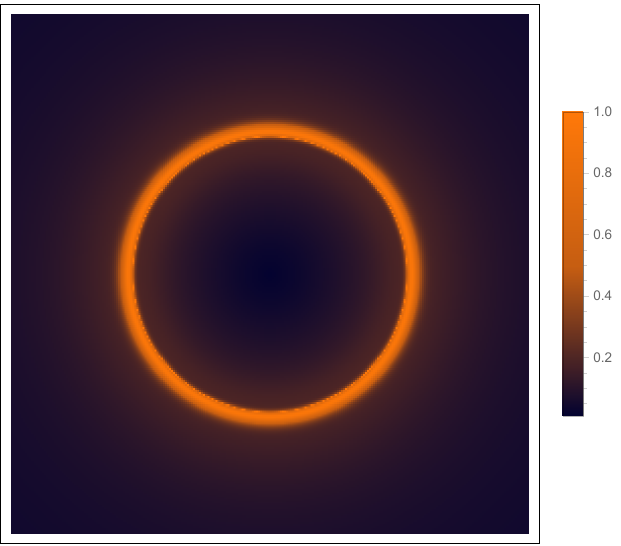}}
	\subfigure[$Q^2=1.0,\theta=0.001^\circ$]{\includegraphics[scale=0.45]{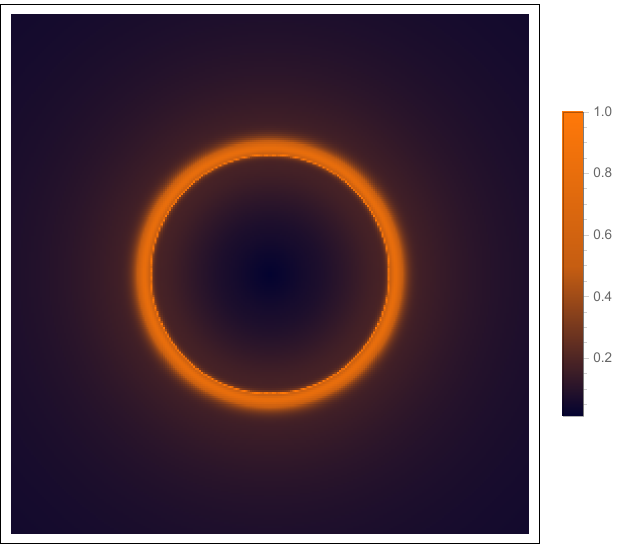}}
	
	\subfigure[$Q^2=0.1,\theta=17^\circ$]{\includegraphics[scale=0.45]{theta=17du,a=0.5hou.pdf}}
	\subfigure[$Q^2=0.5,\theta=17^\circ$]{\includegraphics[scale=0.45]{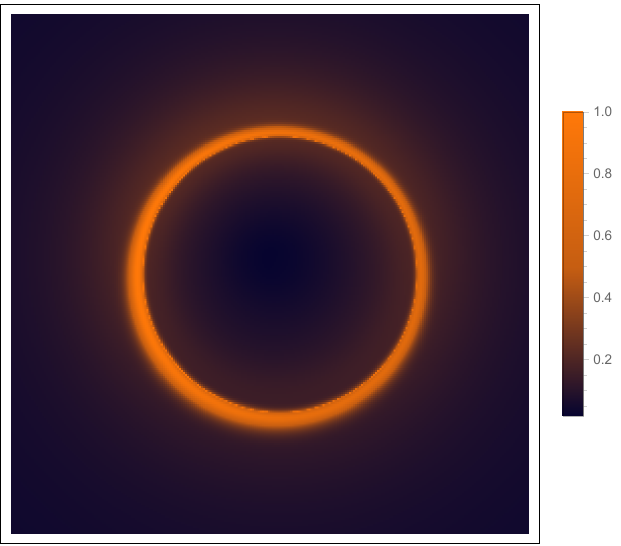}}
	\subfigure[$Q^2=1.0,\theta=17^\circ$]{\includegraphics[scale=0.45]{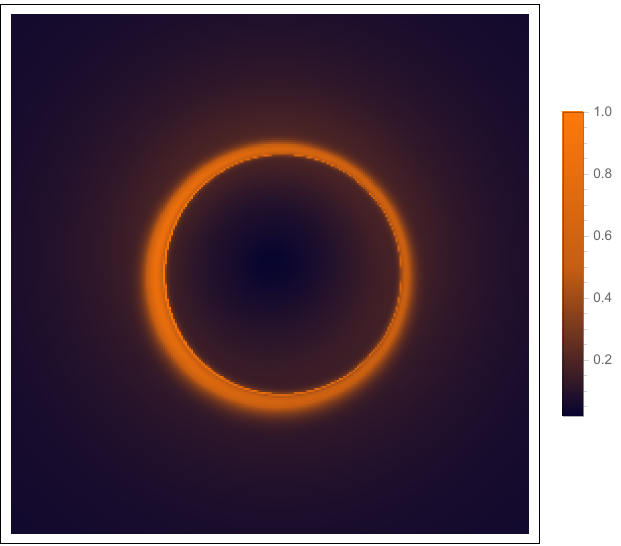}}
	
	\subfigure[$Q^2=0.1,\theta=60^\circ$]{\includegraphics[scale=0.45]{theta=60du,a=0.5hou.pdf}}
	\subfigure[$Q^2=0.5,\theta=60^\circ$]{\includegraphics[scale=0.45]{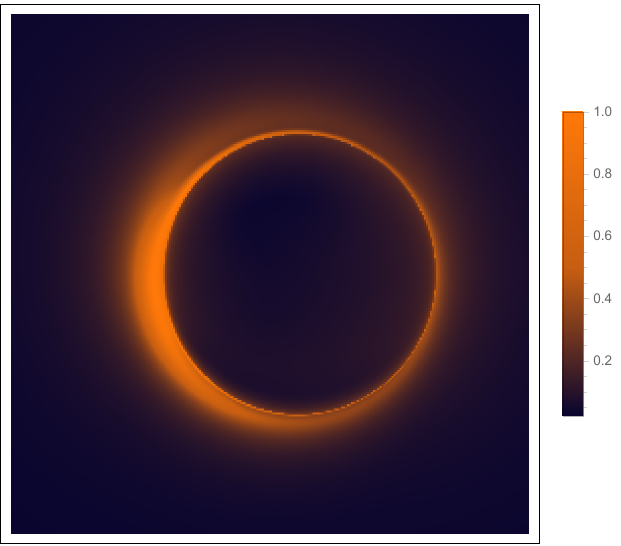}}
	\subfigure[$Q^2=1.0,\theta=60^\circ$]{\includegraphics[scale=0.45]{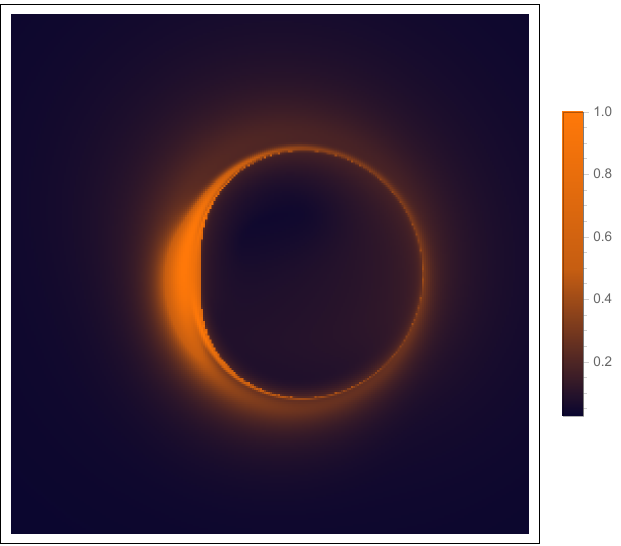}}
	
	\subfigure[$Q^2=0.1,\theta=85^\circ$]{\includegraphics[scale=0.45]{theta=85du,a=0.5hou.pdf}}
	\subfigure[$Q^2=0.5,\theta=85^\circ$]{\includegraphics[scale=0.45]{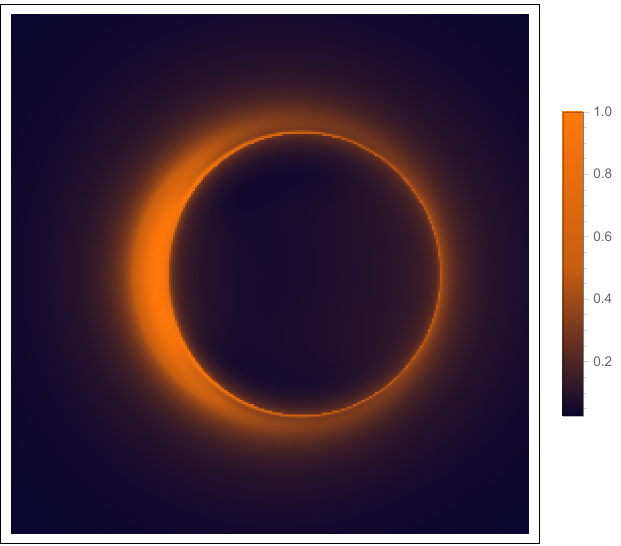}}
	\subfigure[$Q^2=1.0,\theta=85^\circ$]{\includegraphics[scale=0.45]{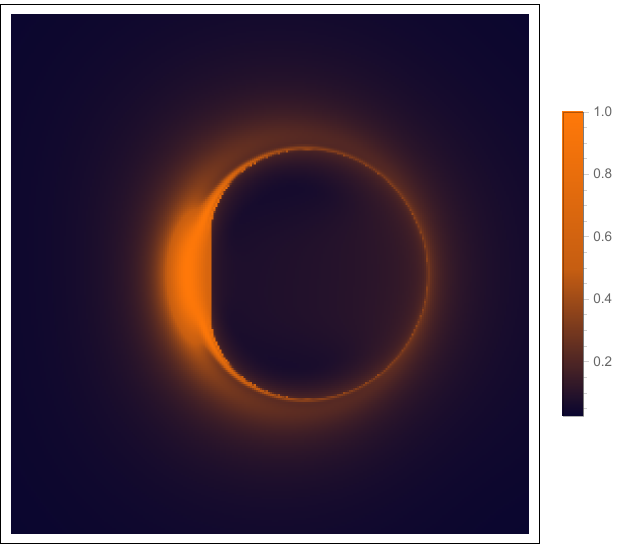}}
	
	\caption{Black hole shadow images of the BAAF model, with $a = 0.5$.}
	
\end{figure}

\begin{figure}[!htbp]
	\centering 
	
	\subfigure[$Q^2=0.1,\theta=85^\circ$]{\includegraphics[scale=0.45]{theta=85du,a=0.5houvague.pdf}}
	\subfigure[$Q^2=0.5,\theta=85^\circ$]{\includegraphics[scale=0.45]{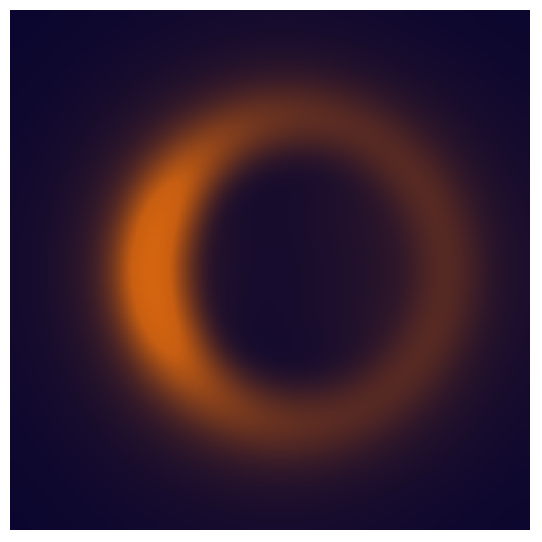}}
	\subfigure[$Q^2=1.0,\theta=85^\circ$]{\includegraphics[scale=0.45]{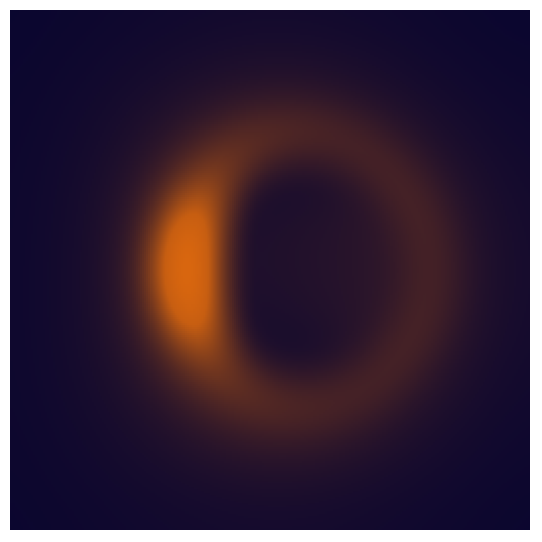}}

	\caption{Blurred images obtained using a Gaussian filter with a standard deviation of $1/12$ of the field of view $\gamma_{\rm fov}$. The parameters are the same as those used in Figure~11.}
\end{figure}

Figure~11 illustrates the effects of the black hole charge $Q$ on the BAAF model images, while Figure~12 shows the corresponding blurred images. Overall, the influence of the charge $Q$ on the higher-order images in the BAAF model is qualitatively similar to that in the RIAF model. However, notable differences are present. In the BAAF model, the region associated with the primary image is significantly reduced, whereas in the RIAF model the intensity outside the higher-order images remains relatively strong. This indicates that, in the BAAF model, the observed intensity is predominantly contributed by the higher-order images.

\subsection{Polarized Images in the BAAF Model}

\begin{figure}[!htbp]
	\centering 
	\subfigure[$Stokes$ $\mathcal I_o$]{\includegraphics[scale=0.25]{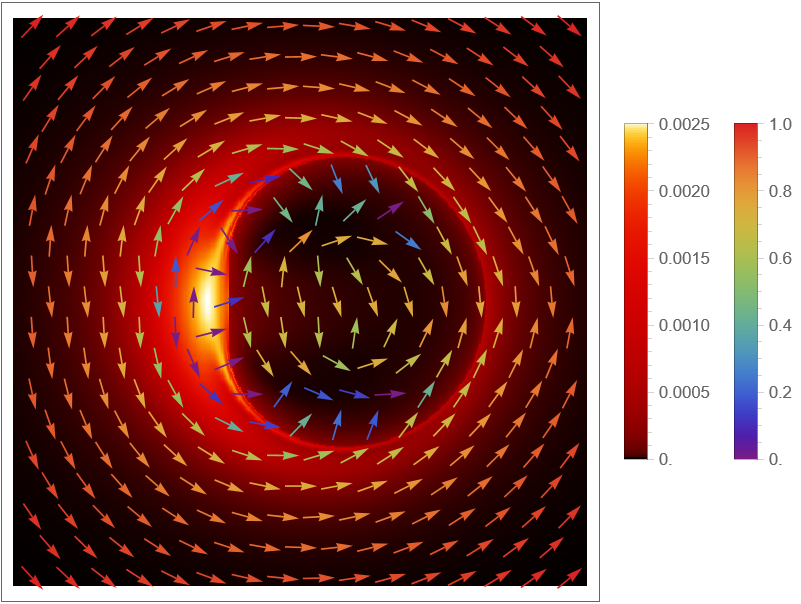}}
	\subfigure[$Stokes$ $\mathcal Q_o$]{\includegraphics[scale=0.25]{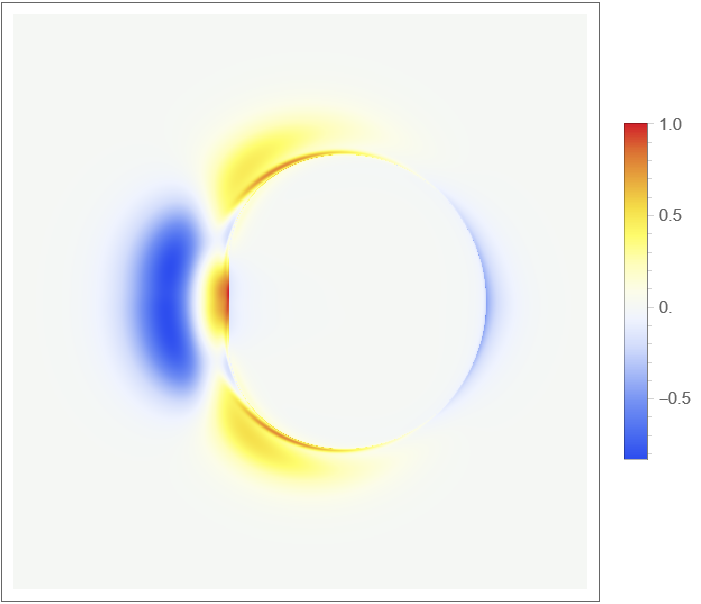}}
	\subfigure[$Stokes$ $\mathcal U_o$]{\includegraphics[scale=0.25]{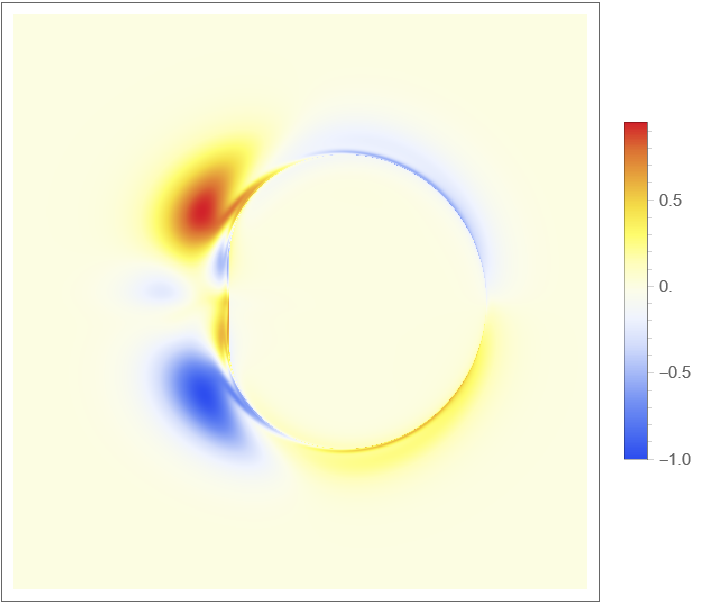}}
	\subfigure[$Stokes$ $\mathcal V_o$]{\includegraphics[scale=0.25]{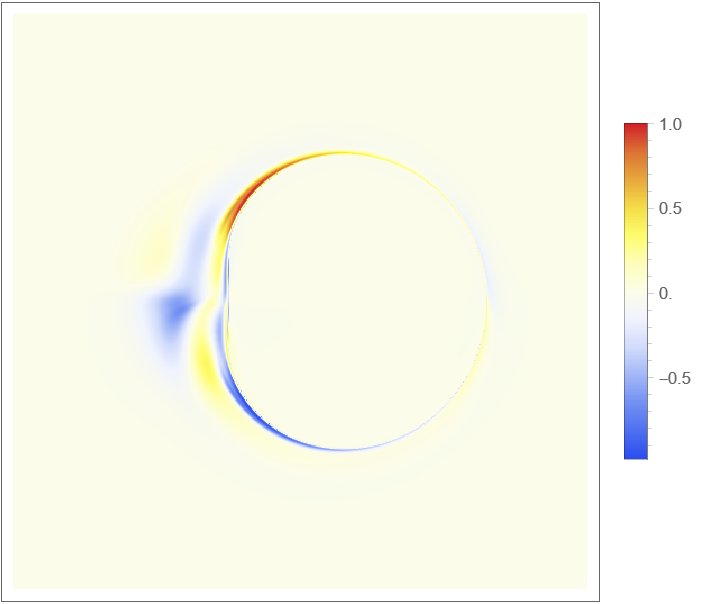}}

	\caption{The resulting Stokes parameters $\mathcal{I}$, $\mathcal{Q}$, $\mathcal{U}$, and $\mathcal{V}$ for the RAAF model. The accretion flow follows the conical solution, with parameters fixed at $a = 0.5$, $Q^2 = 1.0$, $\theta = 85^\circ$, and an observation frequency of $230~\mathrm{GHz}$.}
\end{figure}

Figure~13 presents representative numerical results for the Stokes parameters $\mathcal{I}_o$, $\mathcal{Q}_o$, $\mathcal{U}_o$, and $\mathcal{V}_o$. The accretion flow is modeled using the conical solution, with parameters fixed at $a = 0.5$, $Q^2 = 1.0$, and an observer inclination of $\theta = 85^\circ$. The quantity $\mathcal{I}_o$ describes the intensity distribution. The arrows denote the linear polarization vector $\vec{f}$, with their color indicating the polarization degree $\mathcal{P}_o$ and their orientation representing the EVPA $\Phi_{\mathrm{EVPA}}$. The spatial distributions of $\mathcal{Q}_o$ and $\mathcal{U}_o$ jointly determine the polarization direction, while $\mathcal{V}_o < 0$ corresponds to right-handed circular polarization. Since $\vec{f} \perp \vec{b}$, the magnetic field is inferred to be approximately radial. Closer to the event horizon, frame-dragging effects become dominant, causing the magnetic field to twist toward a more azimuthal configuration. Under the flux-freezing condition of ideal magnetohydrodynamics (MHD), the magnetic field tends to align with the fluid motion. Therefore, the observed rotation of the polarization angle reflects the frame-dragging effect of the black hole on the accreting plasma. The parameters $\mathcal{Q}_o$ and $\mathcal{U}_o$ reach their maxima near the higher-order images and decay rapidly away from this region. The distribution of $\mathcal{V}_o$ indicates right-handed polarization on both sides of the higher-order images, while the remaining regions exhibit left-handed polarization.

\begin{figure}[!htbp]
	\centering

	\subfigure[$a=0.5,\theta=0.001^\circ$]{\includegraphics[scale=0.22]{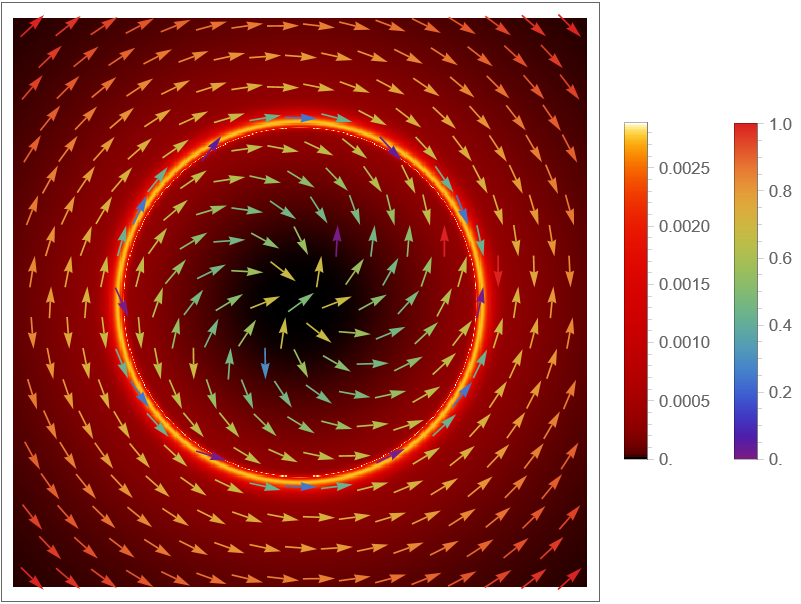}}
	\subfigure[$a=0.65,\theta=0.001^\circ$]{\includegraphics[scale=0.22]{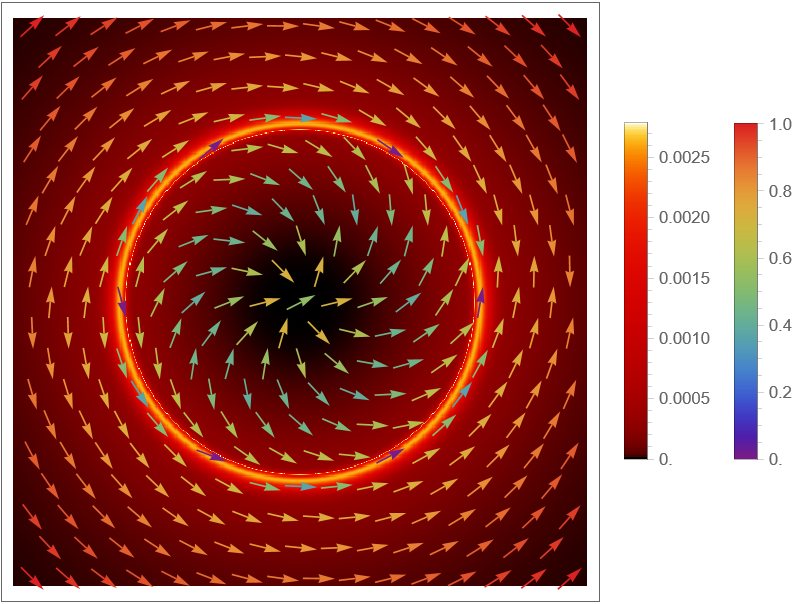}}
	\subfigure[$a=0.8,\theta=0.001^\circ$]{\includegraphics[scale=0.22]{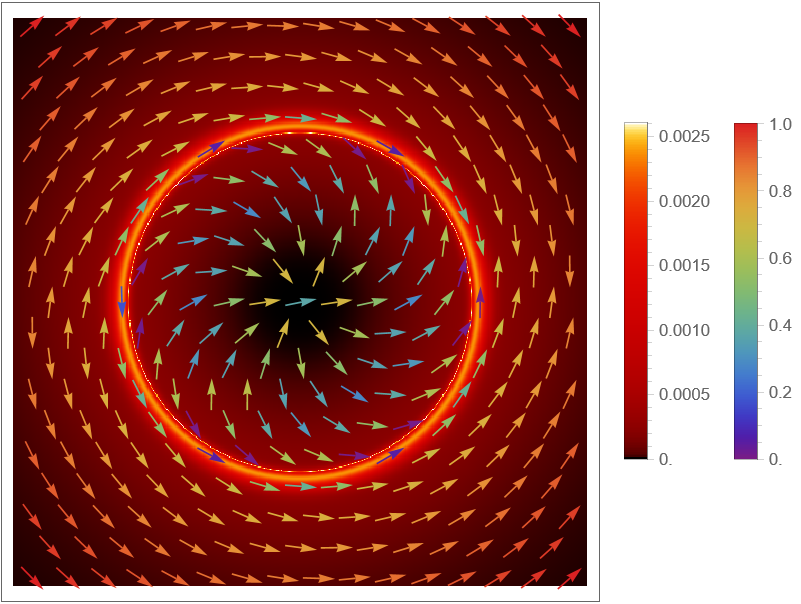}}
	\subfigure[$a=0.95,\theta=0.001^\circ$]{\includegraphics[scale=0.22]{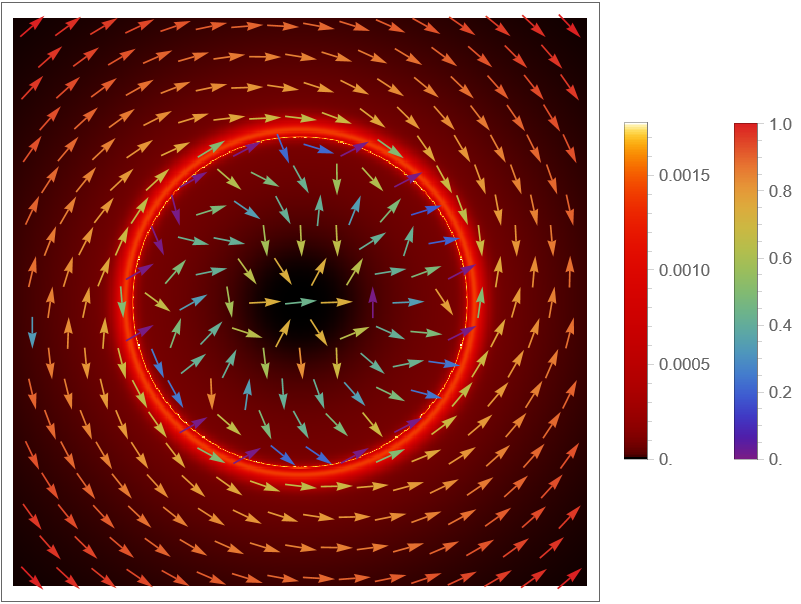}}
	
	\subfigure[$a=0.5,\theta=17^\circ$]{\includegraphics[scale=0.22]{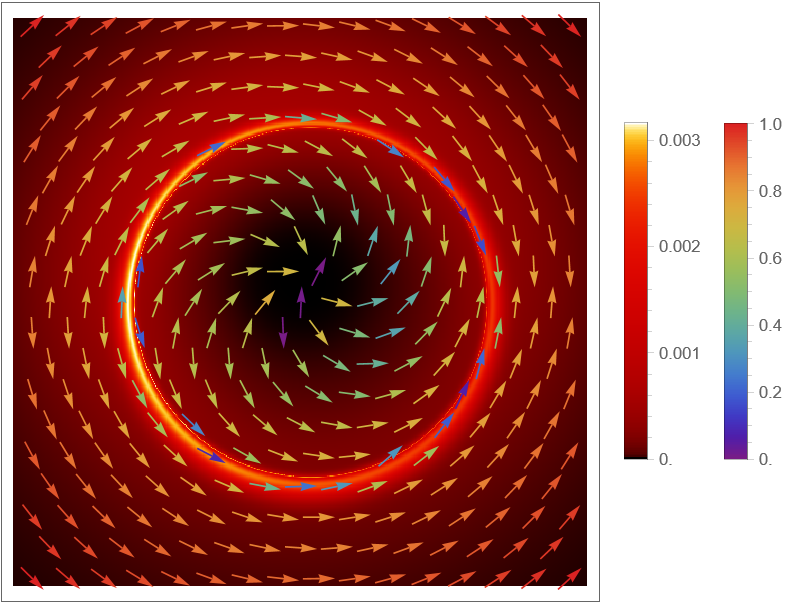}}
	\subfigure[$a=0.65,\theta=17^\circ$]{\includegraphics[scale=0.22]{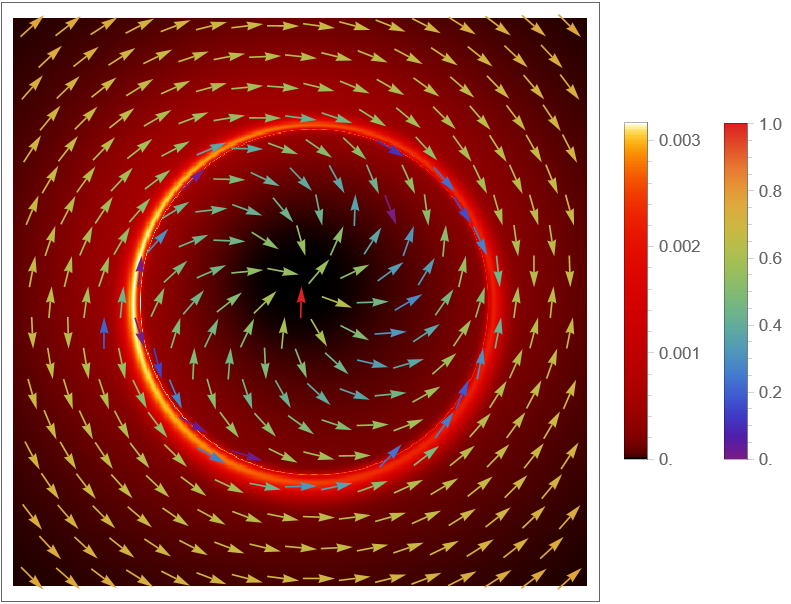}}
	\subfigure[$a=0.8,\theta=17^\circ$]{\includegraphics[scale=0.22]{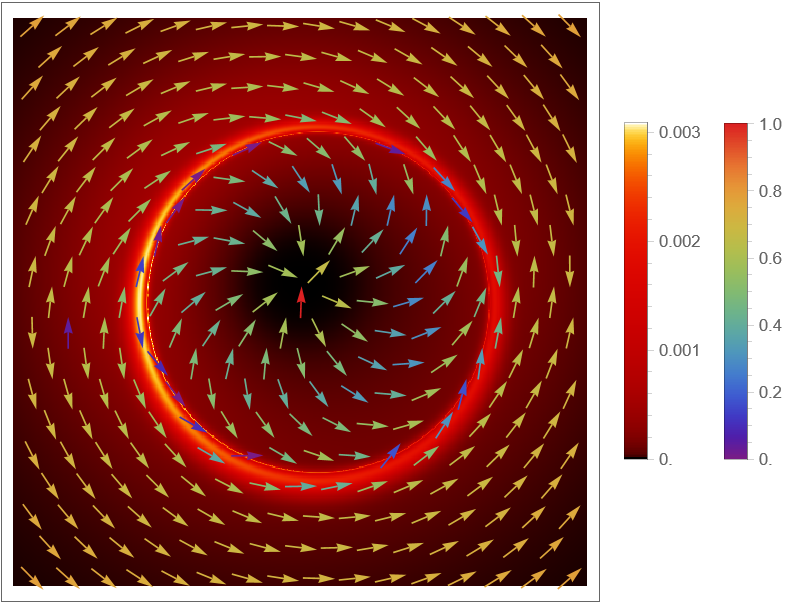}}
	\subfigure[$a=0.95,\theta=17^\circ$]{\includegraphics[scale=0.22]{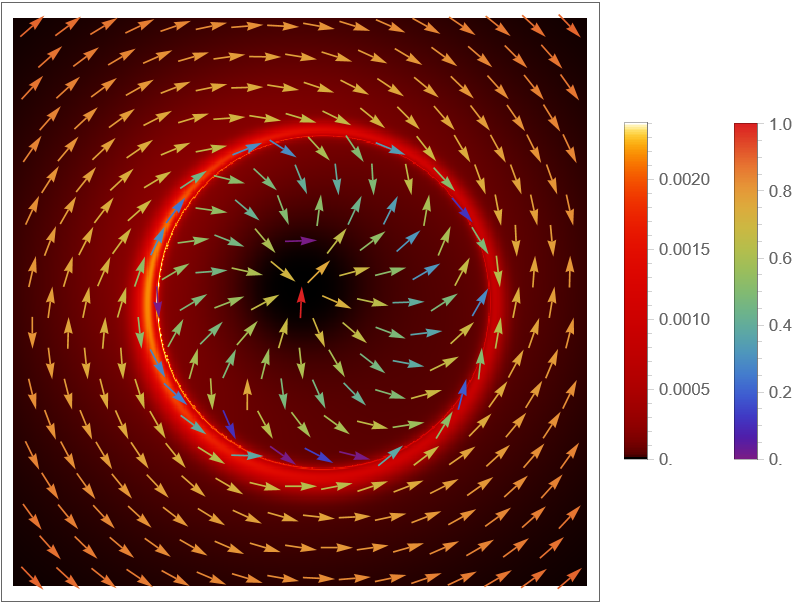}}
	
	\subfigure[$a=0.5,\theta=60^\circ$]{\includegraphics[scale=0.22]{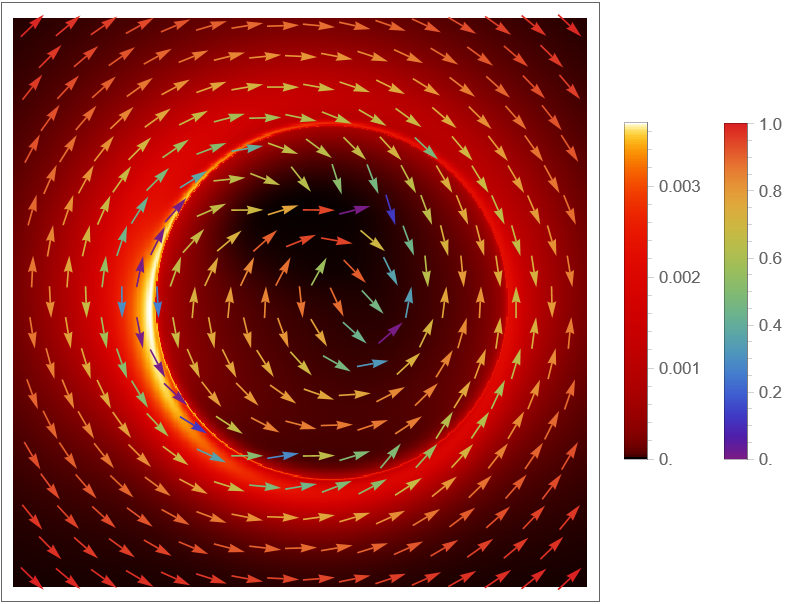}}
	\subfigure[$a=0.65,\theta=60^\circ$]{\includegraphics[scale=0.22]{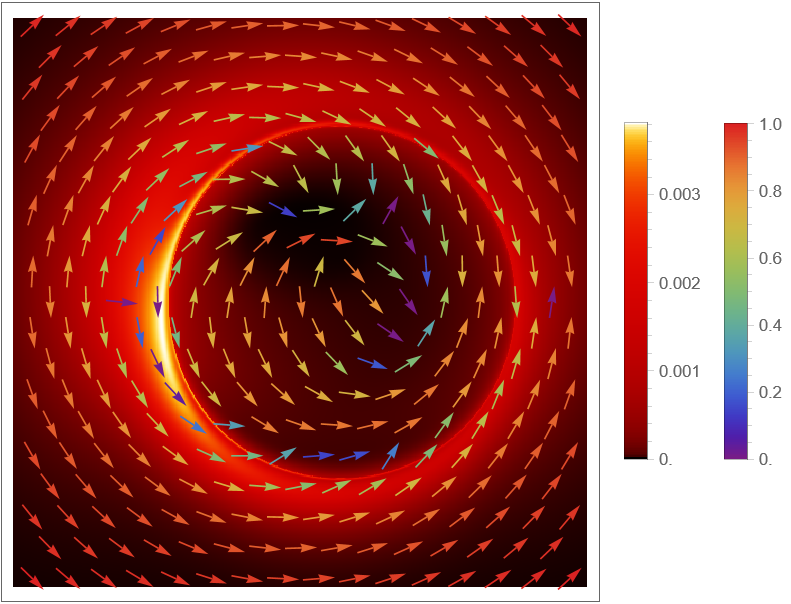}}
	\subfigure[$a=0.8,\theta=60^\circ$]{\includegraphics[scale=0.22]{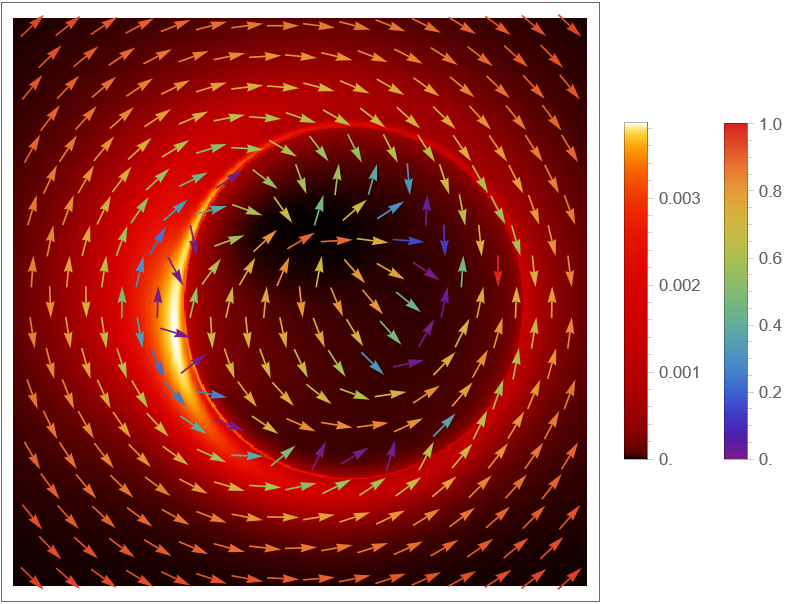}}
	\subfigure[$a=0.95,\theta=60^\circ$]{\includegraphics[scale=0.22]{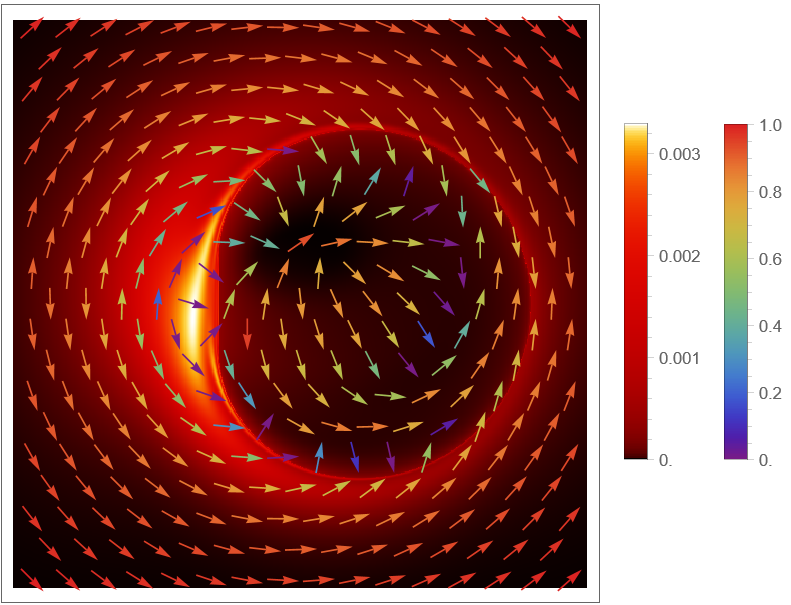}}
	
	\subfigure[$a=0.5,\theta=85^\circ$]{\includegraphics[scale=0.22]{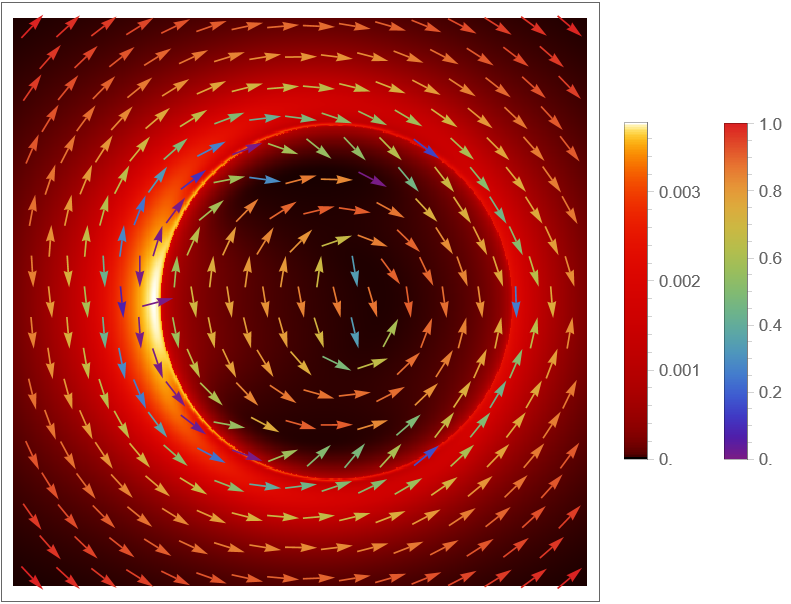}}
	\subfigure[$a=0.65,\theta=85^\circ$]{\includegraphics[scale=0.22]{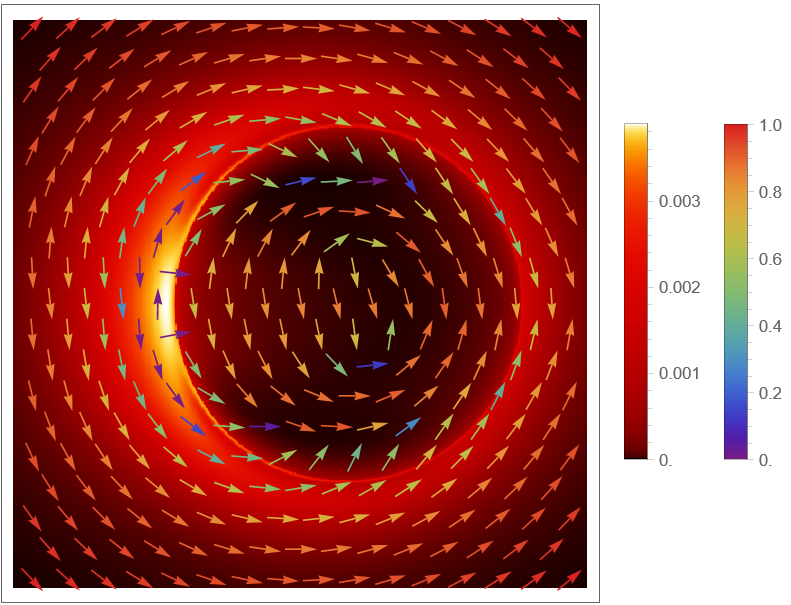}}
	\subfigure[$a=0.8,\theta=85^\circ$]{\includegraphics[scale=0.22]{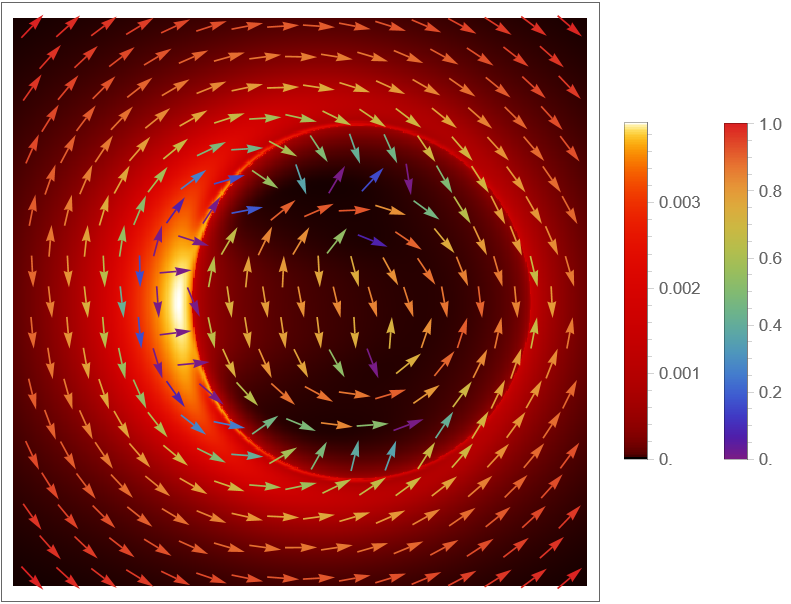}}
	\subfigure[$a=0.95,\theta=85^\circ$]{\includegraphics[scale=0.22]{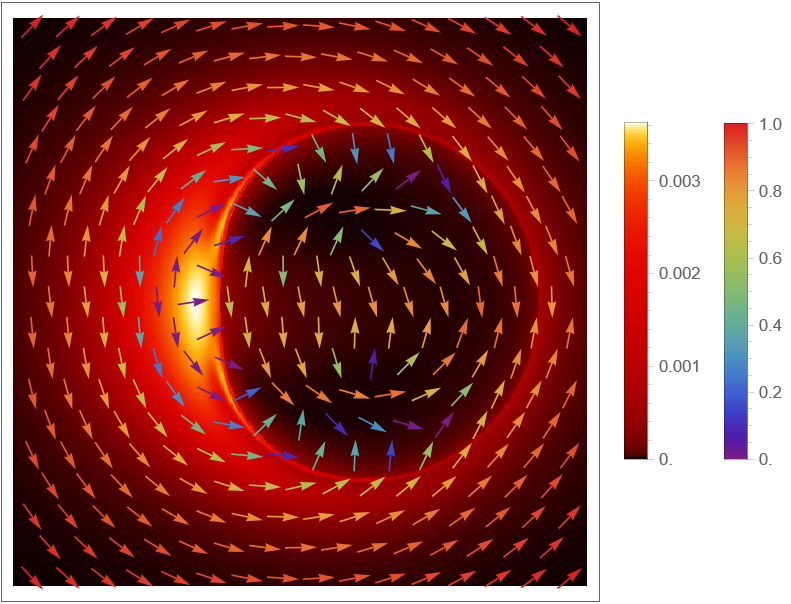}}
	
	\caption{Polarized black hole shadow images for the RAAF disk model. The accretion flow follows the conical solution, with an observation frequency of $230~\mathrm{GHz}$ and $Q^2 = 0.1$.}
\end{figure}

Figure~14 shows the effects of the observer inclination $\theta$ and the spin parameter $a$ on the polarization images, with the black hole charge fixed at $Q^2 = 0.1$. The images indicate that increasing $\theta$ leads to significant changes in the overall polarization morphology. At small inclinations, the polarization vectors are approximately symmetric. As the inclination increases, the distribution of polarization vectors becomes increasingly asymmetric, indicating that both gravitational lensing and frame-dragging strongly influence the observed EVPA $\Phi_{\mathrm{EVPA}}$. Notably, at high inclinations, the profile along the $y$-direction exhibits two distinct dark regions separated by a central bright region. This feature arises from the partial obscuration of the event horizon by the accretion flow. In addition, as the spin parameter $a$ increases, the magnetic field direction near the horizon is progressively distorted by frame-dragging, and the region where the field lines are significantly twisted shifts outward from the black hole center.

\begin{figure}[!htbp]
	\centering

	\subfigure[$Q^2=0.1,\theta=0.001^\circ$]{\includegraphics[scale=0.3]{theta=0du,a=0.5,ll.png}}
	\subfigure[$Q^2=0.5,\theta=0.001^\circ$]{\includegraphics[scale=0.3]{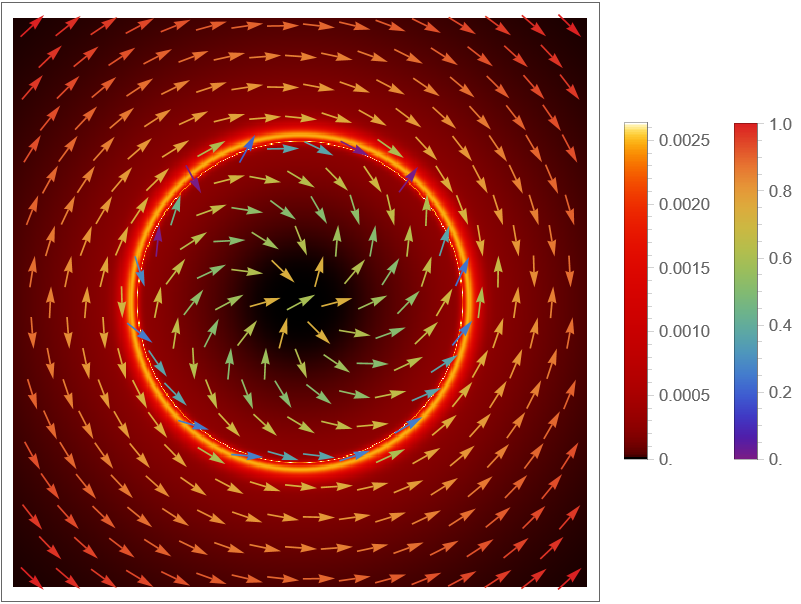}}
	\subfigure[$Q^2=1.0,\theta=0.001^\circ$]{\includegraphics[scale=0.3]{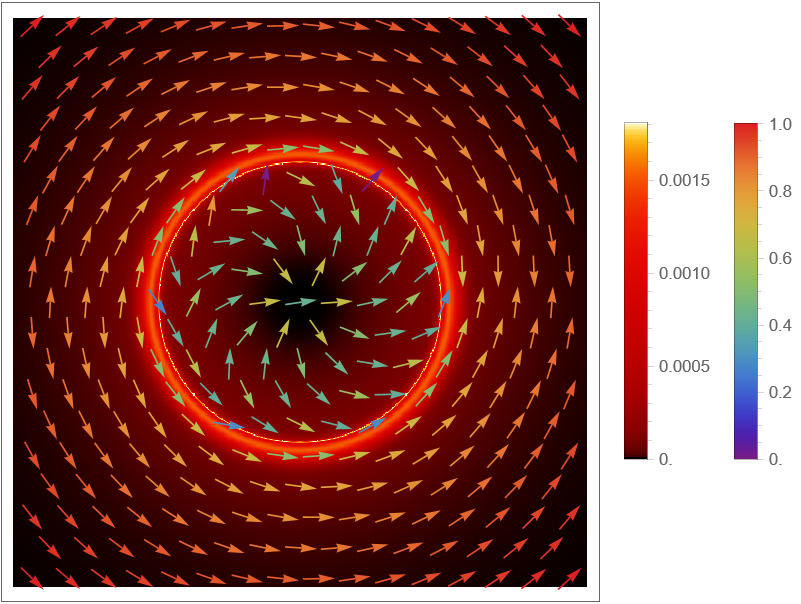}}
	
	\subfigure[$Q^2=0.1,\theta=17^\circ$]{\includegraphics[scale=0.3]{theta=17du,a=0.5,ll.png}}
	\subfigure[$Q^2=0.5,\theta=17^\circ$]{\includegraphics[scale=0.3]{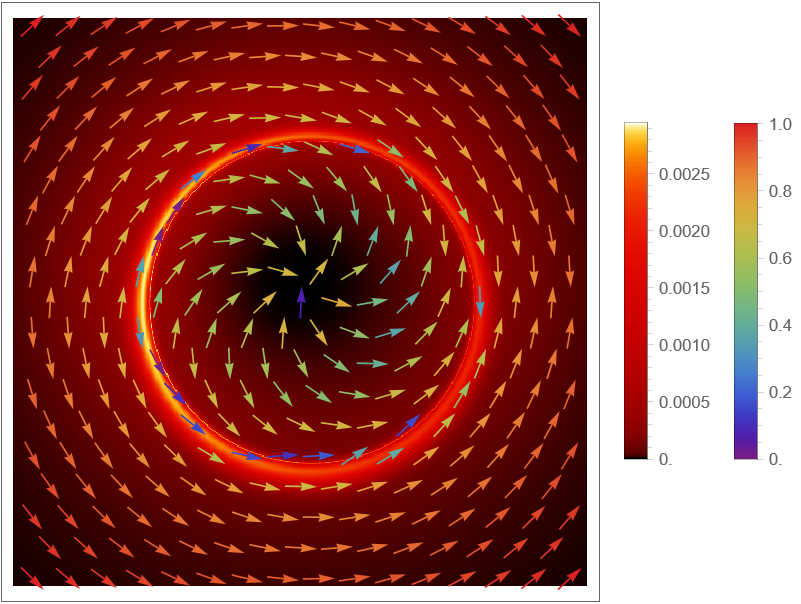}}
	\subfigure[$Q^2=1.0,\theta=17^\circ$]{\includegraphics[scale=0.3]{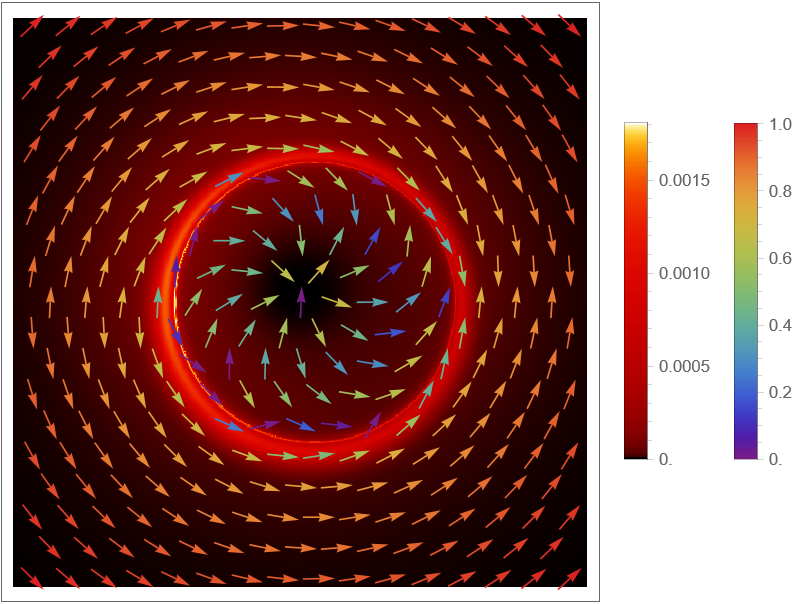}}
	
	\subfigure[$Q^2=0.1,\theta=60^\circ$]{\includegraphics[scale=0.3]{theta=60du,a=0.5,ll.png}}
	\subfigure[$Q^2=0.5,\theta=60^\circ$]{\includegraphics[scale=0.3]{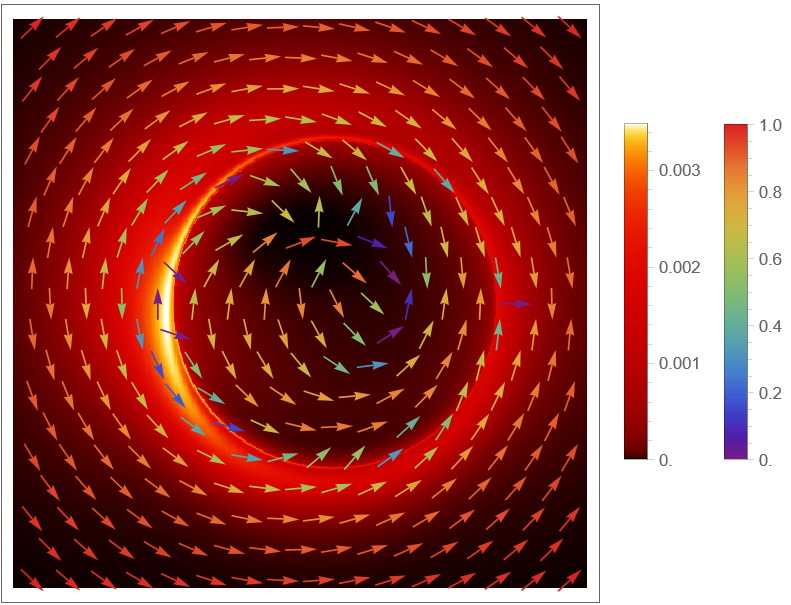}}
	\subfigure[$Q^2=1.0,\theta=60^\circ$]{\includegraphics[scale=0.3]{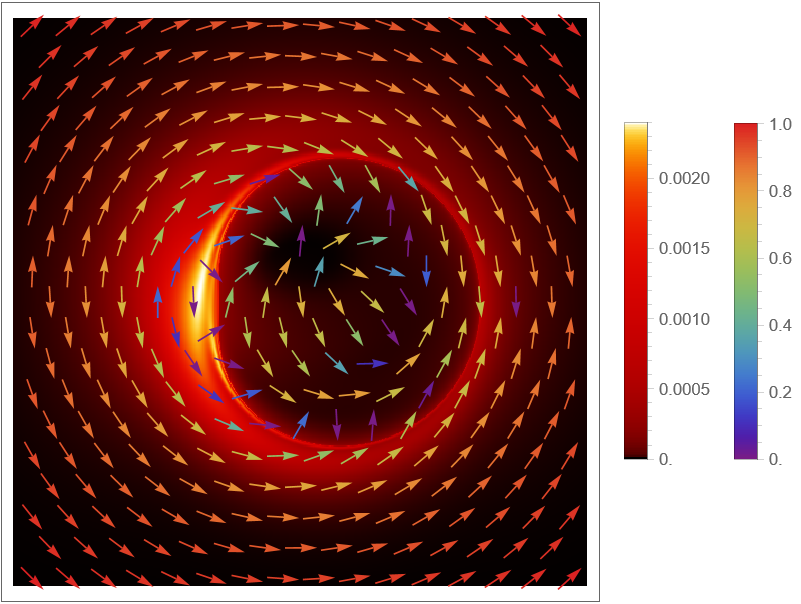}}
	
	\subfigure[$Q^2=0.1,\theta=85^\circ$]{\includegraphics[scale=0.3]{theta=85du,a=0.5,ll.png}}
	\subfigure[$Q^2=0.5,\theta=85^\circ$]{\includegraphics[scale=0.3]{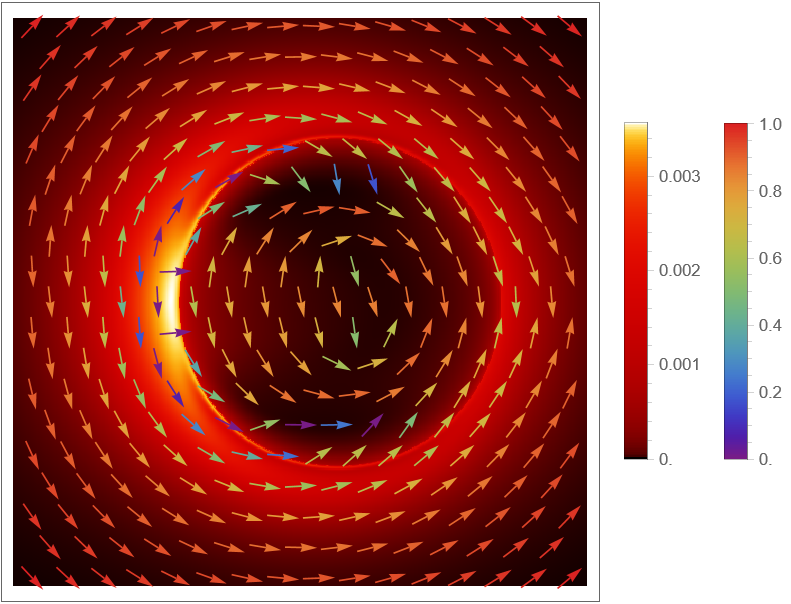}}
	\subfigure[$Q^2=1.0,\theta=85^\circ$]{\includegraphics[scale=0.3]{theta=85du,a=0.5,Q2=1.0,ll.png}}
	
	\caption{Polarized black hole shadow images for the BAAF model. The accretion flow follows the conical solution, with an observation frequency of $230~\mathrm{GHz}$ and $a = 0.5$.}
\end{figure}

Figure~15 illustrates the effects of the black hole charge $Q$ on the polarization images. As $Q$ increases, both the intensity $\mathcal{I}_o$ near the higher-order images and the size of the higher-order images decrease. The polarization patterns exhibit significant differences for different values of $Q$, reflecting the influence of the underlying spacetime structure on the polarization properties.

\section{Summary and discussion}\label{sec7}
In this work, we investigate the optical   images  of a rotating Kerr--Sen black hole surrounded by a geometrically thick accretion flow. We first review the fundamental properties of the Kerr--Sen spacetime, focusing on the event horizon and null geodesics, and provide the definition of the photon sphere, and then summarize two representative geometrically thick accretion flow models, namely the phenomenological RIAF model and the analytical BAAF model. By numerically solving the geodesic equations and the radiative transfer equations, we obtain the corresponding shadows images and polarization structures.

For the RIAF model, we consider the isotropic and anisotropic radiation respectivly. In the isotropic case, the accretion flow follows the ballistic approximation, i.e., the fluid moves along geodesics, and the observation frequency is fixed at $230~\mathrm{GHz}$. The numerical results show that increasing the spin parameter $a$ enhances the frame dragging effect, leading to a pronounced left right asymmetry in the intensity distribution. In contrast, increasing the black hole charge $Q$ reduces the overall size of the photon ring and the central dark region, while having little effect on their shape. Unlike geometrically thin accretion disks, for large observer inclination $\theta$, the inner shadow in thick disk models is partially obscured by radiation from regions away from the equatorial plane, resulting in a splitting of the central dark region into two parts. We also find that the effects of $a$, $Q$, and $\theta$ on anisotropic radiation images are qualitatively similar to those in the isotropic case. However, under anisotropic radiation, the highe order images exhibit enhanced intensity near the polar regions, which arises from the angular dependence of the synchrotron emissivity.

For the BAAF model, the bright ring is typically narrower, and the separation between the primary and higher order images is more pronounced. At large inclinations, the higher-order images do not exhibit the two distinct dark regions seen in the RIAF model. This difference can be attributed to the fact that, for the chosen parameters, the BAAF flow treated under the conical approximation is structurally thinner than the corresponding RIAF configuration. For the polarization images of the BAAF model, we find that both the spin parameter $a$ and the black hole charge $Q$ significantly affect the magnitude and orientation of the polarization vectors. This indicates that the polarization properties of a rotating Kerr-Sen black hole can effectively probe the underlying spacetime structure. In addition, compared with thin disk models, the influence of gravitational lensing on radiation from regions away from the equatorial plane leads to a broader distribution of polarization vectors across the image plane, resulting in more complex polarization patterns in thick-disk models.

Overall, this work demonstrates that the main physical parameters of a black hole have significant effects on its shadow features in geometrically thick and optically thin accretion flow models. Such models provide a more realistic description of astrophysical environments than geometrically thin disks. By combining intensity and polarization imaging, one can obtain a more comprehensive characterization of the radiation properties and spacetime structure around black holes. In future studies, it would be worthwhile to compare the imaging signatures of black holes with those of other compact objects, such as neutron stars and boson stars, in order to explore observable differences among different gravitational sources. These efforts may provide valuable theoretical guidance for forthcoming high resolution astronomical observations.


\cleardoublepage

\vspace{10pt}
\noindent {\bf Acknowledgments}

\noindent
This work is supported by the National Natural Science Foundation of China (Grants Nos. 12375043, 12575069 ), and Chongqing Normal University Fund Project (Grants No. 26XLB001).

\bibliographystyle{utphys} 
\bibliography{biblio} 
	
\end{document}